\documentclass[11pt]{article}
\usepackage{amsmath, xypic, graphicx}
\usepackage{stmaryrd,xspace,amssymb}
\usepackage{url}

\newcommand{\downset}{\ensuremath{\mathop{\downarrow\!}}}
\newcommand{\IR}{\mathbb{IR}}
\newcommand{\Q}{\mathbb{Q}}
\newcommand{\uIR}{\underline{\mathbb{IR}}}

\newcommand{\drie}{\vartriangleleft}

\newcommand{\coveredd}{\ensuremath{\mathop{\blacktriangleleft}}}

\newcommand{\assign}{:=}
\newcommand{\beq}{\begin{equation}}
\newcommand{\eeq}{\end{equation}}
\newcommand{\bea}{\begin{eqnarray}}
\newcommand{\eea}{\end{eqnarray}} \newcommand{\nn}{\nonumber}

\newcommand{\Sets}{\mbox{\textbf{Sets}}}
\newcommand{\sr}{\stackrel}

\newcommand{\ca}{C*-algebra}

 \newcommand{\ovl}{\overline}
 
\newcommand{\raw}{\rightarrow}

 \newcommand{\Raw}{\Rightarrow}
 \newcommand{\Law}{\Leftarrow}

\newcommand{\LRaw}{\Leftrightarrow}

 \newcommand{\wed}{\wedge}

\newcommand{\x}{\times} 
 
\newcommand{\Tr}{\mbox{\rm Tr}\,}

\newcommand{\inv}{^{-1}}

\newcommand{\er}{\eqref}
\newcommand{\al}{\alpha} 
 \newcommand{\Gm}{\Gamma}
\newcommand{\dl}{\delta} \newcommand{\Dl}{\Delta}
\newcommand{\ep}{\epsilon} \newcommand{\varep}{\varepsilon}

\newcommand{\lm}{\lambda} \newcommand{\Lm}{\Lambda}
\newcommand{\rh}{\rho} \newcommand{\sg}{\sigma}
\newcommand{\Sg}{\Sigma}  
 \newcommand{\phv}{\varphi}
\newcommand{\ch}{\chi} \newcommand{\ps}{\psi} \newcommand{\Ps}{\Psi}
\newcommand{\om}{\omega} \newcommand{\Om}{\Omega}
\newcommand{\CA}{{\mathcal A}} 
 \newcommand{\CF}{{\mathcal F}}
 \newcommand{\CI}{{\mathcal I}}
   \newcommand{\CL}{{\mathcal L}}

\newcommand{\CO}{{\mathcal O}} \newcommand{\CP}{{\mathcal P}}
 
\newcommand{\CT}{{\mathcal T}} \newcommand{\CV}{{\mathcal V}}
 
\newcommand{\C}{{\mathbb C}} 
 
 \newcommand{\R}{{\mathbb R}}
\newcommand{\T}{{\mathbb T}} 
\newcommand{\Hom}{\mbox{\rm Hom}}

\newcommand{\alg}[1]{\ensuremath{#1}}
\newcommand{\functor}[1]{\ensuremath{\underline{#1}}}
\newcommand{\after}{\circ}
\newcommand{\cat}[1]{\ensuremath{\mathbf{#1}}}
\newcommand{\Cat}[1]{\ensuremath{\mathrm{\textbf{#1}}}}

\newcommand{\id}[1]{\ensuremath{\mathrm{id}}}
\newcommand{\op}{\ensuremath{^{\mathrm{op}}}}
\newcommand{\Cstar}{\Cat{CStar}\xspace}
\newcommand{\Ccstar}{\Cat{cCStar}\xspace}

\newcommand{\Set}{\Cat{Sets}\xspace}
\newcommand{\Sh}{\ensuremath{\mathrm{Sh}}}
\newcommand{\cpxrat}{\ensuremath{\field{C}_\field{Q}}}

\newcommand{\context}{\ensuremath{\mathcal{C}}}

\newcommand{\asstopos}{\ensuremath{\mathcal{T}}}
\newcommand{\interpretation}[1]{\ensuremath{\llbracket{#1}\rrbracket}}
\newcommand{\Mtwo}{\ensuremath{M_2}(\field{C})}

\newcommand{\sa}{\ensuremath{_{\mathrm{sa}}}}

\newcommand{\prop}[1]{\ensuremath{\mbox{\texttt{#1}}}}

\newcommand{\field}[1]{\ensuremath{\mathbb{#1}}}
\newcommand{\norm}[2][]{\ensuremath{\|#2\|_{#1}}}

\newcommand{\opens}{\ensuremath{\mathcal{O}}}
\newcommand{\uS}{\underline{\Sigma}}
\newcommand{\uA}{\underline{A}}

\renewcommand{\CA}{\mathcal{C}(A)}
\newcommand{\TA}{\mathcal{T}(A)}
\newcommand{\ie}{\textit{i.e.}}
\newcommand{\eg}{\textit{e.g.}}
\newcommand{\ulS}{\functor{\Sigma}}
\newcommand{\ulR}{\underline{\mathbb{R}}}
\newcommand{\ulA}{\underline{A}}
\renewcommand{\TA}{\asstopos(\alg{A})}
\renewcommand{\CA}{\context(\alg{A})}
\newtheorem{theorem}{Theorem}
\newtheorem{lemma}[theorem]{Lemma}
\newtheorem{proposition}[theorem]{Proposition}
\newtheorem{corollary}[theorem]{Corollary}
\newtheorem{definition}[theorem]{Definition}

\newenvironment{proof}[1][Proof]%
{ \begin{trivlist}%
  \item[\hskip \labelsep {\bfseries #1}]%
}%
{ \end{trivlist}%
}
\newcommand{\qed}{\nobreak\hfill$\Box$}

\topmargin = - 0.5 cm
\newcommand{\turndown}[1]{%
  \rotatebox[origin=c]{270}{\ensuremath#1}}
\newcommand{\twoheaddownarrow}{\turndown{\twoheadrightarrow}}
\begin{document}
\title{A topos for algebraic quantum theory}
\author{Chris Heunen\footnote{Radboud Universiteit Nijmegen,
    Institute for Mathematics, Astrophysics, and Particle Physics,
    Toernooiveld 1, 6525 ED NIJMEGEN, THE NETHERLANDS.
    Supported by N.W.O.} \and
  Nicolaas P. Landsman\footnote{Radboud Universiteit Nijmegen,
    Institute for Mathematics, Astrophysics, and Particle Physics,
    Toernooiveld 1, 6525 ED NIJMEGEN, THE NETHERLANDS.} \and
  Bas Spitters\footnote{Eindhoven University of Technology,
    Department of Mathematics and Computer Science, P.O. Box 513, 5600
    MB EINDHOVEN, THE NETHERLANDS. Supported by N.W.O.}}
\maketitle
\vspace*{-0.75cm}
\begin{center}{\it
Dedicated to Ieke Moerdijk, at his 50th birthday}
\end{center}
\smallskip
\begin{abstract}
  The aim of this paper is to relate algebraic quantum mechanics to
  topos theory, so as to construct new foundations for quantum logic
  and quantum spaces. Motivated by Bohr's idea that the empirical
  content of quantum physics is accessible only through classical
  physics, we show how a {\it noncommutative} C*-algebra of observables $A$ induces a
  topos $\TA$ in which the amalgamation of all of its commutative
  subalgebras comprises a single {\it commutative} C*-algebra
  $\uA$. According to the constructive Gelfand duality theorem of
  Banaschewski and Mulvey, the latter has an internal spectrum
  $\ulS(\uA)$ in $\TA$, which in our approach plays the role of the
  quantum phase space  of the system. Thus we associate a locale
  (which is the topos-theoretical notion of a space and which
  intrinsically carries  the intuitionistic logical structure of a
  Heyting algebra) to a  C*-algebra (which is the noncommutative
  notion of a space). In this setting,  states on $A$ become
  probability measures (more precisely, valuations) on $\uS$, and
  self-adjoint elements of $A$ define continuous functions  (more
  precisely, locale maps) from  $\uS$ to Scott's interval
  domain. Noting that open subsets of  $\ulS(\uA)$ correspond to
  propositions about the system, the pairing map that assigns a
  (generalized) truth value to a state and a proposition assumes an
  extremely simple categorical form. Formulated in this way, the
  quantum theory defined by $A$ is essentially turned into a classical
  theory, internal to the topos $\TA$.

  These results were inspired by the topos-theoretic approach to
  quantum physics proposed by Butterfield and Isham, as recently generalized by D\"{o}ring and Isham.
\end{abstract}
\begin{center}\textbf{Motto}
\begin{quote}
  `Ces ``nuages probabilistes'', rempla\c{c}ant les rassurantes
  particules mat{\'e}rielles d'antan, me rappellent {\'e}trangement
  les {\'e}lusifs ``voisinages ouverts'' qui peuplent les topos, tels
  des fant{\^o}mes {\'e}vanescents, pour entourer des ``points''
  imaginaires.'
  (A. Grothendieck~\cite{grothendieck:rs})\footnote{`These
  ``probability clouds'', replacing the reassuring material particles
  of before, remind me strangely of the elusive ``open neighborhoods''
  that populate the topoi, like evanescent phantoms, to surround the
  imaginary ``points''.'}
\end{quote}
\end{center}
\newpage

\section{Introduction}
This introduction is intended for both mathematical physicists and topos theorists. We apologize in advance for
stating the obvious for one or the other of these groups at various points, but we hope that most of it is interesting to both communities.
\subsection{The logic of classical physics}\label{lcp}
In classical mechanics, the logical structure of a physical system is encoded in its phase space $M$.
Indeed, elementary propositions (also called `yes-no questions' in physics) correspond to suitable subsets of $M$
(such as all sets, all measurable sets or all regular open sets),  and the logical connectives
are given by the standard set-theoretic operations. This makes the
logic of the system `spatial'; its realization as a Boolean algebra
(\ie\ a distributive lattice with 0 and 1 in which every element has a
complement) confirms the `classical' nature of the situation \cite{Varadarajan}.

Physicists do not usually describe a system in the above way. Instead, they work with
observables $a:M\raw\R$, like position or energy. (Such functions might be arbitrary, or else required to be measurable or continuous as appropriate.)
 From that perspective, the  elementary propositions  are of the form $a\in \Delta$, where  $\Delta\subseteq \R$ is a (arbitrary, measurable, or regular open) subset of the reals.

 Either way, a state of the system may be construed as a catalogue of answers
 to all yes-no questions about the systems. We concentrate on pure states
  $\rh\in M$, which provide sharp (as opposed to probabilistic)
  answers. In the first description, a proposition $U\subseteq M$ is
  true (equivalently, the answer to the corresponding question is
  `yes') iff $\rh\in U$. In the second description, $a\in\Dl$ is true
  for a state $\rh$ iff $a(\rh)\in\Dl$, \ie\ iff $\rh\in a\inv(\Dl)$.
 Thus propositions of the second type fall into equivalence classes
 $[a\in \Delta]=a\inv(\Dl)$. As these are subsets of $M$, this leads us back to the purely spatial picture of the first description.

This truth assignment has a very simple categorical description. We regard
$M$ as an object in the category \Sets\ of all sets as objects and all
functions as arrows, and interpret $\rho$ as an arrow (in fact, a monomorphism)
$\xymatrix@1{1 \ar^-{\rho}[r] & M}$, where 1 is any singleton.
A subset $U\subseteq M$ may
alternatively be described by its characteristic function
$\ch_U:M\raw \{0,1\}$; anticipating the convention  in topos
theory, we relabel $\{0,1\}$ as $\Om$ and regard this as an
object in \Sets. Composition of  $\xymatrix@1{1\ar^-{\rho}[r]&M}$ and
$\xymatrix@1{M\ar^-{\ch_{a\inv(\Dl)}}[rr]&&\Om}$ then yields an arrow
\beq
  \xymatrix@1{1 \ar^-{\langle \rho,a\in \Dl\rangle}[rr] && \Om}
  \qquad=\qquad
  \xymatrix@1{1 \ar^-{\rho}[r] & M \ar^-{\ch_{a\inv(\Dl)}}[rr] && \Om,}
\eeq
\ie\ we have defined
\beq
\langle \rho,a\in \Dl\rangle=\ch_{a\inv(\Dl)}\circ\rho.\label{circrho}\eeq
The image of $1$ under this map is
a point of $\Om$, which is precisely the above truth value of
the proposition $a\in \Dl$ in the state $\rho$ (provided we identify
$\{0,1\}$ with $\{\mathrm{false, true}\})$.

It is important for what follows to reformulate this description in terms of
the topology $\CO(M)$ of $M$ (\ie\ its collection of opens) instead of $M$ itself.  This makes sense if the subsets $U\subseteq M$ above are open, which in our second description is the case if the observables $a$ are continuous and the value sets $\Dl\subseteq\R$ are open as well.  Hence $a\inv(\Dl)$ is an arrow
$\xymatrix@1{1\ar^-{a^{-1}(\Dl)}[rr] && \CO(M)}$ in \Sets, but $\rho\in M$ is now
represented by the `state subobject' \hbox{$[\rho]\subseteq\CO(M)$} given by
\beq
   [\rho]=\{V\in\CO(M)\mid \rho\in V\} =\{V\in\CO(M)\mid \dl_{\rho}(V)=1\},
   \label{old3}
\eeq
where $\dl_{\rho}$ is the Dirac measure on $M$ concentrated at $\rho$.
We describe this object by its characteristic function
$\ch_{[\rho]}:\CO(M)\raw\Om$. The pairing map then becomes
\beq
   \xymatrix@1{1\ar^-{\langle a\in \Dl,\rho \rangle}[rr]&&\Om}
   \qquad =\qquad
   \xymatrix@1{1\ar^-{a^{-1}(\Dl)}[rr] && \CO(M)
   \ar^-{\chi_{[\rho]}}[rr]&&\Om,} \label{spp}
\eeq
or, in other words,
\beq
\langle a\in \Dl,\rho \rangle=\chi_{[\rho]}\circ a^{-1}(\Dl).\label{circainv}\eeq
The reader may verify that $\langle a\in \Dl,\rho \rangle=\langle
\rho,a\in \Dl \rangle$, so that our second categorical description of the
state-proposition pairing is equivalent to the first. More generally, if $\mu$ is a probability measure on $M$, we might define a state object $[\mu]$ by replacing the Dirac measure $\dl_{\rh}$ in \er{old3} by $\mu$, \ie
\beq
   [\mu] =\{V\in\CO(M)\mid \mu(V)=1\}.
   \label{old3bis}
\eeq
In physics, $\mu$ plays the role of a {\it mixed state} (unless it is a point measure, in which case it happens to be pure). Like the pure state $\rh$ (or rather its associated probability measure
$\dl_{\rh}$), the mixed state $\mu$ defines a characteristic function $\ch_{[\mu]}:\CO(M)\raw\Om$. The latter, however, turns out
not to share the attractive logical properties of $\ch_{[\rh]}\equiv \ch_{\dl{\rh}}$ (unless $\mu$ is pure); see Subsection \ref{subsection:locale3}.

\subsection{Spatial quantum logic}\label{subsec:SQL}
The goal of this paper is to generalize this  situation to
quantum mechanics. In particular, we wish to find a spatial notion of
quantum logic.
This objective will be accomplished by:
\begin{enumerate}
\item  Identifying an appropriate notion of a quantum phase `space' $\Sg$.
\item Defining suitable  `subsets' of $\Sg$  that act as  elementary logical propositions of quantum mechanics.
\item Describing observables and states in terms of $\Sg$.
\item Associating a proposition $a\in\Dl$ (and hence a
`subset' $[a\in\Dl]$ of $\Sg$) to an observable $a$ and an open subset $\Dl\subseteq\R$.
\item Finding a  pairing map between pure states and `subsets' of $\Sg$ (and hence between states and propositions of the type $a\in\Dl$).
\end{enumerate}
In the last step, a state assigns a particular truth value to a given proposition; this is supposed to  give empirical content to the formalism.
The codomain $\Om$ of the pairing map in item 5, which may be called the `truth object' of the theory,
is by no means obvious and identifying it is explicitly part of the question.
Certainly, 20th century physics shows that
the `classical' choice $\Om=\{0,1\}$ is out of the question.

The formulation of these objectives and the associated program goes back to
von Neumann, who also famously proposed the following extremely
elegant solution:
\begin{enumerate}
\item A quantum phase space is a Hilbert space $H$.
\item Elementary propositions correspond to closed linear subspaces of $H$.
\item Observables are selfadjoint operators on $H$ and pure states are unit vectors in $H$.
\item  The closed linear subspace $[a\in\Dl]$  is the image $E(\Dl)H$ of
the spectral projection $E(\Dl)$ defined by  $a$ and $\Dl$ (provided the latter is measurable).
\item The pairing map takes values in $[0,1]$ and is given by the ``Born rule''
 $\langle\Psi,a\in\Dl\rangle=(\Psi,E(\Dl)\Psi)$.
\end{enumerate}

Thus subsets of phase space became closed linear subspaces of Hilbert space,
 which, as Birkhoff and von Neumann \cite{birkhoffvonneumann36}
noticed,  form a  lattice $\CL(H)$
under inclusion as partial order. However, this lattice fails to be Boolean, basically because it is nondistributive.
Nonetheless, Birkhoff and von Neumann interpreted the lattice
operations $\wedge$ and $\vee$  as `and' and `or', as in the
classical case, and argued that the departure from the Boolean
structure (and hence from classical logic) meant that one had to deal
with a new kind of logic, which they aptly called {\it quantum
logic}. This looked highly innovative, but on the other hand it
conservatively preserved the spatial nature of the logic of classical
physics, in that the logical structure of propositions is still
associated with the spatial (\ie\ Hilbert {\it space}) structure of the
theory.

Attractive and revolutionary as this spatial quantum `logic' may appear
\cite{Varadarajan,kalmbach1,kalmbach2,redei}, it
faces severe problems. The main {\it logical} drawbacks are:
\begin{itemize}
\item Due to its lack of distributivity, quantum
`logic' is difficult to interpret as a logical structure.
\item In particular, despite various proposals no satisfactory implication operator has been found (so that there is no deductive system in quantum logic).
\item Quantum `logic' is a propositional language;
no satisfactory generalization to  predicate logic has been found.
\end{itemize}

Quantum logic is also problematic from
a {\it physical} perspective. Since (by various
theorems \cite{Bub} and wide agreement) quantum probabilities do not
admit an ignorance interpretation,  $[0,1]$-valued truth values
attributed to propositions by pure states via the Born rule cannot be
regarded as sharp (\ie\ $\{0,1\}$-valued) truth values muddled by human
ignorance. This implies that, if $x=[a\in\Dl]$ represents  a
quantum-mechanical proposition, it is wrong to say that either $x$ or
its negation holds, but we just do not {\it know} which of these
alternatives applies. However, in quantum logic one has
the law of the excluded middle in the form $x\vee x^{\perp}=1$
for all $x$.
Thus the formalism of quantum logic does not match the probabilistic
structure of quantum theory responsible for its empirical
content.

In fact, the above argument suggests
that it is  {\it intuitionistic} logic rather than quantum logic that
is relevant in quantum mechanics
(\textit{cf.}~\cite{AdelmanCorbett}). More generally, as argued in
particular by Butterfield and  Isham \cite{ButVat,butterfieldisham1}, the fact that pure states fail
to define truth assignments in the usual binary sense (\ie\ true or
false)  renders the entire notion of truth in quantum mechanics
obscure and calls for a complete reanalysis thereof
\cite{doringisham1,doringisham2,doringisham3,doringisham4,doeringisham:review}.
 As also probably first recognized by the same authors,
 such an analysis can fruitfully be attempted using topos theory,
 whose internal logic is indeed intuitionistic.

 From our perspective, another reason why topos theory offers itself on a silver tray in our search for a spatial quantum logic lies
in the  interplay between spatial and logical structures inherent in topos theory, as exemplified by
the opening words of the renowned textbook by Mac Lane and Moerdijk:
\begin{quote}
A startling aspect of topos theory is that it unifies two seemingly wholly distinct mathematical subjects: on the one hand, topology and algebraic geometry and on the other hand, logic and set theory.
\end{quote}

We refer to
\cite{goldblatt84,maclanemoerdijk92,johnstone02a,johnstone02b}  for  accounts of topos theory; see also \cite{bell05,McLarty,kroemer07} for historical details.
Briefly, a topos is a category in which one can essentially reason as in the
category \Sets\ of all sets (with functions as arrows), {\it except for the fact that the logic is intuitionistic and the axiom of choice is generally not available}.  Briefly, the mathematics underlying topos theory is  {\it constructive}.

Specifically,
a topos is a category with the following ingredients:
\begin{enumerate}
\item {\it Terminal object.} This is an object called 1 (unique up to
  isomorphism) such that for each object $A$ there is a unique arrow
  $A\raw 1$, generalizing the singleton set in the category \Sets.
\item {\it Pullbacks.} These generalize the fibered product $B\x_A
  C=\{(b,c)\in B\x C\mid f(b)=g(c)\}$  of $\xymatrix@1{B\ar^-{f}[r]&A}$ and
  $\xymatrix@1{C\ar^-{g}[r]&A}$ in \Sets\ into a pullback square with
  appropriate universality property.  Cartesian products are a special
  case.
\item {\it Exponentials.} These generalize the idea that the class
  $B^A$ of functions from a set $A$ to a set $B$ is itself a set, and
  hence an object in \Sets, equipped with the evaluation map
  $\mathrm{ev}:A\x B^A\raw B$.
\item {\it Subobject classifier.} This generalizes the idea that one
  may characterize a subset $A\subseteq B$ by its characteristic
  function $\ch_A:B\raw \{0,1\}$. Subsets generalize to subobjects,
  \ie\ monic (``injective") arrows $A \rightarrowtail B$, and in a
  topos there exists an object $\Om$ (the subobject classifier) with
  associated arrow $\xymatrix@1{1\ar^-{\top}[r]&\Om}$ (``truth'') such
  that for any subobject $A \rightarrowtail B$ there is a {\it unique}
  arrow $\xymatrix@1{B\ar^-{\chi_A}[r]&\Om}$ for which
  $\xymatrix@1{B&A\ar_-{f}[l]\ar[r]&1}$ is a pullback of
  $\xymatrix@1{B\ar^-{\chi_A}[r]&\Om}$ and
  $\xymatrix@1{1\ar^-{\top}[r]&\Omega}$. Conversely, given any arrow
  $\xymatrix@1{B\ar^-{\ch}[r]&\Om}$ there exists a subobject $A
  \rightarrowtail B$
  of $B$ (unique up to isomorphism) whose classifying arrow $\ch_B$
  equals $\ch$. The subobject classifier in a topos play the role of a
  ``multi-valued truth object'', generalizing the simple situation in
  \Sets, where  $\Om=\{0,1\}=\{\mathrm{false, true}\})$;
  see~\er{circrho} and subsequent text.
\end{enumerate}

We assume that our topoi are cocomplete and have a natural numbers object.
\subsection{Generalized notions of space}\label{subsection:locale1}
Our first objective in the list at the beginning of Subsection \ref{subsec:SQL}, \ie\ the identification of an appropriate notion of a quantum phase `space',
will be met by a combination of two profound notions of generalized space
that have been around for some time.
\begin{enumerate}
\item
First, let us recall the strategy of
noncommutative geometry  \cite{Connes,ConnesMarcolli}.
 One starts
with the replacement of a compact topological space $X$  by the
associated algebra of complex-valued continuous functions $C(X,\C)$. If
$X$ fails to be Hausdorff, this step loses information, but if it is,
one may recover $X$ from the commutative C*-algebra $C(X,\C)$ as its
Gelfand spectrum. This yields a duality between the category
of compact Hausdorff spaces and the category of unital commutative
C*-algebras: nothing is lost, but nothing is gained either by
abstracting spaces as commutative C*-algebras. The thrust of noncommutative geometry, then, is to allow C*-algebras to be noncommutative
without losing the spatial perspective. That this can be done is
impressive enough, but as the logical situation is obscured by moving from
commutative to noncommutative C*-algebras, further ideas are needed (at least if one is interested in quantum logic).
\item
A second  approach to generalizing  topological spaces
 would be to replace $X$  by its topology $\CO(X)$. This has a natural lattice structure under inclusion, and in fact defines a highly structured kind of lattice known as a
{\it frame}. This is  a complete distributive lattice such that
$x\wedge \bigvee_{\lambda}y_{\lambda}=\bigvee_{\lambda}x\wedge
y_{\lambda}$ for arbitrary families $\{y_{\lambda}\}$ (and not just for finite ones, in which case the said property follows from the definition of a distributive lattice).  For  example, if $X$ is a topological space, then the
topology $\CO(X)$ of $X$ is a frame with $U\leqslant V$ if $U\subseteq V$.
A {\it frame homomorphism} preserves finite meets and arbitrary
joins; this leads to the category $\mathbf{Frm}$  of frames and frame homomorphisms.
\end{enumerate}
Abstracting frames  $\CO(X)$ coming from a topological space  to general frames  is a genuine generalization of the concept of a space, as plenty of frames exist that are not of
the form $\CO(X)$. A simple example is the frame $\CO_{\mathrm{reg}}(\R)$ of {\it regular open subsets} of $\R$, \ie\ of open subsets $U$ with the property $\neg\neg U=U$, where $\neg U$ is the interior of the complement of $U$.
This may be contrasted with the situation for unital commutative
C*-algebras, which, as just recalled, are all  of the form $C(X)$. Moreover, far from {\it obscuring} the logical structure of space, the generalization of spaces by frames  rather {\it explains} and {\it deepens}  this structure.

Indeed, a frame is a  {\it complete Heyting algebra}, with its intrinsic
structure of an intuitionistic propositional logic.
Here a  Heyting algebra
is a distributive lattice $\CL$ with a map $\raw:\CL\x\CL\raw\CL$
satisfying $x\leqslant (y\raw z)$
iff $x\wed y\leqslant z$, called implication  \cite{goldblatt84,maclanemoerdijk92,Vic:LocTopSp}.  Every Boolean algebra is
a Heyting algebra, but not {\it vice versa}; in fact, a Heyting
algebra is Boolean iff $\neg\neg x=x$ for all $x$, which is the case
iff $\neg x\vee x=\top$  for all $x$. Here negation is a derived
notion, defined by $\neg x=(x\raw\perp)$. For example,  $\CO_{\mathrm{reg}}(\R)$ is Boolean, but $\CO(\R)$ is not.
In general, the elements of a Heyting algebra
form an  intuitionistic propositional logic under the usual logical  interpretation of the
lattice operations.

A  Heyting algebra is {\it complete} when arbitrary
joins (\ie\ sups) and meets (\ie\ infs) exist.
A complete Heyting algebra is essentially the same thing as a
frame, for
 in a frame one may define $y\raw z=\bigvee\{x\mid x\wed y\leqslant z\}$.
Conversely, the infinite distributivity law in a frame is
automatically satisfied in a Heyting algebra.
The set of subobjects of a given object in a
topos forms a complete Heyting algebra (as long as the topos in question is defined ``internal to \Sets''), generalizing the fact that the set of
subsets of a given set is a Boolean algebra. The subobject classifier
of such a topos is a complete Heyting algebra as well; in fact, these two statements
are equivalent.  (Note, however, frame maps do not necessarily preserve the implication
$\raw$ defining the Heyting algebra structure, as can already be seen in
examples of the type $f\inv:\CO(Y)\raw \CO(X)$, where $f:X\raw Y$ is
continuous \cite{maclanemoerdijk92}. Consequently, negation may not be preserved by frame maps either.)

The category  $\mathbf{Loc}$ of {\it locales} is the opposite category to  $\mathbf{Frm}$, \ie\ it has the same objects but all arrows go in the opposite direction.
Some topos theorists write $X$ for a locale and $\CO X$  or $\CO(X)$ for the
same object seen as a frame  \cite{johnstone82,maclanemoerdijk92,Vic:LocTopSp}.
Apart from the already unfortunate fact that this notation is applied also when $\CO(X)$ does not stand for the opens of a space $X$ but denotes a general frame,
it fails  to distinguish between
a topological space $X$ and the associated locale (\ie\ the frame $\CO(X)$ seen as a locale).
Nonetheless, this notation often leads to elegant expressions and we will heavily use it.

If $X$ and $Y$ are spaces, a continuous
map $f:X\raw Y$ induces a frame map $f\inv:  \CO(Y)\raw \CO(X)$ and hence an arrow
$\CO(X)\raw \CO(Y)$ in $\mathbf{Loc}$, simply defined as
$f\inv$ {\it read in the opposite direction}. We write the latter arrow in $\mathbf{Loc}$
simply as $f:X\raw Y$. In general, an arrow in $\mathbf{Frm}$ is written
as $f\inv:  \CO(Y)\raw \CO(X)$ (whether or not the frames in question come from topological spaces and if so, whether or not $f\inv$ is indeed the pullback of a continuous function between these spaces), and the corresponding arrow in $\mathbf{Loc}$ is denoted by $f:X\raw Y$.
 Similarly,  we will write $C(X,Y)$
 for $\mathrm{Hom}_{\mathbf{Loc}}(X,Y)=\mathrm{Hom}_{\mathbf{Frm}}(\CO(Y),\CO(X))$.
 In particular, for a locale $X$, $C(X,\C)$
 will denote the set of frame maps $\CO(\C)\raw \CO(X)$.
 \subsection{Points and opens of locales}\label{subsection:locale2}
An element of a set $X$ (and hence {\it a fortiori} also a point of a topological space $X$)
may be identified with an arrow $*\raw X$, where $*$ is a given singleton (for simplicity we write $*$ instead of the more usual $\{*\}$). The same goes for locales $X$, so that by definition a {\it point} of a locale $X$ is  a locale map  $p:*\raw X$, hence a frame map
$p\inv: \CO(X)\raw\CO(*)\cong\{0,1\}=\Om$; recall that the subobject classifier in \Sets, seen as a topos, is $\Om=\{0,1\}$ and note that $*$ is precisely the locale associated to $\Om$, as our notation $\Om=\CO(*)$ has indicated.

A point of a locale $X$ being defined as a locale map $*\raw X$ or as the corresponding
frame map $\CO(X)\raw\Omega$, an {\it open} of $X$ is defined as a  locale map $X\raw S$, where $S$ is the locale defined by the so-called {\it Sierpinski space}, \ie\ $\{0,1\}$ with $\{1\}$ as the only open point.
The corresponding frame map $\CO(S)\raw \CO(X)$ is determined by its  value at $1$ (since $\emptyset\mapsto\emptyset$ and $\{0,1\}\mapsto X$), so that we may simply look at opens in $X$ as arrows
 $1\raw \CO(X)$ (where the singleton $1$ is seen as the terminal object in $\Sets$).   Clearly, if $X$ is a genuine topological space with associated frame $\CO(X)$ of opens, then
each such map $1\raw \CO(X)$ corresponds to an open subset of $X$ in the usual
sense. Using this concept, the set $\mathrm{Pt}(X)$ of points of a locale $X$
may be topologized in a natural way, by declaring its opens to be
the sets of the form
\beq
\mathrm{Pt}(U)=\{p\in \mathrm{Pt}(X)\mid p\inv(U)=1\}, \label{openpt}\eeq
 where  $U\in \CO(X)$ is some open.  We say that a locale $X$ is {\it spatial}
if it is isomorphic (in the category of locales) to
$\mathrm{Pt}(X)$ (more precisely, to the locale associated to the frame
$\CO(\mathrm{Pt}(X))$ in the above topology). Conversely, a topological space
$X$ is called {\it sober} if it is homeomorphic to $\mathrm{Pt}(X)$
 (which, with the notation introduced above, really stands for the space of
points of the locale associated to the frame $\CO(X)$). It is useful to know
that $X$ is sober when it is Hausdorff. If $X$ is sober, any frame map
$\phv:\CO(Y)\raw \CO(X)$ is induced by a continuous map $f:X\raw Y$ as
$\phv=f\inv$. This provides additional justification for the notation $f\inv:
\CO(Y)\raw \CO(X)$ for a general frame map, and
$f:X\raw Y$ for the associated locale map. See \cite[\S IX.3]{maclanemoerdijk92}
for a very clear exposition of all this.

For example, referring to Subsection \ref{lcp}, the characteristic function
$\ch_{[\rho]}:\CO(M)\raw\Om$ introduced below \er{old3} is easily checked to
define a frame map. Renaming this map
as $\ch_{[\rho]}\equiv\rh\inv$, the associated locale map $\rh: *\raw M$ is
therefore a point of the locale $M$ in the above sense. In this special case,
such a point may also be described by an arrow $1\raw M$, where 1 is the
terminal object in \Sets\ {\it and $M$ denotes $M$ as a set rather than as a
locale}. This notion of points as elements of sets will be avoided in what
follows.

Thus  frames and locales are two sides of the same coin: the elements
$1\raw\CO(X)$  of the Heyting algebra $\CO(X)$ are the {\it opens} of the
associated locale $X$,
to be thought of as {\it propositions}, whereas the {\it points} of the locale
correspond to {\it models} of the logical theory defined by these propositions.
See \cite{johnstone02b,maclanemoerdijk92} and especially \cite{Vic:LocTopSp} for
a very clear explanation of this perspective.
More precisely, recall that {\it geometric} propositional logic stands for
the following  fragment of  intuitionistic propositional  logic
\cite{maclanemoerdijk92,johnstone02b,Vic:LocTopSp}. A formula $\phv$ in propositional
geometric logic must be  built from
atomic propositions  using the  symbols $\top$ (for ``truth''), $\wed$ (for
``and''), and $\vee$
(for ``or''),  where
$\vee$ {\it but not $\wed$} is allowed to carry an infinite index set. (This may
be motivated by the remark that to verify a proposition $\vee_{\lm\in\Lambda}
p_{\lm}$, one only needs to find a single $p_{\lm}$, whereas
to  verify $\wed_{\lm\in\Lambda} p_{\lm}$ the truth of each $p_{\lm}$ needs to
be established, an impossible task in practice when $\Lm$ is infinite.)
Sequents or axioms must take the form  $\phv\raw\ps$, where $\phv$ and $\ps$ are formulae.

A frame $\CO(X)$, then, defines a geometric propositional theory
 whose propositions
correspond to opens in $X$, combined by logical connectives given by the lattice
operations
in $\CO(X)$ {\it \`a la} Boole.  This quite literally holds in the case of
classical physics discussed in Subsection \ref{lcp},
where the opens of the locale $M$ are just the opens $U$ of $M$ as a topological
space in the naive sense, construed as
propositions ``the system is in a state located within $U$''.
Conversely, a propositional geometric theory $\T$ has an associated Lindenbaum
algebra $\CO([\T])$, defined as the poset  of
formulae of $\T$ modulo provable equivalence, ordered by entailment. This poset
turns out to be a frame, and the (standard) models of $\T$ (that by definition
assign one of the two  standard truth values 0 or 1 to the propositions of $\T$
in a consistent way) bijectively correspond to frame maps $\CO([\T])\raw
\{0,1\}$. Identifying $\{0,1\}$ with $\Om=\CO(*)$ as explained above,
we see that a model of the theory $\T$ is the same thing as a
 point $*\raw [\T]$ of the locale $[\T]$.  More generally, one may consider a
model of
 $\T$ in a frame $\CO(Y)$ (generalizing the standard models where $Y=*$) to be a
 locale map $Y\raw [\T]$.
 \subsection{Locales in topoi}\label{subsection:locale3}
The generalization from topological spaces to frames is
an important step towards our goal, but it is not enough.
Seeking further generality pertinent to quantum theory,
 one may proceed in at least two different ways. First, one may generalize
  locales to {\it
quantales} \cite{Mulvey:andthen}. This step leads to recognizable
logical structures, but it does not relate well to the Copenhagen
Interpretation of quantum mechanics we favour.

 Instead, we
 pass from frames as special objects in the category of sets (as defined above)
to frames in more general topoi. This is indeed possible, as all of the above
concepts
 can be defined in any topos by using its internal
 language \cite{maclanemoerdijk92}; see \cite{borceux3} for details. In
particular,
 in a topos $\CT$ one may consider the category $\mathbf{Frm}_{\CT}$ of internal
frames
 and its opposite category $\mathbf{Loc}_{\CT}$ of internal locales. The
terminal object of the latter is the locale $*$ whose associated frame $\CO(*)$
is the subobject classifier $\Om$ of $\CT$. Opens, points and models are then
defined in exactly the same way as in \Sets, as long as one realizes that the
identification of $\Om$ with $\{0,1\}$ and of $*$ with the singleton
 is peculiar to \Sets.

In particular, a point of a locale $X$ in $\CT$ is a frame map $\CO(X)\raw\Om$,
whereas an open in $X$ may be defined as an arrow $1\raw \CO(X)$.
 The collection $\mathrm{Pt}(X)$ of a locale is still defined as the subobject
of
$\Om^{\CO(X)}$ corresponding to frame maps, its opens being given by
interpreting \er{openpt} in the internal language of $\CT$, where $U\in \CO(X)$
is interpreted as an arrow $1\stackrel{U}{\raw}\CO(X)$
and $p\inv(U)=1$ means that $p\inv \circ U=\top$, \ie\ the truth arrow
$\top:1\raw\Om$ in $\CT$.

In any case, it is reassuring that topos theorists simply refer to `internal'
locales as `spaces' \cite{joyalmoerdijk90, joyaltierney84,moerdijk84}: returning
to the opening words
from Mac Lane and Moerdijk quoted earlier, one might say that the unification in
question
is exemplified by the idea of an internal locale with its associated Heyting
algebra structure.

{\it Our quantum phase spaces $\ulS$, then, will be examples of locales in
topoi.}
Their opens $1\raw \CO(\ulS)$ will correspond to the elementary propositions or
yes-no questions about the system, and each physical state on the system will
define a map
$\CO(\ulS)\raw \underline{\Om}$, where $\underline{\Om}$ is the subobject
classifier in the particular topos in which $\ulS$ is defined.
It is important to note that such maps generally {\it  fail to be
frame maps, \ie\ they do not define models in the above sense}.
This phenomenon already arises in classical physics if one considers mixed
rather than pure states; indeed, the map
$\ch_{[\mu]}:\CO(M)\raw\Om$ introduced below \er{old3bis} fails to be a frame
map (except when $\mu$ happens to be pure).

However, a fundamental difference between classical and quantum physics in this
respect lies in the Kochen--Specker Theorem, which in its topos-theoretic
incarnation (given in different versions in \cite{butterfieldisham1} and in
Theorem~\ref{prop:kochenspecker} below)
states that (generically) the quantum phase space  $\ulS$ has no points at all,
{\it although the quantum system has pure states}
(see Subsection \ref{subsec:qs}). Hence whereas pure states in classical physics
- as defined in the usual sense through convexity  - are also  `pure' in the
logical sense, this is no longer the case in
quantum physics.

Nonetheless, pairing states and propositions into an internal truth value, \ie\
taking the subobject classifier to be the codomain of the pairing map,
is a central goal of this work, which we share with (and adopted from) the work
of Isham et al.\ \cite{butterfieldisham1,doringisham1}. Unlike real-number
valued pairings (which from a logical perspective might be preferable), an
$\Om$-valued pairing
avoids both the problems with the ignorance interpretation of the Born
probabilities (see Subsection
\ref{subsec:SQL}) and the bizarre ontology of the so-called
Many-Worlds interpretation of quantum mechanics
(\textit{cf}.~\cite{Bub,ButVat}). A philosophical defence of this goal
may also be found in \cite{ButProc}. However, the final verdict about its
validity, or rather its relevance to physics, can only be given once
the Born rule has been derived from our  $\underline{\Om}$-valued
pairing, along with an appropriate interpretation of the Born probabilities.
 This derivation will be given in future work, in which the
results of Section \ref{Spp} of this paper will be combined with those
in \cite{NPLBorn}.
\subsection{Basic construction}\label{sec:Bohr}
The two notions of generalized space just described, \ie\ noncommutative \ca s
and  locales in arbitrary topoi,
will be related by one of the main constructions in this paper, which we
 summarize in this subsection.
 This construction associates a certain  internal locale to a noncommutative
\ca\ (assumed unital), and hinges on three ideas:
\begin{enumerate}
\item {\it Algebraic quantum theory}
\cite{Emch,Haag:LQP,landsman98};
\item {\it Constructive Gelfand duality}
\cite{banaschewskimulvey00b,banaschewskimulvey00a,banaschewskimulvey06,
coquand05,CoquandSpitters:cstar};
\item {\it Bohr's doctrine of classical concepts}
\cite{bohr49,scheibe,landsman07}.
\end{enumerate}
From the first, we just adopt the methodology of describing a quantum system by
a
noncommutative \ca\ $A$ (defined in the usual topos \Sets). This move
generalizes the usual Hilbert space framework of quantum theory and has the
advantage of being able to incorporate superselection rules in infinite systems,
as well as other limiting situations like the transition from  quantum to
classical mechanics (and back).

As to the second, it turns out that the
notion of a \ca\ makes sense in an arbitrary topos,
so that one may, in particular, internalize
{\it commutative}  \ca s. Examples of such internal commutative \ca s
arise from compact completely regular locales
(\cite{banaschewskimulvey06, johnstone82}, see also
footnotes~\ref{footnote:compactlocale} and~\ref{footnote:completelyregular}
below):
 if $X$ is such a locale in some topos $\CT$, and if
$\C$ is the locale defined by the complex numbers object in $\CT$ (as
in\cite{banaschewskimulvey06}), then the
object $C(X,\C)$ of all locale maps from
$X$ to $\C$ is a commutative \ca\ in $\CT$ under natural operations.
The Gelfand duality theorem of Banaschewski and Mulvey
\cite{banaschewskimulvey06} states that, like in the case of the topos \Sets, up
to isomorphism these are the only examples of unital
 commutative \ca s:
 if $A$ is a unital commutative \ca\ in a topos $\CT$,
 there exists a  compact completely  regular
 locale $\Sg$ such that
$A\cong C(\Sg,\C)$. Here $\cong$ denotes isomorphism in the category $\CT$ and
the arrows implementing this isomorphism are \ca\ maps.
 Moreover,  this isomorphism extends to a categorical duality
 between  compact completely  regular locales and unital
 commutative \ca s in $\CT$.  We call the locale $\Sg$ or $\Sg(A)$
the {\it Gelfand spectrum} of $A$. It is defined up to isomorphism of locales.

Third, Niels Bohr's ``doctrine of classical concepts''  states that we can only
look at the quantum world through classical glasses, measurement merely
providing a
``classical snapshot of reality''.  The combination of all such
snapshots should then provide a complete picture. In Bohr's own words
(\cite{bohr49}, p.\ 209):
\begin{quote}
However far the phenomena transcend the scope of classical physical
explanation, the account of all evidence must be expressed in
classical terms. (\ldots) The argument is simply that by the word {\it
experiment} we refer to a situation where we can tell others what we
have done and what we have learned and that, therefore, the account of
the experimental arrangements and of the results of the observations
must be expressed in unambiguous language with suitable application of
the terminology of classical physics.
\end{quote}
This doctrine  has a transparent formulation in algebraic quantum
theory, to the effect that the empirical content of a quantum theory
described by a certain noncommutative C*-algebra $A$ is contained in
suitable commutative C*-algebras associated to $A$. In the simplest
case, which we study in this paper, these are simply the (unital)
commutative $C^*$-subalgebras of $A$. (To understand classical
behaviour in general, the  pertinent commutative C*-algebras have to
be extracted from $A$ using  limiting procedures like $\hbar\raw 0$ or
$N\raw\infty$ \cite{landsman07}.)

The following construction  weaves these three threads together.
 Let $A$ be a unital \ca\ (in the usual sense, \ie\ in \Sets)
and let $\CA$ be the collection of its unital
commutative $C^*$-subalgebras, partially ordered by inclusion. We
regard the poset $\CA$ as a category, whose objects are
the  unital $C^*$-subalgebras $C\subseteq A$, and whose Hom-sets
$\mathrm{Hom}_{\CA}(C,D)$ consist of a single arrow if
$C\subseteq D$ and are empty otherwise. The category $\CA$ is a catalogue of all
`classical snapshots of reality' one may take of the quantum system described by
$A$.

Recall that for any category $\mathcal{C}$,
the topos $\Sets^{\mathcal{C}}$ has functors
$\mathcal{C}\raw\Sets$ as objects  and natural
transformations as arrows \cite{maclanemoerdijk92}.
 Put
\beq
  \asstopos(\alg{A})=\Sets^{\CA}.\label{defTA}
\eeq
The philosophical idea is that  as
observers we are confined to the topos $\asstopos(\alg{A})$, whereas
the physical system itself divinely  exists in the ambient topos
$\Set$. According to Bohr and Heisenberg, the system
might seem to behave probabilistically from our limited classical
perspective, but this behaviour  is just a consequence of our confinement to
$\asstopos(\alg{A})$ ({\it cf.}\ Theorem~\ref{thm:statesareintegrals} below).

We will $\underline{\mathrm{underline}}$ entities internal to $\TA$.
It turns out that the tautological functor
$\ulA:C\mapsto C$, which (with some abuse of notation) maps a  unital
commutative $C^*$-subalgebra $C$ of $A$ (seen as an object of the
category $\CA$) into itself (seen as a set), is a unital
commutative \ca\ in $\asstopos(\alg{A})$. We call $\ulA$ the {\it Bohrification}
of $A$.
It has an
associated Gelfand spectrum $\ulS(\uA)$, which is a locale in $\TA$. The map
$A\mapsto \ulS(\uA)$
associates a `space' $\ulS(\uA)$ in the sense of topos theory to a `space' $A$
in the sense of noncommutative geometry.

In principle, this construction leads to the solution of all five problems
listed at the beginning of Subsection~\ref{subsec:SQL}:
\begin{enumerate}
\item  The quantum phase space of the system described by $A$ is the
  locale $\ulS\equiv\ulS(\uA)$ in the topos $\TA$.
\item The  ``subsets'' of the locale $\ulS$  acting as elementary
  propositions about $A$ are simply the `opens' in $\ulS$, defined as
  arrows $1\to\CO(\ulS)$ in $\TA$. Thus the quantum logic
  of $A$ is given by the Heyting algebra underlying $\ulS(\uA)$.
\item Observables $a\in A$ define locale maps $\dl(a):
  \ulS\to\underline{\mathbb{IR}}$, where $\underline{\mathbb{IR}}$ is
  the so-called {\it interval domain}. States $\rho$ on $A$ yield
  probability measures (more precisely, valuations) $\mu_{\rho}$ on
  $\ulS$.
\item An open interval $\Dl\subseteq \mathbb{R}$ defines an arrow
  $\xymatrix@1{\underline{1} \ar^-{\Delta}[r] &
  \CO(\underline{\field{IR}})}$ of $\TA$ (where $\underline{1}$ is the
  terminal object in $\TA$), which, composed with the map
  $\xymatrix@1{\CO(\underline{\mathbb{IR}}) \ar^-{\underline{\dl}(a)\inv}[r] & \CO(\ulS)}$
  underlying $\underline{\dl}(a)$, yields the desired proposition
  \[
    \xymatrix@1{\underline{1} \ar^-{[a \in \Delta]}[r] & \CO(\uS)}
    \qquad =\qquad
    \xymatrix@1{\underline{1} \ar^-{\Delta}[r] &
    \CO(\underline{\field{IR}}) \ar^-{\underline{\dl}(a)\inv}[r] & \CO(\ulS).}
  \]
\item State-proposition pairing is defined exactly as in \er{spp}, \ie
  by
  \begin{equation}
   \xymatrix{\underline{1} \ar^-{\langle a\in \Dl,\rho \rangle}[r] & \underline{\Om}}
   \qquad =\qquad
   \xymatrix@1{\underline{1} \ar^-{[a \in \Delta]}[r] & \CO(\ulS)
   \ar^-{\chi_{[\rho]}}[r] & \underline{\Om},}
   \label{sppbis}
 \end{equation}
  where $\underline{\Om}$ is the subobject classifier of $\TA$ and
  $\chi_{[\rho]}$ is the characteristic map of the subobject $[\rho]$
  of $\CO(\ulS)$ consisting of all opens $U$ of $\ulS$ with
  $\mu_{\rho}(U)=1$ (defined  through the internal language of
  $\TA$).
\end{enumerate}
The construction of $\underline{\dl}(a)$ is inspired by, and partly generalizes,
the {\it Daseinisation} map of D\"{o}ring and Isham (\cite{doringisham2,
doringisham3}, \textit{cf.} also Appendix~\ref{sec:doringisham}).

The subobject classifier $\underline{\Om}$ is the functor
$\CA\raw\Sets$ given by
\beq
  \underline{\Om}(\alg{C}) = \{ S \subseteq \uparrow\! C \mid S
  \mbox{ is an upper set}\}, \label{cosieves}
\eeq
where for any poset $P$  an upper set in $P$ is a subset
 $U\subseteq P$ for which $x\in
U$ and $x\leqslant y$ implies $y\in U$, and one writes  $\uparrow\!  x=\{y\in P\mid
x\leqslant y\}$ for the so-called principal upper set on $x$.
Note that $\underline{\Om}(\alg{C})$ is a poset (and even a frame) under inclusion as partial ordering, with
$\emptyset$ as bottom element, and $\uparrow\! C$ as top element.
 (One might 
think of the principal upper set $\uparrow\! C$ on the  ``classical snapshot of reality'' $\alg{C}$ as the collection of all finer versions of the knowledge present in
$\alg{C}$.) The subobject classifier $\underline{\Om}$ is a
(covariant) functor by stipulating that if $C\subseteq D$, then the induced map $\underline{\Om}(\alg{C})\raw \underline{\Om}(\alg{D})$ is given by $S\mapsto S\, \cap \uparrow\! D$.

In this setup, we have taken $\Set$ as the ambient
  topos.  There are several reasons, however,  one might want to consider other
ambient
  topoi. Leaving the matter to future investigation, let us briefly indicate
 an important application.
  An algebraic quantum field theory (AQFT)~\cite{Haag:LQP} may be defined
as a functor
  \hbox{$(\opens(M), \subseteq) \to \Cstar$} satisfying certain
  separability constraints, where $M$ is Minkowski space-time and
  $\opens(M)$ is its set of
  opens~\cite{BrunettiFredenhagenVerch}. Analogous to
Theorem~\ref{cor:internalalgebra} below, an AQFT may then be shown to be given
by a {\it single}
  C*-algebra in the presheaf topos $\Set^{\opens(M)}$.
\subsection{Internal and external language}\label{Intext}
We have repeatedly used the word `internal' for a construction intrinsic to a
certain topos $\CT$;
for example, $\uA$ and its Gelfand spectrum $\ulS$ are internal to $\TA$, as is
the Heyting algebra
structure of $\ulS$. At this point, confusion may arise, for on the one hand the
propositional logic carried
by $\ulS$ is intuitionistic, while on the other hand all constructions (ranging
from the initial \ca\ $A$ to
the locale $\ulS(\uA)$ as an object in the
associated functor topos $\Sets^{\CA}$)
eventually arise from the topos \Sets, whose underlying logic is classical.

To clarify this, we remark that
it is a very important aspect of topos theory that one may, indeed,  usually
adopt two
points of view: an {\it external} one and an {\it internal} one. External
constructions are carried out using classical mathematics, which (at
least for the topoi used in this paper) takes place in the familiar
topos $\Sets$ (even if the constructions in question are concerned with some
other
topos). Internal constructions, on the other hand, only use concepts
intrinsic to the topos one is studying. This idea is formalized by the {\it
internal}
or {\it Mitchell-B{\'e}nabou} language associated with each topos
\cite{borceux3,johnstone02b,maclanemoerdijk92}. This is a logical language that
for many instances and purposes allows one to reason within a given topos {\it
as if it were} the topos \Sets. For example,
one may employ the usual logical and set-theoretic symbols (the latter even if
an object $X$ has no or few elements in the sense of arrows $1\raw X$), whose
meaning is determined by the so-called Kripke-Joyal semantics associated with
the Mitchell-B{\'e}nabou language. We will actually use this semantics
in our theory of state-proposition pairing. However, using the
internal language one may (in general) neither appeal to the law of
excluded middle $x\vee\neg x=\top$, nor to the Axiom of Choice
(although restricted versions thereof are sometimes valid). These
limitations are a consequence of the fact that the internal language
of a topos happens to be based on {\it intuitionistic} predicate logic
(see  \cite{borceux3,johnstone02b} for the precise rules).
\begin{center}
\begin{figure}
  \includegraphics{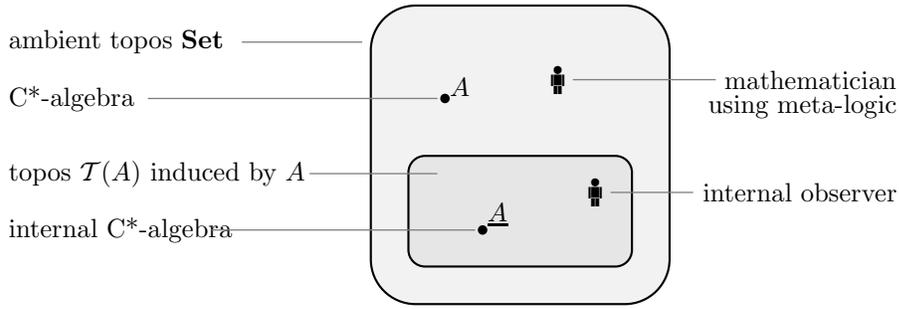}
  \caption{Illustration of  universes of discourse}
  \label{fig:toposlevels}
\end{figure}
\end{center}

Thus a topos can be seen as a universe of discourse, to which a  mathematician
or
observer may wish  to confine himself. On the other hand, even the internal
language and associated logic  can alternatively be studied externally with
classical
meta-logic. The various entities at play in our application of topos theory to
quantum physics are illustrated in Figure \ref{fig:toposlevels}. This
illustrates, in particular, that our quantum logic is meant to be the logic of
an `internal' observer, with all the restrictions this brings with it (whereas
the quantum `logic' of Birkhoff and von Neumann, to the extent it is a logic at
all, rather pertains to a fictitious entity like
Laplace's demon, to whose intellect `nothing would be uncertain and the future
just like the past would be present before its eyes.')

Let us give three closely related examples of internal versus external
descriptions, each relevant to our logical approach to quantum theory.

First, a fundamental fact of topos theory is
that the subobjects $\mathrm{Sub}_{\CT}(A)$ of a given object $A$ in a topos
$\CT$ (with subobject classifier $\Om_{\CT}$) form a (complete) Heyting algebra.
\begin{itemize}
\item
 {\it Externally}, one simply looks at $\mathrm{Sub}_{\CT}(A)$
as a {\it set}, equipped with the structure of a Heyting algebra in the
category \Sets.
\item {\it Internally}, $\mathrm{Sub}_{\CT}(A)$ is described as the
exponential $\Om_{\CT}^A$ (or power `set' $\CP(A)$), which is a Heyting algebra
object in $\CT$. See
\cite[p.\ 201]{maclanemoerdijk92}.
\end{itemize}

Second, as these Heyting algebras are complete, they are frames. The
explicit internal description of a frame or locale is rather complicated as far
as the completeness property of the underlying lattice is concerned
\cite{borceux3}. However, if the topos $\CT = \Sh(X)$ is that of
sheaves on a locale $X$ (which, we recall, consists of
those functors $F$ in $\Sets^{\CO(X)^{\mathrm{op}}}$ that satisfy a
gluing condition stating that $F(U)$ can be computed from the $F(U_i)$
under any open covering $U=\cup_i
U_i$~\cite[Ch.~II]{maclanemoerdijk92}), a simple external description
is available~\cite{johnstoneloc,joyaltierney84} (also {\it cf.}\ \cite[\S
C1.6]{johnstone02b}): a locale $\overline\CL$ in $\Sh(X)$ is externally described by
a locale map $f: \CL\raw X$ in \Sets, with
\beq \CL=\overline\CL(X).\label{extL}\eeq
Furthermore, if $\overline\CL_1$ and $\overline\CL_2$ are locales in
$\Sh(X)$ with external descriptions $f_i: \CL_i\raw X$, then
an internal locale map $\overline{g}:\overline\CL_1\raw \overline\CL_2$ in 
$\Sh(X)$ is externally given by a
locale map
$g:\CL_1\raw \CL_2$ in $\Sets$ such that $f_2\circ g=f_1$.

 To see  that this situation is relevant to our
construction, first recall the  {\it Alexandrov topology} on a poset $P$.
Its opens  are simply the upper sets,
and the special upper sets of the form $U=\mathop{\uparrow}\!  x$
 form a basis of the Alexandrov topology. Equipping $P$ with the Alexandrov
topology, one has an isomorphism of categories
\begin{equation}\label{Al}
  \Sets^P\cong \Sh(P).
\end{equation}
 To understand this, just
note that a sheaf $\overline{F}$  on $P$ is determined by its values on
the basis opens $\mathop{\uparrow}\!  x$; a functor  $\underline{F}:P\raw\Sets$
then corresponds to $\overline{F}$ by 
\beq \underline{F}(x)=\overline{F}(\mathop{\uparrow}\!
x). \label{FF}\eeq
It is, then, immediate from \er{defTA} and  \er{Al} that
\beq
\asstopos(\alg{A})\cong \Sh(\CA),\label{TASh}
\eeq
so that we have the above-mentioned external description of locales to our
avail, with $X=\CA$.

Explicitly,  to describe an internal locale $\CL$ in $\Sh(X)$
externally, \ie\ in terms of the topos \Sets, consider the {\it set}
$\Gm\CO(\CL)=\mathrm{Hom}_{\Sh(X)}(1,\CO(\CL))$  of global sections of the
associated frame $\CO(\CL)$; this set coincides with $\CO(\CL)(X)$ 
(since a
natural transformation in $\mathrm{Hom}_{\Sh(X)}(1,\CO(\CL))$  is determined by its value at $X$)
and defines a frame $\CO(\CL)(X)$  in $\Sets$ under the lattice structure borrowed
from $\CO(\CL)$. For $V\subseteq U$, let $\CL^U_V:
\CO(\CL)(U)\raw \CO(\CL)(V)$ be the
arrow part of the functor $\CO(\CL):\CO(X)^{\mathrm{op}}\raw\Sets$, with
special case $\CL_V\equiv\CL^X_V$. The completeness of $\CO(\CL)$ implies
that $\CL_V$ has a left adjoint $\CL^*_V:\CO(\CL)(V)\raw\CO(\CL)(X)$, which in
turn defines a map $f^*:\CO(X)\raw\CO(\CL)(X)$ by $f^*:V\mapsto
\CL^*_V(\top_{\CL(V)})$, where $\top_{\CL(V)}$ is the top element of
the lattice $\CO(\CL)(V)$. This is a frame map, and if we write  $\CO(Y)=\CO(\CL)(X)$, the
corresponding locale
map $f: Y\raw X$  is the external description
of $\CL$.

Conversely, a locale $L$ in $\Sets$ along with a
locale map $f:L\raw X$ (\ie\ a frame map $f^*:\CO(X)\raw \CO(L)$)
induces a locale $\CL$ in $\Sh(X)$, defined as the sheaf
$\CL(U)=\{V\in L\mid V\leqslant f^*(U)\}$. These constructions are adjoint
to each other, yielding an equivalence of the category
$\mathbf{Loc}(\Sh(X))$ of locales in $\Sh(X)$ and the slice category
$\mathbf{Loc}/X$ relative to the category $\mathbf{Loc}$ of locales
in \Sets.

For us, the external description of locales is useful for  two reasons:
\begin{enumerate}
\item Returning to \er{sppbis}, we are clearly interested in the set
$$\Gm\CO(\ulS)\cong \Hom_{\mathbf{Sets}^{\mathcal{C}(A)}}(1,\CO(\ulS))$$ of opens in
$\ulS$, as it encodes the quantum logic of our \ca\ $A$. Let  $\CO(\ovl{\Sg})$ be the
sheaf on $\mathcal{C}(A)$ that corresponds to
$\CO(\ulS)$ by \er{FF}, so that $\Gm\CO(\ulS)\cong \Gm\CO(\ovl{\Sg})$, with
$\Gm\CO(\ovl{\Sg})=\Hom_{\Sh(\mathcal{C}(A))}(1,\CO(\ovl{\Sg}))
$. Clearly, if a poset $P$ has a bottom element $\perp$ and $Z$ is any object in
$\Sets^P$, then $\Gm Z \cong Z(\perp)$. This applies to $P=\mathcal{C}(A)$ with
$\perp=\C\cdot 1$, so that
\beq \Gm\CO(\ulS)\cong \CO(\ulS)(\C\cdot 1)\cong
\CO(\ovl{\Sg})(\mathcal{C}(A))=\CO(\Sigma),\label{eq8}\eeq
where we have used \er{extL}.
Hence the external description of the quantum logic of the \ca\ $A$ is entirely
given in terms
of the locale $\Sigma$ in $\Sets$.
\item Important internal number systems in $\TA$ that are defined by   geometric propositional theories $\T$
(see  Subsection \ref{subsection:locale2})
may be computed from \er{FF} and their description in $\Sh(\CA)$, which in turn is based on their external
 description  in \Sets. Specifically, if  $[\T]$ is the locale defined by $\T$ in \Sets, then the locale $[\ovl{\T}]\equiv [\T]_{\Sh(X)}$ 
 giving the interpretation of
 $\T$ in $\Sh(X)$ has external description $\pi_1: X\x [\T]\raw X$, where $\pi_1$ is projection on the first component.
 It follows that the frame $\CO([\ovl{\T}])$ in $\Sh(X)$ corresponding to $[\ovl{\T}]$ is given by the sheaf
 $U\mapsto \CO(U\x[\T])$. Applying this to the case at hand, we see that frame $\CO([\underline{\T}])$ corresponding to
 the interpretation $[\underline{\T}]\equiv [\T]_{\CA}$ of $\T$ in $\TA$ is given by the functor
 \beq \CO([\underline{\T}]): C\mapsto \CO(\uparrow\! C \x [\T]). \label{intT}\eeq
 See Subsections \ref{sec:GT} and \ref{subsec:ID} for examples of this procedure.
\end{enumerate}

Our third example applies the second one to points of locales
\cite{maclanemoerdijk92}, and continues the discussion
in Subsection \ref{subsection:locale2}:
\begin{itemize}
\item {\it Internally},  a point of a locale $Y$ in a general topos
 $\CT$ (internal to \Sets\ for simplicity) is a  locale map  $*\raw Y$, which is
the same thing as
  an internal  frame map $\CO(Y)\raw\Om$ (where $\Om$ is the subobject
classifier in $\CT$).
  \item {\it
 Externally}, we look at $\Om$ as the frame $\mathrm{Sub}_{\CT}(1)$ in \Sets\
 of subobjects of the terminal object $1$ in $\CT$. The locale in \Sets\ with
frame  $\mathrm{Sub}_{\CT}(1)$ is called the {\it localic reflection}
 $\mathrm{Loc}(\CT)$ of $\CT$, \ie\
$\CO(\mathrm{Loc}(\CT))=\mathrm{Sub}_{\CT}(1)$.
 For example, in $\CT=\Sh(X)$ one has $\mathrm{Sub}_{\Sh(X)}(1)\cong\CO(X)$ and
hence
 $\mathrm{Loc}(\Sh(X))\cong X$. Applying the second example above, we find that
 the external description of the locale $*$ in $\Sh(X)$ is just
 $\mathrm{id}:X\raw X$, so that  points in a locale $\CL$ in $\Sh(X)$
 with external description $f:Y\raw X$ are given by locale maps
 $\varphi:X\raw Y$ that satisfy $f\circ\varphi=\mathrm{id}$, \ie\
 cross-sections of $f$.
\end{itemize}

The fourth example continues both the previous one and the discussion  of models
in Subsection \ref{subsection:locale2}. We initially defined a standard model
of a geometric propositional theory $\T$ as a locale map $*\raw[\T]$, and
subsequently mentioned more general models
$Y\raw [\T]$, still in \Sets. We now consider even more general models of $\T$
in a topos $\CT$.
\begin{itemize}
\item
{\it Externally}, these are given by  locale maps $\mathrm{Loc}(\CT)\raw [\T]$
in \Sets. This is because the classifying topos of $\T$ is $\Sh([\T])$, and one
has an equivalence
between geometric morhisms $\CT\raw \Sh([\T])$ (which classify $\T$-models in
$\CT$, {\it cf.}\  \cite[Thm.\  X.6.1]{maclanemoerdijk92})
and locale maps $\mathrm{Loc}(\CT)\raw [\T]$ (see
\cite[\S~IX.5]{maclanemoerdijk92}).
\item
{\it Internally}, one may interpret the theory $\T$ in $\CT$ and thus define a
locale
$[\T]_{\CT}$ internal to $\CT$. The points of this locale, \ie\ the locale maps
 $*\raw [\T]_{\CT}$ or frame maps $\CO( [\T]_{\CT})\raw\Om_{\CT}$,
 describe the  models of $\T$ in  $\CT$ internally.
\end{itemize}
 One may explicitly verify the equivalence between the internal and
 the external descriptions for $\CT=\Sh(X)$, for in that case the
 external description of $[\T]_{\CT}$ is the map $\pi_1: X\times
 [\T]\raw X$. Hence locale
 maps  $\varphi:X\raw  X\times [\T]$ that satisfy
 $\pi_1\circ\varphi=\mathrm{id}$ are just (unconstrained) locale
 maps $X\raw  [\T]$.
\subsection{Observation and approximation}\label{subsec:AO}
Our construction of the locale map $\dl(a): \ulS\raw
\uIR$ in Section \ref{sec:Bohr} involves the so-called {\it interval domain}
$\mathbb{IR}$
\cite{scott:interval-domain}.
To motivate its definition, consider  the
approximation of real numbers by nested intervals with endpoints in $\Q$. For
example, the real number $\pi$ can be described by
specifying the sequence
\[
  [3,4],[3.1,3.2],[3.14,3.15],[3.141,3.142],\ldots
\]
Each individual interval may be interpreted as finitary information
about the real number under scrutiny, involving the single observation that the
real
number is contained in the interval. This description of the reals, which goes
back to L.E.J. Brouwer,
 is formalized by the notion of the interval domain. Consider the
 poset  $\mathbb{IR}$ whose  elements are compact intervals $[a,b]$ in $\R$  (including
singletons $[a,a]=\{a\}$),
  ordered by {\it reverse} inclusion (for a smaller interval means that we have
more information about the real number that the ever smaller intervals converge
to). This poset is a so-called dcpo (directed complete partial order); directed
suprema are simply intersections. As such, it carries the
{\it  Scott topology}~\cite{abramskyjung94,scott:interval-domain}, whose closed
sets are
lower sets that are closed under suprema of directed subsets. Here a {\it lower} set in a poset $P$
is a subset $L\subseteq P$ such that $x\in L$ and $y\leqslant x$ implies $y\in L$; equivalently,
$\downarrow\! L\subseteq L$, where $\downarrow\! L=\{y\in P\mid \exists\, x\in L: y\leqslant x\}$. (Lower sets are sometimes called down sets or downward closed sets.)
Consequently,
Scott opens must be upper sets $U$ (defined in the obvious way) with the additional property that  for every
directed set $D$ with $\bigvee D \in U$ the intersection $D\cap U$ is nonempty.
In the case of $\mathbb{IR}$, this means that
each open interval  $(p,q)$ in $\R$ (with $p=-\infty$ and $q=+\infty$ allowed)
corresponds to a Scott open $\{[a,b]\mid p<a, b<q\}$ in  $\mathbb{IR}$, and
these opens form a basis of  the Scott topology. The collection
$\CO_{\mathrm{Scott}}(\mathbb{IR})$ is, of course, a frame, initially defined in
\Sets. The basis opens $(r,s)$ may be reinterpreted as a collection of 
generators for this frame, which from the point of view of generators and relations differs from the frame
$\CO(\R)$ of Dedekind reals in that  the relation $(p,q)=(p,q_1)\vee (p_1,q)$ for $p\leqslant p_1\leqslant q_1\leqslant q$
holds for the reals, but not for the interval domain (see \cite[D4.7.4]{johnstone02b} or Subsection \ref{sec:GT} below
for the other relations for $\CO(\R)$). 
The interval domain admits an internal definition in any topos. Its
realization in $\TA$ will play an important role in this paper; see
Subsection~\ref{subsec:ID}.

A related notion of approximation  appears
when considering an observable $a \in A\sa$ of a quantum system described by a
\ca\ $A$,
{\it as seen from inside its associated topos
$\asstopos(A)$}. Specifically, we should approximate $a$ within
 each  classical snapshot $C$ of $A$, where $C\in \mathcal{C}(A)$ is some
 commutative subalgebra. The difficulty is, of course, that
$a$ need not lie in $C$, but neither is there a single element of $C$ that forms
the `best approximation' of $a$ in $C$. The best one can do is
approximate $a$ by a family of elements of $C$, as follows. 

The self-adjoint part $A\sa$ of a \ca\ $A$ has a natural
 partial order $\leqslant$, defined
 by $a\leqslant b$ iff $b-a=c^*c$ for some $c\in A$. (Equivalently, $a\leqslant
 b$ iff $b-a=f^2$ for some $f\in A\sa$.) This partial order is linear---
 in the sense that $a+c\leqslant b+c$ whenever $a \leqslant b$.  For the
 C*-algebra $A=C(X,\C)$ one just recovers the pointwise order on
 (real-valued) functions, since $A\sa=C(X,\R)$.
 For $A=B(H)$, the
 bounded operator on Hilbert space $H$, one has $a\geq 0$ for $a\in
 A\sa$ iff $(\Psi,a\Psi)\geq 0$ for all $\Psi\in H$. (Indeed, this is
 really a pointwise order as well, if one regards operators $a$ on $H$
 as functions $\hat{a}$ on  $H$
 by $\hat{a}(\Ps)=(\Psi,a\Psi)$. See \eg\cite{landsman98}.) Thus one may
approximate $a$ in $C$ by collections of intervals of the type
 $\{[f,g]\mid f,g \in C\sa,\, f\leqslant a\leqslant g\}$ (note that this is
 inhabited, since $f=-\|a\|\cdot 1$ and $g=\|a\|\cdot 1$ occur). Since
 the intervals $[f,g]$ in $C\sa$ fail to form a dcpo, however, a
 slight adaptation of this idea is needed, for which we refer to
 Subsection~\ref{subsec:ID} below.

\subsection{Structure of this article}\label{mainresults}
Section~\ref{sec:constructivecstaralgebra} reviews the Gelfand
duality theory in a topos due to Banaschewski and Mulvey.
Our original results are as follows.
\begin{enumerate}
\item The  construction of
the `quantum phase space' $\ulS(A)$ from
a C*-algebra $A$ in \Sets\
as the Gelfand spectrum of the Bohrification $\uA$ of $A$ in the topos $\TA$
 is explained in Section~\ref{sec:internalalgebra}. This section also contains
our version of the Kochen--Specker Theorem.
\item Section~\ref{sec:statesasintegrals} first describes
the construction of states on $A$ as probability integrals on the self-adjoint
part of $\uA$. These, in turn, are equivalent to probability valuations on its
Gelfand spectrum
 $\ulS(\uA)$. On this basis, we eventually show that states define subobjects of
the quantum phase space  $\ulS(\uA)$, as in classical physics (see \er{old3}).
\item  The interpretation of observables in $A$ in terms of the
  Bohrification $\uA$ is the subject of
  Section~\ref{sec:interpretationofquantumphysics}. In particular, we
  give our analogue of the {\it Daseinisation} map of D{\"o}ring and
  Isham in Subsection \ref{sec:Das} (and more fully in Appendix \ref{AppA:Das}).
\item The pairing of states and propositions is elucidated in Section~\ref{Spp},
yielding an
element of the subobject classifier of $\TA$ that we explicitly compute. This
pairing  connects the
mathematical constructions to  quantum physics and completes steps 1 to 5 of our
general program mentioned at the beginning of Subsection \ref{subsec:SQL}.
\end{enumerate}
Appendix \ref{AppA} contains a number of technical results that somewhat
distract from the main development of the paper. Finally,
Appendix~\ref{sec:doringisham} discusses related work by
D{\"o}ring and Isham, which partly inspired the present article.
\subsubsection*{Acknowledgement}
The authors are
indebted to Andreas D{\"o}ring, Ieke Moerdijk,  Chris Mulvey, Isar Stubbe, and
Steve Vickers for guidance and
useful feedback on talks and earlier drafts of this article. We are exceptionally
grateful to the referee of this paper for unusually detailed and helpful
comments.
\section{\ca s and Gelfand duality in a topos}
\label{sec:constructivecstaralgebra}

This section recapitulates a constructive version of Gelfand duality, which is
 valid in every topos
  \cite{banaschewskimulvey00b,banaschewskimulvey00a,banaschewskimulvey06}.
Recall that the usual version of Gelfand duality characterises
 unital commutative  C*-algebras as algebras of complex-valued continuous
functions on
a compact Hausdorff space. More generally, the category $\Ccstar$ of  unital
commutative  C*-algebras and unital $\mbox{}^*$-homomorphisms is dual to the
category $\Cat{KHausSp}$ of compact Hausdorff spaces and continuous maps (see
\cite{johnstone82} for a proof aimed at algebraists and \cite{NPLCK} for a proof
in the spirit of \ca s). From a topos-theoretic point of view, this formulation
is internal to the topos $\Sets$, since both categories are defined relative to
it.

To understand the generalization of Gelfand duality to arbitrary topoi, a slight
reformulation
of the situation in  \Sets\ is appropriate: we replace topological spaces $X$ by
the associated locales, and hence replace $\Cat{KHausSp}$ by  the
equivalent category  $\Cat{KRegLoc}$ of compact regular locales
\cite{johnstone82}.\footnote{A locale $\CL$ is {\it compact} if every
subset $S\subseteq \CL$ with $\bigvee S=\top$ has a finite subset $F$
with $\bigvee F=\top$. It is {\it regular} if every element of $\CL$
is the join of the elements well inside itself, where $a$ is well
inside $b$ (denoted  $a \ll b$) if there exists $c$ with $c\wedge
a=\perp$ and $c\vee b=\top$. The (internal) categories
$\Cat{KRegLoc}$ and $\Cat{KHausSp}$ in a topos $\CT$ are equivalent
when the full axiom of choice is available in
$\CT$~\cite{johnstone82}. \label{footnote:compactlocale}
} Consequently, the duality $\Ccstar\simeq \Cat{KHausSp}$ may be replaced by
$\Ccstar\simeq \Cat{KRegLoc}$: the contravariant functor $\Ccstar\raw
\Cat{KRegLoc}$
is still given by $A\mapsto \Sigma(A)$, where $\Sg(A)$ is the locale defined by
the usual Gelfand spectrum of $A$ (\ie\ its pure state space), and in the
opposite direction one has
the familiar expression $X\mapsto C(X,\C)$, in which  the right-hand side now
stands for
the locale maps from $X$ to $\C$.

For technical reasons, in general topoi regular compact locales have to be
replaced by
{\it completely} regular compact
locales,\footnote{See~\cite{johnstone82}
or~\cite{banaschewskimulvey06} for the definition of complete
regularity.  If the axiom of dependent choice  (stating that for any
nonempty set $X$ and any relation $R \subseteq X\times X$ such that
for all $x$ there is an $y$ with $(x,y) \in R$, there is a sequence
$(x_n)$ such that $(x_n,x_{n+1}) \in R$ for each $n \in \field{N}$)
is valid in a topos, then compact regular locales are automatically
completely regular. This is the case in  \Sets, for example (where, of
course, the full axiom of choice holds), and also in topoi like $\TA$
consisting of functors whose codomain validates dependent
choice~\cite{fourmanscedrov82}.\label{footnote:completelyregular}}
  but otherwise one has a direct generalization of the above reformulation of
Gelfand duality in \Sets. The following theorem is predicated on an internal
definition of the category $\Ccstar$, which we shall give in
Subsection~\ref{subsec:cstaralgebras}. Here and in what follows, all
mathematical symbols are to be interpreted in the internal language of the topos
$\CT$ at hand.
\begin{theorem}[Gelfand duality in a topos]\emph{\cite{banaschewskimulvey00b,
banaschewskimulvey00a, banaschewskimulvey06}}.
\label{thm:gelfandrepresentation}
  In any topos $\CT$, there is a categorical duality (\ie\ contravariant
equivalence)
  \[\xymatrix{
       \Ccstar \ar@<1ex>^-{\Sg}[rr] \ar@{}|-{\perp}[rr]
    && \Cat{KRegLoc}, \ar@<1ex>^-{C(-,\field{C}_{\CT})}[ll]
  }\]
where the categories in questions are defined internally to $\CT$.
\end{theorem}
  For $A\in \Ccstar$, the locale $\Sg(\alg{A})$ is called the \emph{Gelfand
spectrum} of
  \alg{A}.
     Here the symbol $\C_{\CT}$ stands for the locale of Dedekind complex
     numbers in $\CT$.
\subsection{C*-algebras in a topos}
\label{subsec:cstaralgebras}
In any topos $\CT$ (with natural numbers object), the rationals $\Q$ can be
interpreted
\cite[\S~VI.8]{maclanemoerdijk92}, as can  the  {\it Gaussian integers}
\hbox{$\cpxrat = \{ p+qi \;:\; p,q \in\field{Q} \}$}.
For example, the interpretation of $\cpxrat$ in a functor topos
$\Set^{\cat{C}}$ (where, in our case, $\cat{C}$ is a poset) is the
constant functor that assigns the set  $\cpxrat$ to every $C \in
\cat{C}$.

 A \emph{*-algebra} in $\CT$ is a vector space ${A}$ over $\cpxrat$
  that carries an associative bilinear map
$\cdot\colon{A}\times{A}\to{A}$,
and is furthermore equipped with a map
$(-)^*\colon{A}\to{A}$ satisfying
\begin{align*}
    (a+b)^* & = a^* + b^*,
  & (z\cdot a)^* & = \overline{z} \cdot a^*,
  & (a\cdot b)^* & = b^* \cdot a^*,
  & a^{**} & = a,
\end{align*}
for all $a,b \in {A}$ and $z \in \cpxrat$. $A$ is called
\emph{commutative} if $a \cdot b = b \cdot a$ for all $a,b \in
{A}$, and \emph{unital} if there is a neutral element $1$ for the
multiplication.

To define an internal C*-algebra, we define a {\it seminorm} on such an algebra;
in
general, a norm may not actually be definable in the internal language of a
topos. This is a relation  $N \subseteq
{A} \times \field{Q}^+$, which in \Sets\  would have the meaning that
$(a,q)\in N$ iff $\|a\|< q$. In general, $N$ must satisfy
\begin{align*}
   &\;(0,p) \in N, \\
   &\;\exists_{q \in \field{Q}^+}[ (a,q) \in N ], \\
   (a,p) \in N \rightarrow &\; (a^*,p) \in N, \\
   (a,q) \in N \leftrightarrow &\;\exists_{p<q}[ (a,p) \in N ], \\
   (a,p) \in N \;\wedge\; (b,q) \in N \rightarrow&\;
     (a+b,p+q) \in N, \\
   (a,p) \in N \;\wedge\; (b,q) \in N \rightarrow&\;
     (a \cdot b,p \cdot q) \in N, \\
   (a,p) \in N \rightarrow &\; (z\cdot a,p \cdot q) \in N & (|z|<q), \\
\intertext{for all $a,b \in {A}$, $p,q \in \field{Q}^+$, and $z
\in \cpxrat$. For a unital *-algebra, we also require}
   &\;(1,p) \in N & (p>1).\\
\intertext{If the seminorm relation furthermore satisfies }
  (a^* \cdot a, q^2) \in N \leftrightarrow&\; (a,q) \in N
\end{align*}
for all $a \in {A}$ and $q \in \field{Q}^+$, then ${A}$ is said
to be a \emph{pre-semi-C*-algebra}.

To proceed to a C*-algebra, one requires $a=0$ whenever
$(a,q)\in N$ for all $q$ in $\field{Q}^+$, making the seminorm into a
norm, and subsequently one requires this normed space to be complete in a
suitable sense (see \cite{banaschewskimulvey06} for details).
As a  consequence of its completeness, a C*-algebra is automatically an algebra
over
a suitable completion of $\cpxrat$ (and not just over $\cpxrat$ itself, as baked
into the definition).
Note that in general topoi one has to distinguish certain real and complex
number objects that coincide in \Sets.
From $\Q$, one may construct  the locale  $\R_d\equiv\R$ of Dedekind real numbers
\cite[\S~VI.8]{maclanemoerdijk92} (see also
Subsection~\ref{sec:GT} below); we will drop the suffix $d$ for simplicity.  The object $\mathrm{Pt}(\C)$ (which
is the completion of $\cpxrat$ meant above)
 comprises the points of the
complexified locale $\C=\R+i\R$; see also \cite{banaschewskimulvey06} for
a direct description that avoids $\R$.
 In $\Sets$, $\C$ is the locale with frame $\CO(\C)$, where (abusing notation) $\C$ are the usual
complex numbers. In any topos,
the one-dimensional \ca\ $C(*,\C)$
is nothing but $\mathrm{Pt}(\C)$ and has Gelfand spectrum $*$ (\ie\ the locale
with frame $\Om$).

A \emph{unital *-homomorphism} between C*-algebras ${A}$
and ${B}$ is, as usual,  a linear map $f:{A}\to{B}$ satisfying
$f(ab)=f(a)f(b)$, $f(a^*) = {f(a)}^*$ and $f(1_A)=1_B$.  Unital C*-algebras
with unital *-homomorphisms form a category $\Cstar$ (internal to $\CT$);
commutative unital C*-algebras form a full subcategory $\Ccstar$ thereof.
\subsection{Spectrum}\label{subsec:spectrum}
The definition of the category $\Cat{KRegLoc}$ of completely regular compact
locales
can be internalized without difficulty. The next step is to explicitly describe
the Gelfand spectrum $\Sg({A})\equiv\Sg$ of a given commutative \ca\ ${A}$. We
will do so following  the reformulation in
\cite{coquand05,CoquandSpitters:cstar} of the pioneering work of
Banaschewski and Mulvey~\cite{banaschewskimulvey06}.

To motivate the description, note that even in \Sets\ the spectrum is now
described
(with the usual notational ambiguity explained in Subsection
\ref{subsection:locale1}) as the {\it locale} $\Sg$ defined by the frame
$\CO(\Sg)$ of open subsets of the usual Gelfand spectrum $\Sg$ of $A$ (defined
as
   the subset of the dual $A^*$ consisting of space of nonzero multiplicative
functionals on ${A}$ in the relative weak$\mbox{}^*$ topology).
    The topology on the space $\Sg$ can be described by giving a sub-base, for
which one often takes
$U_{(a,\rh_0,\varep)}=\{\rh\mid | \rh(a)-\rh_0(a)|<\varep\}$ for $a\in {A}$,
$\rh_0\in\Sg$, $\varep>0$.  However, a much simpler
choice of sub-base
would be
\beq \prop{D}_a=\{\rh \in \Sg \mid \rh(a)>0 \},\label{subb}\eeq
  where $a\in {A}\sa$.
Both the property that the $\rh$ are multiplicative and the fact that
the $\prop{D}_a$ form a sub-base of the Gelfand topology may then be
expressed lattice-theoretically by saying that $\CO(\Sg)$ is
the frame $F_{A\sa}$ freely generated by the formal symbols $\prop{D}_a$,
$a\in{A}\sa$,
subject to the relations
\begin{align} \prop{D}_1  & =\top, \label{eq:spectrum1} \\
  \prop{D}_a\wedge \prop{D}_{-a} & =\bot, \label{eq:spectrum2} \\
                \prop{D}_{-b^2} & =\bot, \label{eq:spectrum3} \\
             \prop{D}_{a+b} & \leqslant \prop{D}_a\vee
                                  \prop{D}_b, \label{eq:spectrum4} \\
              \prop{D}_{ab} & = (\prop{D}_a \wedge \prop{D}_b) \vee
                                    (\prop{D}_{-a} \wedge
                                    \prop{D}_{-b}), \label{eq:spectrum5}
\intertext{supplemented with the `regularity rule'}
  \prop{D}_{a} & \leqslant \bigvee_{r\in \field{Q}^+}\prop{D}_{a-r}.
    \label{eq:continuity}
\end{align}

This turns out to be a correct description of the spectrum of ${A}$
also in an arbitrary topos $\CT$, in which case
\er{eq:spectrum1}--\er{eq:continuity} have to be interpreted in $\CT$,
of course.\footnote{ See \cite{vickers89} and the Appendix to this paper for the procedure of constructing a frame from generators and relations. 
Equivalently, in the spirit of \cite{banaschewskimulvey06}
one could
rephrase the above definition by saying that $\Sg$ is the locale $[\T]$
corresponding to the
propositional geometric theory $\T$ (in the sense explained in Subsection
\ref{subsection:locale2})
determined by the collection of propositions $\prop{D}_a$, $a\in A\sa$, subject
to the axioms
\er{eq:spectrum1}--\er{eq:continuity}, with $\leqslant$ replaced by $\vdash$.}

\subsection{Gelfand transform}\label{sec:GT}

Classically, for a commutative unital \ca\ $A$ the Gelfand transform
$A\stackrel{\cong}{\raw} C(\Sigma,\C)$
 is given by $a\mapsto \hat{a}$ with $\hat{a}(\rh)=\rh(a)$. In our setting it is
 convenient to restrict the Gelfand transform to $A\sa$, yielding an isomorphism
 \beq
 A\sa\cong C(\Sg,\R)  \label{GTisoR}.
 \eeq
 In a topos $\CT$,
 the Gelfand transform of an internal commutative unital \ca\ ${A}$ in $\CT$
associates a locale map
\beq \hat{a}:\Sg\raw \R_{\CT},\label{GTlm}\eeq
 to each $a\in{A\sa}$, where
$\Sg$ is the spectrum of ${A}$ and
 $\R_{\CT}$ is the locale of internal Dedekind real numbers in
$\CT$; see below. Recalling from Subsection \ref{subsection:locale1} that
 $\hat{a}$ is by definition a frame map
 \beq \hat{a}\inv: \CO(\R_{\CT})\raw\CO(\Sg), \label{GTfm}\eeq
 and using the ``$\lm$-conversion rule'' $\frac{Y\raw Z^X}{Y\times X\raw Z}$
  \cite[\S I.6]{maclanemoerdijk92},
 we note that  the Gelfand transform may alternatively  be regarded as a map
\beq
\hat{\cdot}: A\sa\times \CO(\R_{\CT})\raw \CO(\Sg).\label{GTlc}\eeq
 Thus
 the use of the symbol $a\in A\sa$ in the internal language of $\CT$ may be
avoided in principle.
 In practice, however, we will often use the notation \er{GTlm} or \er{GTfm},
and hence the formal symbols $\prop{D}_a$.
For example, in the description \er{eq:spectrum1}--\er{eq:continuity} of
the spectrum $\Sg$ in terms of generators and relations, it is
sufficient to define the frame map \er{GTfm} on basic opens $(-\infty,r)$ and
$(s,\infty)$ in $\R_{\CT}$. In the classical case (\ie\ in \Sets) discussed above,
one has
$\hat{a}\inv(0,\infty)=\prop{D}_a$ from \er{subb}, and this remains true in
general if $\hat{a}\inv$ has the meaning \er{GTfm}.
 Using \er{eq:spectrum1}--\er{eq:spectrum4}, one then finds
\begin{eqnarray}
\hat{a}\inv : (-\infty,s) &\mapsto&  \prop{D}_{s-a};\label{GT0}\\
 (r,\infty) &\mapsto & \prop{D}_{a-r}. \label{GT}
\end{eqnarray}
As $\hat{a}\inv$ is a frame map, for bounded open intervals $(r,s)$
 we therefore obtain\footnote{Banaschewski and Mulvey
\cite{banaschewskimulvey06} work with such intervals $(r,s)$ as basic opens, in
terms
of which they write the Gelfand transform as $\hat{a}\inv : (r,s) \mapsto a\in
(r,s)$. Here the role of generators of the locale $\Sg$ is played by elementary
propositions of the logical theory generating $\Sg$ as its Lindenbaum algebra,
our generator $\prop{D}_a$
corresponding to their proposition $a\in(0,\infty)$. Classically,  the
proposition
$a\in (r,s)$ may be identified with the open
$a\inv(r,s)$ in the spectrum $\Sg$; {\it cf.}\ Subsection~\ref{lcp}.}
\beq \hat{a}\inv : (r,s) \mapsto \prop{D}_{s-a}\wedge \prop{D}_{a-r}.\label{GT2}
\eeq

We now  recall an explicit construction of the Dedekind reals
\cite{fourmangrayson}, \cite[D4.7.4 \&\ D4.7.5]{johnstone02b}.
Define the propositional geometric theory $\T_{\R}$
 generated by formal symbols $(p,q)\in\Q\x\Q$ 
 with $p<q$, ordered as $(p,q)\leqslant (p',q')$ iff $p'\leqslant p$ and $q\leqslant q'$,
  subject to the following axioms (or relations):
 \begin{enumerate}
\item $(p_1,q_1)\wed (p_2,q_2)=(\max\{p_1,p_2\},\min\{q_1,q_2\})$ if
$\max\{p_1,p_2\}<\min\{q_1,q_2\}$, and $(p_1,q_1)\wed (p_2,q_2)=\bot$ otherwise;
\item $(p,q)=\bigvee \{(p',q')\mid p<p'<q'<q\}$;
\item $\top=\bigvee \{(p,q)\mid p<q\}$;
\item $(p,q)=(p,q_1)\vee (p_1,q)$ if $p\leqslant p_1\leqslant q_1\leqslant q$.
\end{enumerate}

This theory may be interpreted in any topos $\CT$, defining an internal locale
$(\T_{\R})_{\CT}\equiv \R_{\CT}$ with associated frame $\CO(\R_{\CT})$.
Points $m$ of  $\R_{\CT}$, \ie\  frame maps
$m\inv:\CO(\R_{\CT})\raw\Om_{\CT}$,  correspond bijectively to Dedekind cuts $(L,U)$ of
$\Q$ ({\it cf.}\ \cite[p.\ 321]{maclanemoerdijk92}) in the following way: a
model $m$ determines a Dedekind cut by
\begin{eqnarray}
L&=& \{p\in\Q\mid m\models (p,\infty) \};\label{eq21}\\
U&=& \{q\in\Q\mid m\models  (-\infty,q)\},\label{eq22}
\end{eqnarray}
where $(p,\infty)$ and $(-\infty,q)$ are defined in terms of the formal generators of the frame
$\mathcal{O}(\Q)$ by $(p,\infty)=\bigvee \{(p,r)\mid p<r\}$ and $(-\infty,q)=\bigvee \{(r,q)\mid r<q\}$. 
The notation $m\models (p,q)$ used here means that $m\inv (p,q)=\top$, where
$\top:1\raw\Om_{\CT}$ is the truth element of $\Om_{\CT}$ and $(p,q)$ is seen as an arrow
$(p,q):1\raw \Q\x\Q\raw \CO(\R_{\CT})$.
Conversely, a Dedekind cut $(L,U)$ uniquely determines a point $m$
that maps a  generator $I=(p,q)$ to
$m(I)=\top$ iff $I\cap U\neq \emptyset$ and $I\cap L\neq \emptyset$.
The Dedekind reals  $\mathrm{Pt}(\R_{\CT})$, then,
 are defined in any topos $\CT$
 as the subobject of \hbox{$\CP(\Q_{\CT})\x\CP(\Q_{\CT})$} consisting of those
$(L,U)$ that are points of $\R_{\CT}$  \cite{maclanemoerdijk92}.

We mention four examples:
\begin{enumerate}
  \item In $\CT= \Sets$, 
   a point $m$ of $\R\equiv \R_{\Sets}$ corresponds to a real
    $x$ described in the usual calculus way, so that $L= \{p\in\Q\mid
    p<x\}$ and $U=\{q\in\Q\mid x>q\}$. Hence  $\mathrm{Pt}(\R)$ may
    be identified with $\R$ in the usual sense, and $\R$ is spatial
    as a locale; its frame  $\CO(\R)$ is just the usual topology of $\R$
        \cite[D4.7.4]{johnstone02b}.
    From this perspective, the first
    condition in the definition of $\T_{\R}$  enforces that $L$ and $U$ are lower and upper
    sections of $\Q$, respectively, the second implies that they are
    open, and the third means that $L$ and $U$ are both inhabited. The fourth -- Dedekind --
    relation says that $L$ and $U$ `kiss' each other.\footnote{The collection of $L$ satisfying only the first three relations
    forms the locale of \emph{lower reals}, which we denote by
    $\field{R}_l$. Locale maps to $\field{R}_l$ are, classically,
    lower-semicontinuous real-valued functions. Analogously, there is
    a locale $\field{R}_u$ of \emph{upper reals}. See \cite{johnstone02b}.}

  \item If $X$ is a topological space (or, more generally, a locale), the structure of the locale
  $\R_{\Sh(X)}$ and its associated sheaf of Dedekind reals $\mathrm{Pt}(\R_{\Sh(X)})$
     in the topos $\Sh(X)$ of
    sheaves on $X$ follows from the argument above \er{intT} in Subsection \ref{Intext}.
    First, the frame of of Dedekind reals is given by the sheaf
\beq \CO(\R)_{\Sh(X)}: U\mapsto \CO(U\x \R),\eeq
whereas the Dedekind real numbers object is the sheaf (See also \cite{maclanemoerdijk92})
\beq \mathrm{Pt}(\R)_{\Sh(X)}: U\mapsto C(U,\R).\label{DRN}\eeq
 \item
 Consequently, using \er{Al} and \er{FF} we infer that
    in our functor topos  $\TA=\Sets^{\CA}$,   the frame  of Dedekind reals is the functor
    \beq
    \CO(\underline{\R}):
    C\mapsto \CO((\uparrow\! C)\x \R); \label{ddr}\eeq
   the set on the right-hand side may be identified with the set of monotone functions from $\uparrow\! C$ to $\CO(\R)$.\footnote{\label{monotone} This identification proceeds in two steps. First, for any topological space $X$ one has a bijection $\CO(X)\cong C(X, S)$, where $S=\{0,1\}$ carries the Sierpinski topology, see Subsection \ref{subsection:locale2}; explicitly,
 $U\in\CO(X)$ is mapped to  $\chi_U$, whereas in the opposite direction $g\in C(X, S)$
   is sent to  $g^{-1}(\{1\})$.
Hence $\CO(\uparrow\!C \x \R) \cong C(\uparrow\!C \x \R, S)$ (with apologies for the double use of $C$, first for `continuous' and second for $C\in\CA$). Second, in general $\lambda$-conversion or `currying' gives a  bijection between functions $Y \x \R\raw S$ and  functions 
$Y\raw S^{\R}$; with $Y=\uparrow\!C$ equipped with the Alexandrov topology and $C(\R,S)\cong \CO(\R)$,  continuity then translates into monotonicity.}
    
Perhaps surprisingly, the associated functor of points 
 $\mathrm{Pt}(\ulR)$ may be
    identified with the constant functor 
 \beq    
     \mathrm{Pt}(\ulR): C\mapsto \R;  \label{cf} \eeq
this follows from \er{DRN} and the fact that Alexandrov-continuous functions $U\raw\R$
(or, indeed, into any Hausdorff space) must be locally constant on any open $U\subseteq \CA$.\footnote{We take $X=\CA$,
equipped with the Alexandrov topology, and prove that in this topology
any $f \in C(U,\field{R})$ must be locally constant. 
 Suppose $C \leqslant D$ in $U$, take and $V \subseteq
\field{R}$ open with $f(C)\in V$. Then 
 tautologically $C\in f\inv (V)$ and $f\inv(V)$ is open  by continuity of $f$.
 But the smallest open set containing $C$ is $\uparrow\! C$, which contains $D$, so that
 $f(D)\in V$. 
 Taking $V=(f(C)-\ep,\infty)$ gives $f(D)> f(C)-\ep$ for all $\ep>0$, whence
$f(D)\geqslant f(C)$, whereas  $V=(-\infty,f(C)+\ep)$ yields $f(D)\leqslant f(C)$.
Hence $f(C)=f(D)$.}
  \item If $\Sg$ is the Gelfand spectrum of a commutative \ca\ ${A}$
    in $\CT$,  in the sheaf topos $\Sh(\Sg)$ internal to $\CT$  we
    similarly have
    \begin{equation}
      \mathrm{Pt}(\R)_{\Sh(\Sg)}:U\mapsto C(U,\R).\label{ShSg}
    \end{equation}
    Here we identify the open $U$ of $\Sg$ with its associated
    sublocale $\{V\in\Sg\mid V\leqslant U\}$ of $\Sg$. This locale, as well
    as $\R$, is to be interpreted in the
    ambient topos $\CT$ as explained in the above items.
\end{enumerate}

Example 4 leads to an elegant reformulation of
the isomorphism \er{GTisoR} given by the Gelfand theory: since
\beq C(\Sg,\R)=\Gm(\mathrm{Pt}(\R)_{\Sh(\Sg)}),\label{preelegant}\eeq where
$\Gm$ is the global sections functor, one infers from \er{GTisoR} that
 \beq
 A\sa\cong \Gm(\mathrm{Pt}(\R)_{\Sh(\Sg)}).\label{elegant}\eeq
 In other words, the self-adjoint part of a unital commutative \ca\ $A$ in a
topos is
 isomorphic to the global sections of the Dedekind reals in the internal topos
of sheaves on its spectrum (and $A$ itself
 ``is'' the complex numbers in the same sense).
\section{The internal C*-algebra and its spectrum}
\label{sec:internalalgebra}
In this section we explain the association  of a particular commutative \ca\
$\uA$, which is
internal to a certain functor topos $\TA$, to a (generally) noncommutative \ca\
$A$.
As mentioned in the Introduction, this  construction is motivated by Bohr's
doctrine of classical concepts, so that we call $\uA$ the {\it Bohrification} of
$A$.
\subsection{The topos associated to a C*-algebra}
We first construct the topos $\TA$ in which $\uA$ resides and draw attention to
the functoriality of the map $A\mapsto \TA$.
We denote the category of partially ordered sets and monotone
functions  by $\Cat{Poset}$.

\begin{proposition}
\label{prop:contextfunctor}
  There is a functor $\context : \Cstar \to \Cat{Poset}$,  defined
  on objects as
  \[
    \CA = \{ \alg{C} \subseteq \alg{A} \mid
                           \alg{C} \in \Ccstar \},
  \]
  ordered by inclusion.
  On a morphism $f:\alg{A} \to \alg{B}$ of \Cstar,
  it acts as $\mathcal{C}(f) : \CA \to
  \mathcal{C}(\alg{B})$ by the direct image $\alg{C} \mapsto f(\alg{C})$.
\end{proposition}
As announced in \er{defTA} in the Introduction, the collection of
functors  $\CA \to \Set$ forms a topos
$\asstopos(\alg{A})=\Sets^{\CA}$. This is the \emph{topos associated
to $A$}. We recall our convention to  $\underline{\mathrm{underline}}$
entities internal to $\TA$. The subobject classifier $\functor{\Omega}$ in
$\asstopos(\alg{A})$ has already been given in \er{cosieves}.

Recall that a geometric morphism $f:\mathcal{S} \to \mathcal{T}$ between
topoi is a pair of adjoint functors, consisting of a direct image
part $f_*:\mathcal{S}\to\mathcal{T}$ and an inverse image part $f^*:\mathcal{T}
\to \mathcal{S}$, of which $f^*$ is required to preserve finite limits.
Denote the category of elementary topoi and geometric
morphisms  by \Cat{Topos}.\footnote{We will not worry about the fact that
\Cat{Topos}, like \Cat{Poset} and \Cstar, is a large category;
when pressed one can limit these categories to a chosen universe to
make them small.}
\begin{proposition}
\label{prop:asstopos}
  There is a functor $\asstopos : \Cstar \to \Cat{Topos}$, defined
  on objects by $\asstopos(\alg{A})=\Set^{\CA}$, the
  category of functors from $\CA$ to the ambient topos.
\end{proposition}
This immediately follows from Theorem VII.2.2 in
\cite{maclanemoerdijk92} (p.\ 359) and Proposition
\ref{prop:contextfunctor}.

To close this subsection, note that instead of
initially regarding $\CA$ as a
 poset as in the main text, we could have considered it as a category from the
start, having the same
objects, but with (equivalence classes of) monomorphisms as arrows (instead of
inclusions). The functor in
Proposition~\ref{prop:contextfunctor} would then have the category
$\Cat{Cat}$ of categories as its codomain. This would still have
allowed us to define the associated topos, and also the internal
C*-algebra we will define below. From then on, most
constructions will be within the associated topos, and hence go
through as well.
\subsection{Bohrification}
Whereas the previous subsection considered the topos
$\asstopos(\alg{A})$ associated to a C*-algebra $\alg{A}$, this
one is devoted to a particular object $\functor{\alg{A}}$ in this
topos.
In fact, the definition of $\functor{\alg{A}}$ is `tautological' in a literal
sense.
\begin{definition}
\label{def:internalalgebra}
  Let $\alg{A}$ be a C*-algebra in $\Sets$.
The functor $\functor{\alg{A}} :
  \CA \to \Sets$ is given on objects by
  \[
    \functor{\alg{A}}(\alg{C}) = \alg{C},
  \]
and on morphisms $\alg{D} \subseteq \alg{C}$ of
  $\CA$ as the inclusion $\functor{\alg{A}}(\alg{D})
  \hookrightarrow \functor{\alg{A}}(\alg{C})$.
\end{definition}
Note that the functor
 $\functor{\alg{A}} : \CA \to\Sets$ factors through $\Cstar$ or $\cat{cCstar}$
 via the  forgetful embedding of $\Cstar$ or $\cat{cCstar}$ in the ambient topos
$\Sets$.
\begin{theorem}
\label{cor:internalalgebra}
  $\functor{\alg{A}}$ is a commutative C*-algebra in
  $\asstopos(\alg{A})$ under the operations inherited from $\alg{A}$.
  More precisely, $\uA$
is a vector space over the internal complex numbers
$\mathrm{Pt}(\underline{\C})$
(given simply by the constant functor
$\mathrm{Pt}(\underline{\C}):C\mapsto\C$)
by
\begin{align*}
  0 & \colon \functor{1} \to \functor{\alg{A}}
    & 0_C(*) & = 0, \\
  + & \colon \functor{\alg{A}} \times \functor{\alg{A}} \to
      \functor{\alg{A}}
    & a +_C b & = a + b, \\
  \cdot \;
    & \colon\mathrm{Pt}(\underline{\C}) \times \functor{\alg{A}} \to
      \functor{\alg{A}}
    & z \cdot_C a & = z \cdot a,
\intertext{and an involutive algebra through
}
  \cdot \;
    & \colon \functor{\alg{A}} \times \functor{\alg{A}} \to
      \functor{\alg{A}}
    & a \cdot_C b & = a \cdot b, \\
  (-)^*
    & \colon \functor{\alg{A}} \to \functor{\alg{A}}
    & (a^*)_C & = a^*.
\intertext{
The norm relation is given by}
  N & \colon \functor{\alg{A}} \times \functor{\field{Q}^+} \to
      \functor{\Omega}
     & N_C(a,q) & \mbox{ iff }\norm{a} < q.
\end{align*}
\end{theorem}
\begin{proof}
One easily checks that the arrows are natural transformations (and
hence morphisms in $\asstopos(\alg{A})$) and that this structure
satisfies the requirements for $\functor{\alg{A}}$ to be a
 pre-semi-C*-algebra in $\asstopos(\alg{A})$. Since each
 $\functor{\alg{A}}(C)$ is a commutative C*-algebra in the ambient
  topos, $\functor{\alg{A}}$ is commutative as well. (Alternatively, since the
definition of a commutative pre-semi-C*-algebra consists only of
  geometrically definable objects
  (\eg \cpxrat) and geometric formulae (see Appendix \ref{AppA} and
  Section~\ref{sec:constructivecstaralgebra}), it follows from
  Lemma ~\ref{prop:geometricformulae} that $\functor{\alg{A}}$ is a
  commutative pre-semi-C*-algebra in $\asstopos(\alg{A})$, because every
  $\functor{\alg{A}}(C)$ is a commutative C*-algebra in the ambient
  topos.)

  In fact, $\functor{\alg{A}}$ is a pre-C*-algebra, \ie\ internally the
  semi-norm is a norm: if for all $q>0$ we have $(a,q)\in N$, then
  $a=0$. To prove this, we need to show that $C\Vdash \forall_{a \in
  \uA\sa} \forall_{q \in \underline{\Q}^+}. (a,q)\in N \rightarrow
  a=0$, where we are using the internal language of $\TA$. In other words:
  \begin{align*}
      & \mbox{for all }C'\supseteq C\mbox{ and }a \in C', \mbox{ if
      }C'\Vdash \forall_{q \in \underline{\Q}^+}.(a,q)
     \in N,\mbox{ then }C'\Vdash a=0, \\
     \ie\ & \mbox{for all }C'\supseteq C\mbox{ and }a \in C', \mbox{ if
     for all }C''\supseteq
     C'\mbox{ and }q \in \Q^+\mbox{ we have }C''\Vdash (a,q)\in N,\\
         & \phantom{\mbox{ for all }C'\supseteq C\mbox{ and }a \in C',
         }\mbox{ then }
         C'\Vdash a=0, \\
     \ie\ & \mbox{for all }C' \supseteq C\mbox{ and }a \in C', \mbox{
     if }\|a\|=0, \mbox{ then }a=0.
  \end{align*}
  But this holds, since every $C'$ is a C*-algebra.

  Finally, $\functor{\alg{A}}$ is in fact a C*-algebra, \ie\ internally
  we have Cauchy completeness. By the axiom of dependent choice
  (which holds because $\TA$ is a functor topos whose codomain
 validates   dependent choice~\cite{fourmanscedrov82})
   it
  suffices to prove that every \emph{regular} Cauchy sequence (\ie\
 a sequence $(x_n)$ such that
  $||x_n-x_m||\leqslant 2^{-n}+2^{-m}$ for all $n,m$)   converges. Thus we need to
prove
  \begin{align*}
     & C\Vdash \forall_{n,m}. ||x_n-x_m||\leqslant
2^{-n}+2^{-m}\rightarrow \exists_{x\in
       A}. \forall_n. ||x-x_n||\leqslant 2^{-n}, \\
     \ie\ & \mbox{for all }C'\supseteq C, \mbox{ if }C'\Vdash (\forall_{n,m}.
           \|x_n-x_m\|\leqslant 2^{-n}+2^{-m}), \\
         & \phantom{\mbox{for all }C'\supseteq C, }\mbox{ then }C'\Vdash
           \exists_{x\in A}. \forall_n.||x-x_n||\leqslant 2^{-n}, \\
     \ie\ & \mbox{for all }C'\supseteq C, \mbox{ if }C'\Vdash\mbox{``$x$ is
           regular'', then } C'\Vdash\exists_{x\in
           A}. \forall_n.\|x-x_n\|\leqslant 2^{-n}.
  \end{align*}
 Once again, this holds because every $C'$ is a C*-algebra.
  \qed
\end{proof}

\emph{The functor $\functor{\alg{A}}$ is our internal
C*-algebra}. By changing the universe of discourse from the ambient
topos \Sets\ to $\asstopos(\alg{A})$, the (generally) noncommutative
C*-algebra $\alg{A}$ has become a commutative C*-algebra $\functor{\alg{A}}$.
Multiplication of two non-commuting operators is no longer
defined, since they live in different commutative subalgebras.\footnote{Kochen
and Specker refer to such a structure as a partial
algebra~\cite{kochenspecker67} and stress its relevance for
the foundations of quantum theory; in a partial algebra both  addition
and multiplication need only be defined for commuting operators.}
\subsection{The Kochen--Specker Theorem}
\label{sec:spectrum}
Combining the material in Sections~\ref{sec:constructivecstaralgebra}
and~\ref{sec:internalalgebra} so far,
we obtain a mapping $A\mapsto\ulS(\uA)$, which associates a certain internal
locale
to a (generally) noncommutative \ca. As argued in the Introduction,  $\ulS(\uA)$
describes the quantum logic of the physical system whose algebra of observables
is $A$.

An important property of the  internal spectrum $\underline{\Sigma}$  is that it may
typically be
highly non-spatial from an external point of view. First, recall (see
Subsection \ref{subsection:locale2}) that a point of a locale
$X$ in a topos $\CT$ is a frame map $\CO(X)\raw\Om$, where $\Om$ is the
subobject classifier in $\CT$.
\begin{theorem}
  \label{prop:kochenspecker}
  Let $H$ be a Hilbert space with $\dim(H)>2$ and let $\alg{A}$ be the
  C*-algebra of bounded operators on $H$. Then the locale $\ulS(\uA)$
  has no  points.
\end{theorem}
\begin{proof}
We reason internally. 
 A point $\underline{\rh}:\underline{*}\raw \ulS$  of the locale $\ulS$ (see Subsection \ref{subsection:locale3})
 may be combined with $a\in\uA\sa$ with Gelfand transform $\hat{a}:\ulS\raw \ulR$ (see \er{GTlm}),  so as to produce a point
 $\hat{a}\circ \underline{\rh}:\underline{*}\raw\ulR$ of the locale $\ulR$. This yields a map 
 $\underline{V}_{\rh}: \uA\sa \to\mathrm{Pt}(\underline\R)$, which can be shown to be an internal multiplicative functional;
see~\cite{banaschewskimulvey00a, banaschewskimulvey06, coquand05}.\footnote{This map
may explicitly be given in the internal language of $\TA$, by noting that
 for each  $a\in \uA\sa$ the expression $
    \tilde{\rh}(a)
  = (L_{\rh,a},U_{\rh,a})
  = (\{r \in \field{Q} \mid \rho \models \prop{D}_{a-r}\},
     \{s \in \field{Q} \mid \rho\models \prop{D}_{s-a} \})$ is a Dedekind
cut in $\TA$.} Being an arrow in $\TA$, the map $\underline{V}_{\rh}$
 is a natural transformation, with components $\underline{V}_{\rh}(C): \uA\sa(C) \raw\mathrm{Pt}(\underline\R)(C)$;
 by Definition \ref{def:internalalgebra} and \er{cf}, this is just  $\underline{V}_{\rh}(C):C\sa\raw \R$. 
 Hence one has a multiplicative functional $\underline{V}_{\rh}(C)$ for each $C\in\CA$ in
the usual sense,
with the property (which follows from naturality) that if $C\subseteq D$, then the restriction of  $\underline{V}_{\rh}(D)$
to $C\sa$ coincides with  $\underline{V}_{\rh}(C)$.
But this is precisely a {\it valuation}\footnote{This terminology is to be
distinguished from the one used in Subsection \ref{ss:th-integrals} below. The naturality property just mentioned is often called {\it noncontextuality} in the philosophy of physics literature.}  on
$B(H)$, whose nonexistence was proved by Kochen and Specker
\cite{kochenspecker67}. \qed
\end{proof}

This is a localic reformulation of the original topos-theoretic version of the
Kochen-Specker theorem due to Butterfield and Isham \cite{butterfieldisham1}. As
in their work, the proof relies on the original version, but in being a statement about the lack of models of a certain theory,  our reformulation has a logical thrust that both the original version by Kochen and Specker and the
reformulation by Butterfield and Isham lack. 

The  theorem certainly holds for more general C*-algebras than
just the collection of all bounded operators on a Hilbert space;
see~\cite{doering:KS} and \cite{HLSSyn} for results on von Neumann algebras.
For C*-algebras, one has the result that a simple infinite unital C*-algebra
does not admit a dispersion-free quasi-state~\cite{Hamhalter}. Evidently,
Theorem~\ref{prop:kochenspecker} holds for such extensions as well.

One way of looking at such results is to see them as
 illustrations of the failure of the Krein-Milman theorem in a constructive
context~\cite{Mulvey:Krein-Milman}. Indeed, recall that the classical
Krein-Milman
theorem states that a compact convex set is the closed convex hull of its
extreme points. The state space of $\uA$ is still a compact convex set in an
appropriate localic sense (see Section~\ref{sec:statesasintegrals}), and the
pure states on $\uA$
would be its extreme boundary. These points,
however, fail to exist, as we have just seen.
\section{(Quasi-)states as integrals}
\label{sec:statesasintegrals}
This section about states, and the next one about observables, are both
concerned with connections between
the two levels we have developed (see Figure~\ref{fig:toposlevels}):
\begin{enumerate}
  \item the ambient topos $\Set$, containing the C*-algebra
    $\alg{A}$;
  \item the associated topos $\asstopos(\alg{A})$, containing the internal
    commutative C*-algebra $\functor{\alg{A}}$ and its
    spectrum $\ulS$.
\end{enumerate}
The main result of this section is Theorem~\ref{thm:statesareintegrals}, which
gives
 an isomorphism
between quasi-states on $\alg{A}$ at level 1 and,  at level 2,  either
probability integrals on $\uA\sa$, or, equivalently, probability
valuations on the Gelfand spectrum $\ulS$.
Subsequently,  we show that probability  valuations
 define subobjects of  $\ulS$, as in classical physics.

 All this requires some preparation, firstly in the theory of quasi-states on
\ca s
 (Subsection \ref{subsec:qs}) and secondly in
 abstract constructive integration theory (Subsection \ref{ss:th-integrals}).
\subsection{States and quasi-states}\label{subsec:qs}
A linear functional $\rho:A \to
\field{C}$ on a C*-algebra $A$ is called \emph{positive} when
$\rho(a^*a) \geq 0$ for all $a \in \alg{A}$. It is a {\it state} when
it is positive and satisfies $\rho(1)=1$. A state $\rh$ is \emph{pure} when
$\rh=t \sg + (1-t)\om$ for some $t\in (0,1)$ and some states $\sg$ and $\om$
implies
$\om=\sg$. Otherwise, it is called \emph{mixed}.
For example, if $A\subseteq B(H)$ for some Hilbert space $H$ (which we may
always assume by the Gelfand--Naimark Theorem), then each unit vector $\Psi\in
H$ defines a state $\ps$ on $A$
by $\ps(a)=(\Ps, a\Ps)$. If $A=B(H)$, such states are pure. (If $H$ is
infinite-dimensional, not all pure state arise in this way, though.) Mixed
states $\rh$ on $B(H)$
arise from countable sequences $(p_i)$, $0\leqslant p_i\leqslant 1$, $\sum_i p_i=1$,
coupled with
an orthonormal family of vectors $(\Ps_i)$, through $\rh(a)=\sum_i p_i
\ps_i(a)$.
(By the spectral theorem, one may equivalently say that such states
are given by positive operators $\hat{\rh}$ on $H$ with unit trace,
through $\rh(a)=\Tr (\hat{\rh} a)$.)
  A state $\rho:\alg{A}\to\field{C}$ is called
\emph{faithful} when $\rho(a^*a)=0$ implies $a=0$. For example, if, in the
situation just described, the $\Ps_i$ comprise an orthonormal basis of $H$ and
each  $p_i>0$, then
the associated state $\rh$ is faithful.
The states of a
C*-algebra form a compact convex set, the extremal points of which are by
definition the
pure states. States are automatically
hermitian, in the sense that $\rho(a^*) = \overline{\rho(a)}$, or
equivalently, $\rho(a) \in \field{R}$ for self-adjoint
$a$.

In algebraic quantum physics, {\it mathematical} states as defined above are
often used to model
the {\it physical} states of the quantum system. However, when taking Bohr's
doctrine of classical concepts seriously, one should
take into account that two observables can only be added in a physically
meaningful way when they are
jointly measurable, \ie\ when the corresponding operators
commute. Thus one may relax the definition of a quantum state, which
 ought to be linear only on
commutative parts. This leads to the notion of a
{\it quasi-state}~\cite{Aarnes}:\footnote{Axiom VII of Mackey's foundation
  of quantum mechanics~\cite{Mackey}
states that a measure on the projections of a von
Neumann algebra extends to a state on the von Neumann algebra. Mackey stresses
that, in contrast to his other axioms, Axiom VII does not have a physical
justification. One can prove that a measure extends to a quasi-state, so one is
led to ask whether every quasi-state is a state. This is not the
case when the von Neumann algebra has a summand of type $I_2$, but it
holds for all other von Neumann algebras~\cite{BunceWright:vonNeumann}.
For C*-algebras the question is more difficult. The main result
seems to be the following~\cite{BunceWright}. Consider a C*-algebra
with no quotient isomorphic to $\Mtwo$ and let $\rho$ be a
quasi-linear functional. Then $\rho$ is linear iff $\rho$ restricted
to the unit ball is uniformly weakly continuous.}
\begin{definition}
\label{def:quasi-state}
  A \emph{quasi-linear} functional on a C*-algebra $\alg{A}$ is a map
  $\rho:\alg{A}\rightarrow \field{C}$ that is linear on all
  commutative subalgebras and satisfies $\rho(a+ib)=\rho(a)+i\rho(b)$
  for all self-adjoint  $a,b \in \alg{A}$ (possibly non-commuting). It
  is called \emph{positive} when $\rho(a^*a) \geq 0$ for all $a \in
  \alg{A}$. When $\alg{A}$ is unital, a positive quasi-linear
  functional is called a \emph{quasi-state} when $\rho(1)=1$.
\end{definition}

This kind of quasi-linearity also determines when some property $P$
of the C*-algebra $A$ descends to a corresponding property $\underline{P}$ of
the
internal C*-algebra $\uA$, as the following lemma shows. To be precise,
for $P \subseteq A$, define a subfunctor of $\uA$
by $\underline{P}(C) = P \cap C$. Let us call a property $P
\subseteq A$ \emph{quasi-linear} when $a \in P$ and $b \in P$ imply $\mu
a+i\lambda b \in P$ for all $\mu,\lambda \in \field{R}$ and
$a,b \in A\sa$.
\begin{lemma}
\label{cor:propertyonjlbextends}
  Let $A$ be a C*-algebra, and let $P \subseteq A$ be a quasi-linear
  property. Then $P=A$ if and only if $\underline{P} = \uA$.
\end{lemma}
\begin{proof}
  One implication is trivial; for the other, suppose that
  $\underline{P} = \uA$. For $a \in \alg{A}$, denote by $C^*(a)$
  the sub-C*-algebra generated by $a$. When $a$ is self-adjoint,
  $C^*(a)$ is commutative. So $\alg{A}\sa \subseteq P$, whence by
  quasi-linearity of $P$ and the unique decomposition of elements in a
  real and imaginary part, we have $\alg{A} \subseteq P$.
  \qed
\end{proof}

\subsection{Algebraic integration theory}\label{ss:th-integrals}
The well-known correspondence between states on commutative \ca s $A$ and
probability measures
on the underlying Gelfand spectrum $\Sg$ is an immediate consequence of the
Gelfand isomorphism
$A\cong C(\Sg,\C)$ and the Riesz-Markov representation theorem in measure
theory.
In the present  topos-theoretical setting, it turns out to be more natural
to work with integrals and valuations rather than measures.
 Recall the {\it a priori} difference between these three concepts:
\begin{itemize}
\item {\it measures} are defined on {\it Borel} subsets of some space $X$;
\item  {\it valuations}  are
defined only on the {\it open} subsets of $X$;
\item  {\it integrals}
are positive linear functionals on the (ordered) vector space $C_c(X,\R)$.
\end{itemize}
Classically, if $X$ is locally compact Hausdorff and the measures in
question are suitably regular, there are isomorphisms between these
notions. From a constructive point of view, however, there is a
subtle difference between valuations and integrals.\footnote{The
integral $I(f)$ of a function $f \in C(X)$ is a Dedekind real, so that
it can be approximated by rationals. This may not be the case for the
valuation $\mu (U)$ of an open $U$, as the `kissing' property
(if $r<s$ then $\mu (U) < s$ or $r<\mu (U)$) may fail. Accordingly,
$\mu (U)$ is only a lower real, and can be thought of as a
predicate $r < \mu (U)$ on the rationals. This predicate is downward
closed: if $r < \mu (U)$ and $s \leqslant r$, then $s < \mu (U)$. But in
general, given $\varepsilon > 0$ one cannot approximate $\mu(U)$ up to
$\varepsilon$ with rationals. Given an integral $I$, we can define a
corresponding valuation $\mu_I (U)$ by taking the sup of $I (f)$ over
all $0 \leqslant f \leqslant 1$ with support in $U$. It is remarkable that for
\textit{any} valuation $\mu$ one can
conversely find a (unique) integral $I$ such that $\mu = \mu_I$. So
despite the fact that one may not be able to compute $\mu (U)$, it is
still possible to compute $\int fd \mu$ as a Dedekind real, which \textit{a
priori} is only a lower real.}
In any case, the fundamental role  locales play in this paper as the
Gelfand spectra of the internal C*-algebras $\uA$ makes it quite
natural to assign probabilities to opens (rather than Borel subsets)
of the spectrum.

The following string of definitions gives an abstract (and
constructive) version of integration theory based on ordered vector spaces,
abstracting from the Riemann, Lebesgue and Daniell
integrals~\cite{Coquand/Spitters:integrals-valuations,
coquandspitters05, spitters:caitnoac}.  Several
axiomatizations are possible, of which the one in terms of so-called f-algebras
is the most convenient for our purposes.
\begin{definition}\label{def:falgebra}
  A \emph{Riesz space} or \emph{vector lattice} is a partially ordered
  vector space $(R,\leqslant)$ over $\R$  (\ie\ a real vector space $R$
  with partial ordering $\leqslant$ such that $f \leqslant g$ implies $f+h \leqslant
  g+h$ for all $h$ and $f \geq 0$ implies $rf \geq 0$ for all
  $r\in\R^+$) that is a distributive lattice with respect to its partial
  order~\cite[Definition~11.1]{LuxemburgZaanen}.

  An \emph{f-algebra} is a commutative, unital, real algebra $R$
  whose underlying vector space is a Riesz space in which $f,g \geq
  0$ implies $fg \geq 0$, and $f \wedge g = 0$ implies $hf \wedge g =
  0$ for all $h\geq 0$. Moreover, the multiplicative unit $1$ has to be
  \emph{strong} in the sense that for each $f\in R$ there exists a
  natural number $n$ such that $-n1 \leqslant f \leqslant
  n1$~\cite[Definition~140.8]{Zaanen:RieszII}.
\end{definition}
Note that although f-algebras are {\it a priori} defined over the
real numbers, they can {\it a fortiori} be defined over $\Q$ as well.

The self-adjoint part of any commutative C*-algebra $\alg{A}$ is an
example of an f-algebra by defining $a\leqslant b$ in the usual way (\ie
iff  $\exists_{c \in A}[b-a=c^*c]$); one has $f\vee g=\max\{f,g\}$ and
$f\wed g=\min\{f,g\}$. Conversely,  by the Stone-Yosida representation
theorem every f-algebra can be densely embedded in a space of real
continuous functions on a compact space.
\begin{definition}
\label{def:probabilityintegral}
  An \emph{integral} on an ordered vector space $R$
  is a linear functional $I:R \to \field{R}$ that is positive,
  \ie\ if $f \geq 0$ then also $I(f) \geq 0$.
  If $R$ has a strong unit $1$ (\emph{e.g.}, the multiplicative unit in the
  case of f-algebras), then an integral $I$ satisfying
  $I(1)=1$ is called a \emph{probability integral}. An integral is
  \emph{faithful} when its kernel is $\{0\}$, \ie, when
  $I(f)=0$ and $f \geq 0$ imply $f=0$.
\end{definition}
Except in the degenerate case $I(1)=0$, any integral can obviously be
normalised to a probability integral.
The prime example of an integral is the Riemann or Lebesgue
integral on the ordered vector space $C[0,1]$. More generally, any
positive linear functional on a commutative C*-algebra provides an example,
states yielding probability integrals.

We wish to use a certain generalization of the Riesz-Markov theorem that
can be proved
constructively~\cite{Coquand/Spitters:integrals-valuations} and hence
can be used within our topos $\TA$. This requires a localic reformulation of
Definition \ref{def:falgebra}, as well as a similar approach to
valuations.

Let $R$ be an f-algebra (in \Sets, for the moment).  In defining the following
frame it is technically convenient to define $R$ as a vector space over $\Q$.
Define $\prop{Integral}(R)$ as the distributive lattice freely
generated by $\prop{P}_f$, $f\in R$, subject to the
relations
\begin{align*}
                \prop{P}_1&=\top,\\
  \prop{P}_f\wedge \prop{P}_{-f} & =\bot, \\
  \prop{P}_{f+g}            & \leqslant \prop{P}_{f}\vee \prop{P}_g,\\
  \prop{P}_f            & =\bot			& (\mbox{for }f\leqslant 0).
\end{align*}
This lattice generates a
frame $\mathcal{O}(\CI(R))$ by adding the regularity condition
\beq\prop{P}(f)=\bigvee_{\Q\ni q>0}\prop{P}(f-q)\eeq to the relations above,
just
like \eqref{eq:continuity} in the case of the spectrum. It can be shown ({\it
cf.}\ \er{subb})
that
\beq \prop{P}_f=\{\rho:R\raw \R \mid \rho(f) > 0\},\label{propP}\eeq
where each $\rh$ is understood to be a positive linear functional.
Models of this theory, \ie\ points of the associated locale,  precisely
correspond to probability integrals
on $R$; if $I$ is such an integral, the associated model $m_I$ is given by
$m_I(\prop{P}_f)=1$ iff $I(f)>0$.
Conversely, a model $m$ defines an integral $I_m$ by (compare with the
proof of Theorem~\ref{prop:kochenspecker})
 $$
I_m(f):=(\{p\mid
m\models \prop{P}_{f-p}\}, \{q\mid
m\models \prop{P}_{q-f}\}), $$
where the right-hand side is seen to be a Dedekind real from the
relations on $\prop{P}_\bullet$.
All this may be internalized to any topos, where, of course, there is
no {\it a priori} guarantee that points of the locale with frame
$\mathcal{O}(\CI(R))$ exist (and hence that expressions like \er{propP} make
good
sense).

The final ingredient of the constructive  Riesz-Markov theorem is the
definition of a locale of valuations. These were studied in~\cite{heckmann}
and~\cite{vickersintegration}.
\begin{definition}
  A {\it probability valuation} on a locale $X$ is a monotone map
  $\mu:\CO(X)\raw [0,1]_l$ that satisfies the usual additivity and
  regularity conditions for measures, \ie\ $\mu(U)+\mu(V)=\mu(U\wed
  V)+\mu(U\vee V)$ and $\mu(\bigvee_{\lm} U_{\lm}) =\bigvee_{\lm}\mu(
  U_{\lm})$ for any directed family. (Here, $[0,1]_l$ is the
  collection of lower reals between 0 and 1.)
\end{definition}

Like integrals, probability valuations on $X$ organize themselves in a locale
$\CV(X)$.

The generalized Riesz-Markov Theorem, then, is as follows.
\begin{theorem}
\label{thm:CSI}
  {\normalfont\textbf{\cite{Coquand/Spitters:integrals-valuations}}}
  Let $R$ be an f-algebra and let $\Sigma$ be its
  spectrum.\footnote{See \cite{coquand05} for the notion of the
  spectrum of an f-algebra, which is described exactly as in
  Subsection \ref{subsec:spectrum}. If the f-algebra is the
  self-adjoint part of a commutative C*-algebra, then its spectrum as
  an f-algebra coincides with  its spectrum as a \ca.}
  Then
the locales $\CI(R)$ and $\CV(\Sigma)$
  are isomorphic. To obtain an integral from a valuation we define:
\[\begin{array}{lll}
  I_{\mu} f & \assign & (\sup_{(s_i)} \sum s_i \mu (s_i < f < s_{i+1}),
  \inf_{(s_i)} \sum s_{i + 1} (1-\mu (s_i > f) - \mu (f > s_{i+1})),
\end{array}\]
where $(s<f)$ is a notation for $\prop{D}_{f-s}$ and
$(s<f<t)$ denotes $\prop{D}_{f-s}\wedge \prop{D}_{t-f}$ and $s_i$ is a partition
of
$[a,b]$ such that $a\leqslant f\leqslant b$.
Conversely, to obtain a valuation from an integral $I$ we define:
\[\mu_I (\prop{D}_a) := \sup \left\{ I (n a^+ \wedge 1) | n \in \mathbb{N}
\right\}.\]

\end{theorem}
Note that  both locales in question are compact
regular~\cite{Coquand/Spitters:integrals-valuations}. Logically
speaking, the theorem follows from the existence of a
 bi-interpretation between the geometric theories $\prop{Integral}(R)$
and $\prop{Valuation}(\Sg)$  (\ie\ there are interpretation maps in two
directions which are each other's inverses) and the equivalence of
the category of
propositional geometric theories with interpretations
to the category of frames.

\subsection{From states on $A$ to subobjects of $\mathcal{O}(\ulS)$}\label{SSS}

We return to our main topic. Since everything in this section so far may be
interpreted in the internal language of a topos and the proof of Theorem
\ref{thm:CSI} is constructive, we have:
\begin{corollary}\label{thm:11}
  Let $A$ be a \ca\ with Bohrification $\ulA$ and associated Gelfand
  spectrum $\uS$ in the topos $\TA$. Then the locale $\CI(\ulA)$ of
  probability integrals on $\uA\sa$ is homeomorphic to the locale
  $\CV(\ulS)$ of probability valuations on $\ulS$.
\end{corollary}

As announced at the beginning of this section,  the next theorem crosses two
levels of Figure \ref{fig:toposlevels}.
\begin{theorem}
\label{thm:statesareintegrals}
  There is a bijective correspondence between quasi-states
  on $\alg{A}$ and either probability integrals on $\functor{\alg{A}}\sa$, or,
  equivalently, probability valuations on its Gelfand spectrum $\ulS$.
\end{theorem}
This theorem may actually be extended to a correspondence between
(faithful) positive quasi-linear functionals on $A$ and (faithful)
integrals on $\functor{\alg{A}}$, etc.
\begin{proof}
  Every positive quasi-linear functional $\rh$ gives a natural transformation
  $I_{\rh}:\uA\sa\raw\underline\R$ if we define its components
   $(I_{\rh})_C:C\sa\raw\R$ to be $\rh_{|C\sa}$ (\ie\ the restriction of $\rh$
to $C\sa\subseteq A\sa$).

Conversely, let $I:\functor{\alg{A}}\sa \raw\underline\R$
   be an integral. Define $\rho:A\sa \to \R$
  by
  \[ \rho(a) = I_{C^*(a)}(a). \]
   For commuting $a,b \in A\sa$,
\begin{align*}
   \rho(a+b)
    & = I_{C^*(a+b)}(a+b)\nn \\
    & = I_{C^*(a,b)}(a+b) \nn \\
    & =  I_{C^*(a,b)}(a) + I_{C^*(a,b)}(b)\nn \\
    & = I_{C^*(a)}(a) + I_{C^*(b)}(b) \nn \\
    & = \rho(a) + \rho(b),
\end{align*}
  because $I$ is a natural transformation, $C^*(a) \cup C^*(b)
  \subseteq C^*(a,b) \supseteq C^*(a+b)$, and $I$ is
  locally linear.  Moreover, $\rho$ is
  positive because $I$ is locally positive (see
  Lemma~\ref{cor:propertyonjlbextends}). Hence we have defined $\rh$
  on $A\sa$ and may extend it to $A$ by complex linearity. It is clear
  that the two maps $I\mapsto \rh$ and $\rh\mapsto I$ and are inverses
  of each other and that if $I$ is a probability integral, then $\rh$
  is a quasi-state, and {\it vice versa}.
  \qed
\end{proof}
In the Introduction, we have seen that in the classical case a (pure)
state $\rh$ defines a subobject $[\rho]$ of the frame of opens of the
classical phase space; see~\er{old3}. As we shall now show, this
remains true, {\it mutatis mutandis}, in the quantum case. The main
technical difficulty is to adapt the condition $\dl_{\rho}(V)=1$
in~\er{old3}.

Theorem \ref{thm:statesareintegrals} yields a bijective correspondence
between quasi-states $\rh$ on $A$ and probability valuations $
\mu_{\rh}$ on $\ulS$. Fix a state, or quasi-state, $\rh$ on $A$.  The
logical formula $\mu_{\rho}(-)=1$ (of the Mitchell-B{\'e}nabou
language of $\TA$) is a predicate on $\CO(\ulS)$ and hence defines a
subobject $[\rh]$ of $\CO(\ulS)$ with  characteristic arrow
$\ch_{[\rh]}:\CO(\ulS)\raw\underline{\Om}$. This arrow is just the
interpretation of $\mu_{\rho}(-)=1$, \ie\
\beq
  \ch_{[\rh]}= \interpretation{\mu_{\rho}(-)=1}.\label{rhoint}
\eeq
Compare with \er{old3}; beyond mimicking the notation, we see that we
have been able to transfer the classical description of states to the
quantum situation in every respect.
\section{Observables and propositions}
\label{sec:interpretationofquantumphysics}

In this section and the next we give the details of  steps 2 to 5 of
our five-step program for spatial quantum logic; {\it cf.}\
Subsections~\ref{subsec:SQL} and~\ref{sec:Bohr}. We start with the
locale map  $\dl(a):\ulS\raw\uIR$, then turn to the description of
elementary propositions $a\in\Dl$ as opens in the spectrum $\ulS$, and
finally consider the pairing of states and propositions to arrive at a
suitable notion of (multi-valued) truth in quantum theory.

\subsection{Interval domain}\label{subsec:ID}

For a commutative unital \ca\ $A$ with Gelfand spectrum $\Sg$ in
\Sets, the Gelfand transform of $a\in A\sa$ is a continuous function
$\hat{a}:\Sg\raw\R$. Equivalently, it is a locale map \er{GTlm}. As we
have seen  in Subsection~\ref{sec:GT}, {\it mutatis mutandis} the
description \er{GTlm} still applies when $A$ is a commutative unital
\ca\ $A$ with Gelfand spectrum $\Sg$  in a topos $\CT$. In particular,
one has the Gelfand transform
\beq
  \hat{a}:  \ulS\raw \ulR \qquad (a\in \uA\sa).\label{beyond}
\eeq
Our problem, however, is to express an element $a\in A\sa$ of a {\it
noncommutative} \ca\ $A$ in $\Sets$ in terms of some locale map
$\dl(a)$ defined on the spectrum $\ulS$ of the Bohrification $\uA$
of $A$ in $\TA$. As we shall see, this problem can be solved if we
introduce some fuzziness, in that $\dl(a)$ no longer takes values
in the internal Dedekind reals $\underline\R$ in $\TA$, like
$\hat{a}$, but in the so-called {\it interval domain} $\IR$,
internalized in $\TA$ as $\underline{\IR}$. Thus, apart from
\er{beyond} we are dealing with a second locale map
\beq
\underline{\dl}(a):  \ulS\raw \underline{\IR} \qquad (a\in A\sa).\label{Da1}
\eeq
In honour of D\"{o}ring and Isham, we refer to $\underline{\dl}(a)$ as the {\it
Daseinisation} of $a$ (although our map differs from theirs, {\it
cf.}\ Appendix~\ref{sec:doringisham}).

We have already encountered Scott's interval domain $\IR$ in
Subsection~\ref{subsec:AO} as the poset of compact intervals in $\R$,
ordered by inverse inclusion. Like the Dedekind real numbers, the interval
domain is easily internalized and hence definable in any topos. In
fact, the construction of the Dedekind real numbers in
Subsection~\ref{sec:GT} only requires a single modification so as to
obtain the interval domain: the corresponding frame  $\CO(\IR)$ is
defined by the very same generators $(p,q)$ and relations as
$\CO(\R)$, except that the fourth relation (\ie\ $(p,q)=(p,q_1)\vee (p_1,q)$ if $p\leqslant p_1\leqslant q_1\leqslant q$)
is dropped. The models of
$\CO(\IR)$ or points of the associated locale $\IR$ again correspond
to pairs $(L,U)$ given by~\er{eq21} and~\er{eq22}, but this time such
a pair may fail to define a Dedekind
cut; axiomatically, only the `kissing' requirement no longer holds.
In any topos $\CT$, we denote the locale defined by the geometric propositional  theory given by 
the first three axioms in the list following \er{GT2}   in
Subsection~\ref{sec:GT} --- interpreted in $\CT$ --- by $\CO(\IR)_{\CT}$, with the usual special case $\underline{\IR}\equiv\IR_{\TA}$. 
Similarly, the  subobject of $\CP(\Q)\x\CP(\Q)$
consisting of models  of $\CO(\IR)_{\CT}$ is denoted by
$\mathrm{Pt}(\IR)_{\CT}$, with $\mathrm{Pt}(\IR)_{\TA}\equiv \mathrm{Pt}(\underline{\IR})$.

The examples in Subsection~\ref{sec:GT} now read
as follows:
\begin{enumerate}
  \item  In \Sets\ (or, more generally, when classical logic applies in
    $\CT$), a cut $(L,U)$ defines a compact interval $[\sup L,\inf U]$
    (where $\sup$ and $\inf$ are taken in $\R$), so that
    $\mathrm{Pt}(\IR)$ may be identified with the classical Scott interval domain
    $\IR$. In that case, a generator $(p,q)\in\CO(\IR)$ may be
    identified with the Scott open in $\IR$ that contains all
    intervals $[a,b]$ such that $p<a\leqslant b<q$.

  \item In a topos $\Sh(X)$ of sheaves one has 
  \beq \CO(\R)_{\Sh(X)}: U\mapsto \CO(U\x \IR),\eeq
  but  its points are not as easily described as \er{DRN}; instead, one has
  \beq
  \mathrm{Pt}(\IR): U\mapsto
     \{(f,g) \mid f,g:U \to \field{R} \mid f \leqslant g,
         f \mbox{ lower-semicont.},  
        g \mbox{ upper-semicont.}\}
.\eeq
    This follows by carefully adapting the proof
    of~\cite[Theorem~VI.8.2]{maclanemoerdijk92} for $\R$. 
  \item
    In particular, for  $\TA=\Sets^{\CA}$, one has 
       \beq
    \CO(\underline{\IR}):
    C\mapsto \CO((\uparrow\! C)\x \IR), \label{ddrr}\eeq
   which may be identified with the set of monotone functions from $\uparrow\! C$ to $\CO(\IR)$.\footnote{The argument is the same as for $\R$, see footnote \ref{monotone}.}
     
   The object  $\mathrm{Pt}(\underline{\IR})$ will not be used in this paper.\footnote{For completeness, we mention that $\mathrm{Pt}(\underline{\IR})(C)$ is the set of all pairs $(L,U)$, where $L$ and $U$ are subfunctors of the constant functor $\mathbb{Q}$, truncated to $\uparrow\! C\subset\CA$, such that for all $D\supseteq C$, $(L(D),U(D))$ is a pair 
   of the form \er{eq21}--\er{eq22}.}
 \end{enumerate}

\subsection{Daseinisation}\label{sec:Das}
After this preparation, we turn to the
{\it Daseinisation} \er{Da1}, or rather to the corresponding frame map
 \beq
\underline{\dl}(a)\inv:  \CO(\underline{\IR}) \raw \CO(\ulS).\label{Da22}\eeq
A complete description of this map, based on the technique of generating (semi)lattices for frames, may be found in Appendix \ref{AppA:Das}. Here, we 
 just look at  the special case
\beq (\xymatrix@1{\underline{1}\ar^-{\underline{\dl}(a)\inv \underline{(r,s)}}[rr] &&  \CO(\ulS)})
= (\underline{1}\sr{\underline{(r,s)}}{\longrightarrow}\CO(\uIR)\sr{\underline{\dl}(a)\inv}{\longrightarrow}\CO(\ulS)),
\label{simplify}
\eeq
where the arrow $\underline{(r,s)}:\underline{1}\raw \CO(\uIR)$ maps into the monotone function  with constant value $\downarrow\!(r,s)$.\footnote{Here $(r,s)$ is seen as an element of the generating semilattice $\Q\x_{<}\Q$, whereas
$\downarrow\!(r,s)$ is its image in the frame $\CO(\IR)$ through the canonical map \er{canmap}; see the Appendix.} 
We may even simplify  \er{simplify} even further by localizing it  at $\C\cdot 1$; this,  however, entails no loss of generality, 
for $\CO(\ulS)(\C\cdot 1)$ is the frame in \Sets\ that (together with the frame map \er{JTmap}) provides the external description
of the internal locale $\ulS$ in $\TA$ (see Subsection \ref{Intext} and Appendix  \ref{AppA:Das}).

The quantity $\underline{\dl}(a)\inv \underline{(r,s)}(\C\cdot 1)$
 is a global element $\underline{U}$ of $\CO(\ulS)(\C\cdot 1)$ as described by Theorem \ref{finaltheorem}
 in the Appendix. Briefly, this theorem states that $\CO(\ulS)(\C\cdot 1)$ may be seen as the set of all subfunctors $\underline{U}$ of the functor $C\mapsto L_C$ that satisfy a certain regularity condition,
 where $L_C$ is the  distributive lattice  freely generated by the formal symbols $\prop{D}_c$, $c\in C\sa$ subject to the relations \er{eq:spectrum1}--\er{eq:spectrum5} (simply interpreted in \Sets).\footnote{See Appendix \ref{App:localization}
 for a detailed description of the functor $C\mapsto L_C$.}
 Abbreviating
\beq
\dl(a)\inv(r,s)=\underline{\dl}(a)\inv \underline{(r,s)}(\C\cdot 1),
\eeq
 the ensuing element $\dl(a)\inv(r,s)$ of $\CO(\ulS)(\C\cdot 1)$ turns out to be the functor
\beq
\dl(a)\inv(r,s): C \mapsto \{{D}_{f-r} \wedge
 {D}_{s-g}\mid f,g\in C\sa, f \leqslant a \leqslant g\}.
\label{Da}
\eeq
This follows from \er{defdapp} and the definition of $\underline{\dl}(a)\inv$
 in Appendix \ref{AppA:Das}, combined with the equality
$$ \bigcup_{\{p<q\mid (p,q)\subseteq (r,s)\}}  \{{D}_{f-p} \wedge
 {D}_{q-g}\mid f,g\in C\sa, f \leqslant a \leqslant g\}=\{{D}_{f-r} \wedge
 {D}_{s-g}\mid f,g\in C\sa, f \leqslant a \leqslant g\}.$$

An alternative description of $\dl(a)\inv$ is as follows.
For fixed $a\in A\sa$, define functors $\underline{L_a}\in\TA$ and
$\underline{U_a}\in\TA$ by
 \begin{eqnarray}
\underline{L_a}(C)&=& \{f\in C\sa\mid f\leqslant a\}; \nn\\
\underline{U_a}(C)&=& \{g\in C\sa\mid a\leqslant g\}.
\end{eqnarray}
Each of these defines a subobject of $\uA\sa$. In fact, the pair $(\underline{L_a},
\underline{U_a})$ is a
\emph{directed} subobject of $\uA\sa\times \uA\sa\op$.
Now take a  generator $(r,s)$ of $\CO(\IR)$, and write $(r,s)=(-\infty,s)\wedge (r,\infty)$.
The Gelfand transform \er{GTlc} defines subobjects $\hat{\cdot}(\underline{L_a},
(r,\infty))$
and $\hat{\cdot}(\underline{U_a}, (-\infty,s))$ of $\CO(\ulS)$. We then put
\beq
  \dl(a)\inv: (r,s)\mapsto \bigvee \hat{\cdot}(\underline{L_a}, (r,\infty))
\wedge \hat{\cdot}(\underline{U_a}, (-\infty,s)).
  \eeq
Using \er{GT0} and \er{GT}, this gives\footnote{Using
a generic point $\sigma$, we may even define\[\dl(a)(\sigma):=(\sup
\sigma(\underline{L_a}),\inf \sigma(\underline{U_a})).\] Analoguously, one can view $\dl(a)$ as an interpretation of
the geometric theory $\Sigma$ in the geometric theory of the intervals;
see~\cite{Coquand/Spitters:integrals-valuations}.}
\beq
  \dl(a)^{-1}(r,s)
  = \bigvee_{f \in \underline{L_a}, g \in \underline{U_a}} \prop{D}_{f-r} \wedge \prop{D}_{s-g}.
  \label{dasbasbis}
\eeq

To illustrate what is going on, it is helpful to
 compute the right-hand side of \er{Da} or \er{dasbasbis} in \Sets\ for $A=C=C(\Sigma,\C)$.  In that case the meaning of $\prop{D}_a$ is given by \er{subb}, so that with $\rh(f)=f(\rh)$ one finds
 $\prop{D}_{f-r}=\{\rh\in\Sg\mid f(\rh)>r\}$ and $\prop{D}_{s-g}=\{\rh\in\Sg\mid g(\rh)<s\}$.
One then obtains (with $\wedge$ for `and')
  \begin{eqnarray}
\dl(a)_A^{-1}(r,s)&=& \bigcup_{ f,g\in C\sa, f \leqslant a \leqslant g}\{\rh\in\Sg\mid f(\rh)>r \wedge
g(\rh)<s\} \nn \\ &=&
 \{\rho \in \Sg \mid \exists_{f \leqslant a}[f(\rho)>r \wedge f(\rho)<s]
     \wedge\exists_{g \geq a}[g(\rho)> r \wedge g(\rho)<s]\}\nn \\
  & =& \{\rho \in \Sg \mid r < a(\rho) < s\}\nn \\
  & = &a^{-1}(r,s). \label{ainverse}\end{eqnarray}

 To close this subsection, we note the following.
\begin{proposition}
\label{prop:selfadjointsascuts}
  The map $\dl:A\sa\raw C(\ulS,\uIR)$ is injective, and
  $a\leqslant b$ iff $\dl(a)\leqslant\dl(b)$.
\end{proposition}
\begin{proof}
  Suppose that $\dl(a)=\dl(b)$. Then for all $\alg{C} \in\CA$,
  the sets $L_a(C)=\{f\in C\sa\mid f\leqslant a\}$ and $U_a(C)=\{g\in
  C\sa\mid a\leqslant g\}$ must coincide with $L_b(C)$ and $U_b(C)$, respectively.
  Imposing these equalities at $C=C^*(a)$ and at $C=C^*(b)$ yields $a=b$.
  The order in $A\sa$ is clearly preserved by $\dl$, whereas the
  converse implication can be shown by the same method as the first
  claim of the proposition.
  \qed
\end{proof}

\subsection{Propositions}
It immediately follows from the existence of the {\it Daseinisation} map \er{Da22} (see
Subsection \ref{sec:Das} and  Appendix \ref{AppA:Das})
that, as in the classical case, elementary propositions $a\in\Dl$
define opens in phase space. For an open in the `quantum phase space'
$\ulS$ is simply defined as a global element
$\underline{1}\raw\CO(\ulS)$ ({\it cf.}\ Subsection
\ref{subsection:locale2})), so that given an observable $a\in A\sa$ and
a Scott open $\Delta\in\CO(\uIR)$, we may combine the corresponding arrows
$\underline{\dl}(a)\inv:\CO(\uIR)\to\CO(\ulS)$ and
$\underline{\Delta}:\underline{1} \to \CO(\uIR)$
 into
\beq
 ( \xymatrix@1{\underline{1} \ar^-{[a\in\Dl]}[rr] && \CO(\ulS)})
  \qquad = \qquad
  (\xymatrix@1{\underline{1} \ar^-{\underline{\Delta}}[r] & \CO(\uIR)
  \ar^-{\underline{\dl}(a)\inv}[rr] && \CO(\ulS)}).
  \label{pijl}
\eeq
This generalises \er{simplify}; in particular, $\underline{\Delta}:\underline{1} \to \CO(\uIR)$ is
defined at $C$
as  the monotone function $\uparrow\!C\raw  \CO(\IR)$ taking constant value $\Delta$.
 In other words, 
\beq
  \label{eq:physicalproposition}
  [a\in\Dl]= \underline{\dl}(a)\inv\circ\underline{\Delta}.
\eeq

\section{State-proposition pairing}\label{Spp}

In Subsection~\ref{SSS} we have shown how a state $\rh$ on $A$ gives
rise to a subobject $[\rh]$ of $\CO(\ulS)$ defined by the predicate
$\mu_{\rho}(-)=1$, and hence to an arrow $\xymatrix@1{\CO(\ulS)
\ar^-{\ch_{[\rh]}}[r] & \functor{\Omega}}$ related to the predicate in
question by~\er{rhoint}.

Also, we have just seen the description~\er{pijl} of propositions
$a\in\Dl$ as opens in $\ulS$. Hence we can pair a physical state
$\rho$ and a physical proposition $a \in \Delta$ by composition, to
end up with a `truth value' $\langle a \in \Delta, \rho \rangle$  in
the subobject classifier $\functor{\Omega}$ of $\TA$. Explicitly, one
has
\beq
  \xymatrix@1{\underline{1} \ar^-{\langle a \in \Delta, \rho
  \rangle}[rr] && \functor{\Omega}}
  \qquad = \qquad
  \xymatrix@1{\underline{1} \ar^-{[a \in \Delta]}[rr] && \CO(\ulS)
  \ar^-{\ch_{[\rh]}}[r] & \functor{\Omega},}
  \label{eq:pairing}
\eeq
 or
 \beq \langle a \in \Delta, \rho \rangle=\ch_{[\rh]}\circ \underline{\dl}(a)\inv\circ\underline{\Delta}.\eeq

In what follows, we need the basic definitions of  Kripke-Joyal semantics. If $\phv$ is some formula
interpreted in a  topos $\CT$ as an arrow $\interpretation{\phv}:F\raw\Om$, and $\al:B\raw F$ is any arrow in $\CT$ (defining a `generalized element' of $F$), then the notation  $B\Vdash\phv(\al)$, or, less precisely,  $B\Vdash\phv$ (for `$B$ forces $\phv$') means that the composite arrow
$B \stackrel{\al}{\raw} F\stackrel{\interpretation{\phv}}{\longrightarrow} \Om$ factors through
$\top:1\raw\Om$. In a functor topos $\Sets^{\mathbf{C}}$, where $\mathbf{C}$ is some category, the notation $C\Vdash \phv$ for some $C\in\mathbf{C}$ is shorthand for
$y(C)\Vdash \phv$, where $y(C):D\mapsto \mathrm{Hom}_{\mathbf{C}}(D,C)$ is the Yoneda functor. In our case $\CT=\TA$, the interpretation $\interpretation{\phv}$ is
a natural transformation $\underline{F}\raw\underline{\Om}$,
given by its components $\interpretation{\phv}(C): \underline{F}(C)\raw
\underline{\Om}(C)$, where $C\in\CA$. In that case the forcing condition
$C\Vdash \phv$ turns out to be
equivalent to  $\interpretation{\phv}(C)(\underline{F}(C))=\top_C$, where $\top_C$ is the maximal upper set on $C$.

Using the Kripke-Joyal semantics of $\TA$, we now explicitly
compute the state-proposition pairing in case that $\Dl=(r,s)$ is a rational interval.
 The computation is straightforward
when using generating lattices (see Appendix). From here on, $\rh$ is a fixed state on $A$ and we abbreviate
$\mu_{\rho}$ by $\mu$. For $D \in \CA$,
\begin{align*}
  (\langle a \in (r,s), \rho \rangle)_D(*)
  & \stackrel{\eqref{eq:pairing}}{=}
    (\ch_{[\rh]} \after [a \in (r,s)])_D(*) \\
  & \stackrel{\eqref{eq:physicalproposition},\eqref{rhoint}}{=}
    \interpretation{\mu(\underline{\dl}(a)^{-1}(r,s))=1}(D).
\end{align*}
Being a global element $\underline{1} \to \underline{\Omega}$ of
the subobject classifier $\functor{\Omega}$ of $\TA$, the right-hand side
is an element of  the set $\functor{\Omega}(D)$, and hence an upper set on $D$.
With slight abuse of notation, we simply call the latter
$\langle a \in (r,s), \rho \rangle(D)$.
It follows that
\beq  \langle a \in (r,s), \rho \rangle(D)=
 \{ C\in \CA \mid C\supseteq D,  C \Vdash \mu(\underline{\dl}(a)^{-1}(r,s))=1 \},\label{truth}
\eeq
where $\mu\circ\underline{\dl}(a)^{-1}(-)=1$ is the obvious predicate on $\CO(\underline{\IR})$
defined by $\mu(-)=1$ on $\CO(\ulS)$ and the {\it Daseinisation} map \er{Da22}.
Since $\langle a \in (r,s), \rho \rangle(D)$ is the truncation to $\uparrow\! D$ of the corresponding
upper set at $\C\cdot 1$, we may use  \er{Da} or \er{dasbasbis}, from which  we see that the forcing condition \hbox{$C \Vdash
\mu(\underline{\dl}(a)^{-1}(r,s))=1$} is equivalent to
\[
 \mu_C\left( \bigvee_{f \leqslant a \leqslant g,\, f,g\in C\sa} \prop{D}_{f-r} \wedge
  \prop{D}_{s-g} \right) = 1.
\]
Here $\mu_C$ is the valuation defined as $\mu_{\rh}$, but with $\rho$
restricted to $C$. Similarly, $\prop{D}_{f-r}$ refers to an open in
the spectrum of $C$ ({\it cf.}\ Theorem \ref{thm:internalspectrum},
according to which the $\prop{D}_a$ with $a\in C\sa$ may be seen as
generators of the spectrum of $C$). Since the measure of the
intersection of two opens equals one if the measures of both opens do,
this means (for $f,g\in C\sa$)
\[
 \mu_C\left(\left( \bigvee_{f \leqslant a} \prop{D}_{f-r} \right) \wedge
            \left( \bigvee_{g \geq a} \prop{D}_{s-g} \right)\right) = 1,
\]
which happens if and only if
\[
 \mu_C\left( \bigvee_{f \leqslant a} \prop{D}_{f-r} \right) = 1
  \mbox{  and  }
 \mu_C\left( \bigvee_{g \geq a} \prop{D}_{s-g} \right) = 1.
\]
The left conjunct means
\begin{equation}
 \forall_{n \in \field{N}}\, \exists_{f\in C\sa,\,  f \leqslant
  a}\, [ \mu_C(\prop{D}_{f-r}) > 1-\frac{1}{n}]\label{equals-one},
\end{equation}
  since $\TA$ is a functor topos and hence the quantifiers
above are interpreted locally.
The construction of $\mu_{\rh}$ from $\rh$ (see Section \ref{sec:statesasintegrals}) implies
$$\mu_C(\prop{D}_h) = \lim_{m \to \infty} \rho((mh^+)\wedge 1),$$
where the limit is a lower real. In other words, $\mu_C(\prop{D}_h)>q$ iff
there exists $m$ in $\field{N}$ such that $\rho((mh^+) \wedge 1) > q$.
So $C \Vdash \mu_{\rho}(\bigvee_{f \leqslant
a}\prop{D}_{f-r})=1$ means that
for each $n \in \field{N}$ there exists $f \in C$ with $f \leqslant a$ and
$\mu_C(\prop{D}_{f-r})>1-\frac{1}{n}$.

Hence at the end of the day the
state-proposition pairing $\langle a \in (r,s), \rho \rangle$
explicitly yields the upper set at $D$ given by
\begin{eqnarray}\label{eq:51}
\lefteqn{ \langle a \in (r,s), \rho \rangle(D) =}  \\ & &
\hspace*{-0.75cm}
\left\{C \in \CA \mid C\supseteq D,
      \mu_C  \left(\bigvee_{f\leqslant a, f\in C\sa} \prop{D}_{f-r}\right)=1 \mbox{ and }
      \mu_C  \left(\bigvee_{a\leqslant g, g\in C\sa} \prop{D}_{s-g}\right)=1 \right\}.\nonumber
\end{eqnarray}
This formula can be put in a slightly more palpable form when
$A$ and each $C\in\CA$ are von Neumann algebras (in
the ambient topos $\Sets$). In that case, it can be shown \cite{HLSSyn} that the open $\prop{D}_{f-r}$ in the spectrum
corresponds to a projection operator $[\prop{D}_{f-r}]$, to which we can directly apply the state $\rh$. Moreover, unlike for general \ca s,
 the supremum $P=\bigvee \{[\prop{D}_{f-r}] \mid f\leqslant a, f\in C\sa  \}$
  exists. One then simply has $\mu_C(P)=1$
when $\rho(P)=1$. Similarly, the projection
$Q=\bigvee \{[\prop{D}_{s-g}] \mid a\leqslant g, g\in C\sa  \}$ exists and
 $\mu_C(Q)=1$
when $\rho(Q)=1$.\\

To close, we remark that one might  consider a proposition $\mu_{\rho}(-)>p$, for some rational
number
$p$, instead of the proposition $\mu_{\rho}(-)=1$ as in this paper. This would simplify the
computations above slightly. For instance, \eqref{equals-one} would become
\[
 \exists_{f\in C\sa,\,  f \leqslant
  a}\, [ \mu_C(\prop{D}_{f-r}) > p].
\]
This eliminates a universal quantification, but otherwise the computations
would continue \emph{mutatis mutandis} as before.
\appendix
\section{Generating lattices for frames}
\label{AppA}
At various places in this article we refer to a presentation
of a frame (or locale) by a generating lattice with a covering
relation. This technique has been developed in the context of formal topology \cite{Sambin87, sambin}, and extends an
analogous construction due to Johnstone \cite{johnstone82}. Note that
formal topology may be developed in the framework of constructive set theory
\cite{aczel}, and hence may be internalized in topos theory.

Let $(L,\leqslant)$  be a meet semilattice (\ie\ a poset in which any pair of elements has a meet = g.l.b.\ = infimum; in most of our applications $(L,\leqslant)$ is actually a distributive lattice).
\begin{definition}\label{def:cover}
A {\it covering relation} on $L$ is a relation $\drie\,\subseteq L\x\CP(L)$ -
equivalently, a function $L\raw \CP(\CP(L))$ -
 written $x\drie U$ when $(x,U)\in\,\drie$, such that:
 \begin{enumerate}
\item If $x\in U$ then $x\drie U$;
\item If $x\drie U$ and $U\drie V$ (\ie\ $y\drie V$ for all $y\in U$) then $x\drie V$;
\item If $x\drie U$ then $x\wed y\drie U$;
\item If  $x\in U$ and  $x\in V$, then $x\drie U\wed V$ (where $U\wed V=\{x\wed y\mid x\in U, y\in V\}$).
\end{enumerate}
\end{definition}
For example, if $(L,\leqslant)=(\CO(X),\subseteq)$ one may take $x\drie U$ iff $x\leqslant\bigvee U$, \ie\ iff $U$ covers
$x$.

Let  $DL$ be the poset of all lower sets in $L$, ordered by inclusion; this is a frame \cite[\S 1.2]{johnstone82}. 
The structure $\drie$ gives rise to a closure operation\footnote{As a map, $\mathcal{A}$ is also defined on $\CP(L)$. 
Let $\ch_{\drie}:L\x\CP(L)\raw\Om$
be the characteristic function of the subset $\drie\,\subseteq L\x\CP(L)$. Then $\mathcal{A}=\hat{\ch}_{\drie}$ is just the `curry' or `$\lm$-conversion' of $\ch_{\drie}$.}
$\mathcal{A}: DL\raw DL$,  given by
\beq \mathcal{A} U=\{x\in L\mid x\drie U\}, \label{clop}\eeq
which has the following properties: $\downset U\subseteq \mathcal{A}U$,
$U\subseteq \mathcal{A}V\Rightarrow \mathcal{A}U\subseteq \mathcal{A}V$,
  $\mathcal{A}U\cap \mathcal{A}V\subseteq \mathcal{A}(\downset U \cap \downset V)$.   The frame $\CF(L,\drie)$ generated by such a structure is then defined by
\beq \CF(L,\drie)=\{U\in DL\mid \mathcal{A}U=U\}=\{U\in \CP(L)\mid x\drie U\Raw x\in U\};
\label{defCFL}
\eeq
the second equality follows because firstly  the property $\mathcal{A}U=U$ guarantees that $U\in DL$, and secondly one has $\mathcal{A}U=U$ iff  $x\drie U$ implies  $x\in U$.
An equivalent description of  $\CF(L,\drie)$ is
\beq \CF(L,\drie)\cong
\CP(L)/\sim,\label{eq:47} \eeq
 where $U\sim V$ iff $U\drie V$ and  $V\drie U$.  Indeed, the map $U\mapsto [U]$ from
 $ \CF(L,\drie)$ (as defined in \er{defCFL}) to $\CP(L)/\sim$ is a frame map with inverse
 $[U]\mapsto \mathcal{A}U$; hence the idea behind the isomorphism \er{eq:47} is that the map $\mathcal{A}$ picks a unique representative 
 in the equivalence class $[U]$, namely $\mathcal{A}U$.
 
 The frame $\CF(L,\drie)$ comes equipped with a canonical map 
 \begin{eqnarray}
 f:L& \raw & \CF(L,\drie) ; \\
  x&\mapsto&  \mathcal{A}(\downset x), \label{canmap}
\end{eqnarray}
which satisfies
 $f(x)\leqslant \bigvee f(U) \mbox{ if } x\drie U$.
 In fact, $f$ is universal with this property, in that any homomorphism $g:L\raw \mathcal{G}$ of meet semilattices into a frame $\mathcal{G}$ such that $g(x)\leqslant \bigvee g(U)$ whenever $x\drie U$
has  a factorisation $g=\phv\circ f$ for some unique frame map $\phv: \CF(L,C)\raw \mathcal{G}$.
This suggests that the point of the construction is that  $\CF(L,\drie)$ is (isomorphic to)
a frame defined by generators and relations, provided the covering relation is suitably defined in terms 
of the relations. More precisely \cite[Thm.\ 12]{aczel}:
\begin{proposition}\label{preciseprop}
Suppose one has a frame $\CF$ and a meet semilattice\footnote{This even works  in case that $L$ is just a set preordered by $x\leqslant y$ when $f(x)\leqslant f(y)$.} $L$ with a map $f:L\raw \CF$ of meet semilattices
that generates $\CF$ in the sense that for each $U\in\CF$ one has $U=\bigvee\{f(x)\mid x\in L, f(x)\leq U\}$.
Define a cover relation $\drie$ on $L$ by
\beq  x\drie U \mbox{ iff }  f(x)\leqslant \bigvee f(U). \label{unicover}
\eeq
Then one has a frame isomorphism $\CF\cong \CF(L,\drie)$.
\end{proposition}

We now turn to maps between frames.
\begin{definition}\label{def:contmap}
Let $(L,\drie)$ and $(M,\coveredd)$ be meet semilattices with covering relation as above,
and let $f^*:L\raw\CP(M)$ be such that:
\begin{enumerate}
\item $f^*(L)=M$;\footnote{If $L$ and $M$ have top elements $\top_L$ and $\top_M$, respectively, then this condition may be replaced by $f^*(\top_L)=\top_M$.}
\item $f^*(x)\wed f^*(y)\coveredd f^*(x\wed y)$;
\item  $x\drie U\Raw f^*(x)\coveredd f^*(U)$ (where $f^*(U)=\bigcup_{u\in U} f(U)$).
\end{enumerate}
 
Define two such maps $f^*_1, f^*_2$ to be equivalent if $f^*_1(x) \sim f^*_2(x)$ (\ie\ 
$f^*_1(x) \coveredd f^*_2(x)$ and $f^*_2(x) \coveredd f^*_1(x)$)
for all $x \in L$. A
 \textit{continuous map} $f:(M,\coveredd) \raw (L,\drie)$ is an equivalence class of such maps
$f^*:L\raw\CP(M)$.\footnote{
Instead of taking equivalence classes, one could demand as a fourth
condition that $f^*(x) = \mathcal{A}f^*(x)$ for all $x \in L$.}
\end{definition}

Our main interest in continuous maps lies in the following result.\footnote{
In fact, one may extend this into an equivalence $\CF$ between the category of formal topologies
and the category of frames. A formal topology is a generalization of the above triples
$(L,\leqslant,\drie)$, where $\leqslant$ is merely required to be a preorder. In
this more general case, the axioms on the cover relation $\drie$ take
a slightly different form. See \cite{battilottisambin,negri}.}
\begin{proposition}\label{approxmap}
  Each continuous map  $f:(M,\coveredd)\raw (L,\drie) $ is equivalent  to a
  frame map \\  $\CF(f): \CF(L,\drie)\raw  \CF(M,\coveredd)$, given by
  \beq \CF(f):U\mapsto \mathcal{A}f^*(U).\label{apeq}\eeq
\end{proposition}

All results in this subsection may be internalized in any topos; for example,
a covering relation on an internal meet semilattice $L$ in a topos
$\CT$ is simply a subobject $\drie$ of $L\x\Om^L$, where $\Om$ is the
subobject classifier in $\CT$. The defining properties of a covering
relation are then interpreted in the internal language of $\CT$.
 Proposition \ref{approxmap} holds in this generality,  since
its proof is constructive; see especially \cite{aczel}.
\subsection{Localization of the spectrum}\label{App:localization}
We now consider some applications pertinent to the main body of the paper.
First, we return to the Gelfand spectrum in
Subsection~\ref{subsec:spectrum}. In its presentation by means of generators and relations,
eqs.\  \er{eq:spectrum1}--\er{eq:spectrum5} play a different role from the regularity rule \er{eq:continuity}, and we will treat the latter separately. First, for an arbitrary unital commutative \ca\ $A$ in some topos, consider
the  distributive lattice $L_A$ freely generated by the formal symbols $\prop{D}_a$, $a\in A\sa$
(\ie\ $a$ is a variable of type $A\sa$), 
subject to the relations \er{eq:spectrum1}--\er{eq:spectrum5}. As shown in \cite{coquand05,
coquandspitters05}, $L_A$ can be described more explicitly, as follows.

Let $A^+:=\{a \in A\sa \mid a \geq 0\}$. Define $p \preccurlyeq
q$ iff there exists $n\in\mathbb{N}$ such that $p\leqslant
n q$. Define $p \approx q$ iff $p\preccurlyeq q$ and $q\preccurlyeq p$. The
lattice operations on $A$ respect $\approx$ and hence $A^+/\approx$ is a lattice.
We then have
\beq 
L_A \cong A^+/\approx. \label{Papp}
\eeq
The image of the generator  $\prop{D}_a$ in $L_A$, seen as an element of $A^+/\approx$, may also be described explicitly:
 decomposing $a\in A\sa$ as $a=a^+-a^-$ with $a^{\pm}\in A^+$ in the usual way, under the isomorphism \er{Papp} this image coincides with the equivalence class $[a^+]$ in $A^+/\approx$.
In explicit computations \cite{HLSFP,HLSSyn}, one may  therefore simply  identify $L_A$ with $A^+/\approx$ and $\prop{D}_a$ (seen as an element of $L_A$) with $[a^+]$, respectively. Such computations are also greatly facilitated by the following `locality' theorem.
\begin{theorem}\label{thm:internalspectrum}
For each $C\in\CA$ one has
 \beq \underline{L}_{\uA}(C)=L_C, \label{Llocal}\eeq
 where the right-hand side is simply defined in \Sets\ (where it may be computed through \er{Papp}). 
 Furthermore,  if $C\subseteq D$, then the map $\underline{L}_{\uA}(C)\raw \underline{L}_{\uA}(D)$  given by the functoriality of $\underline{L}_{\uA}$
simply maps each generator $D_c$ for $c\in C\sa$  to the same generator
for the spectrum of $D$ (this is well defined because $c\in D\sa$,
and this inclusion  preserves the relations
\er{eq:spectrum1}--\er{eq:spectrum5}); we write this as $L_C\hookrightarrow L_D$.
\end{theorem}

A proof of this theorem by explicit computation may be found in \cite[Thm.\ 5.2.3]{caspers}. Here, we give an alternative proof, which
requires some familiarity with geometric logic
\cite{maclanemoerdijk92,johnstone02b,Vic:LocTopSp}.\footnote{Further to our
remarks in Subsection \ref{subsection:locale2}
on geometric {\it propositional} logic, we recall that a geometric
{\it predicate} logic is a theory whose formulae are as described there (where the atomic
formulae may now involve relations and equalities and all the usual structures
allowed in first-order logic as well),
 now also involving finitely many free variables $x=(x_1,\ldots, x_n)$, and the existential quantifier $\exists$, 
with
axioms taking the form $\forall x: \phv(x)\raw\ps(x)$. 
Geometric formulae form
an important class of logical formulae, because they are precisely the ones
whose truth value is preserved by inverse images of geometric morphisms
between topoi. From their syntactic form alone, it follows that their
interpretation
in the external language is determined locally.} It
 relies on the following lemmas.
\begin{lemma}
\label{prop:geometricformulae}
  Let  $\mathbb{T}$ be a geometric theory. For 
  any category   $\cat{C}$, there is an
  isomorphism of categories $\Cat{Mod}(\mathbb{T},\Sets^{\cat{C}}) \cong
  \Cat{Mod}(\mathbb{T},\Sets)^{\cat{C}}$.
\end{lemma}
  Here $\Cat{Mod}(\mathbb{T},\CT)$ is the category of
  $\mathbb{T}$-models in $\CT$.\footnote{This lemma is, in fact, valid for any topos $\mathcal{E}$ replacing \Sets; Johnstone's proof just relies on the
fact that the functor
  $\mathrm{ev}_C : \mathcal{E}^{\cat{C}} \to \mathcal{E}$ that evaluates at
  $C \in \cat{C}$ is (the inverse image part of) a geometric
  morphism. The stated generalization follows because the
  functor $(\mathrm{ev}_C)_* : \mathcal{E} \to \mathcal{E}^{\cat{C}}$ given by
  $(\mathrm{ev}_C)_*(S) = S^{\cat{C}(-,C)}$ determines the direct
  image part~\cite[Exercise~VII.10.1]{maclanemoerdijk92}.}
  This lemma may be found in~\cite[Corollary
  D.1.2.14]{johnstone02b}.
\begin{lemma}\label{invlemma}
The  lattice $L_A$  generating the
spectrum of an internal commutative C*-algebra $A$ is preserved under inverse
images of geometric
morphisms.
 \end{lemma}

 To prove the second lemma, we first use the  characterization of
the real part $A\sa$ of a commutative C*-algebra $A$ as an f-algebra over the rationals
(see Definition \ref{def:falgebra}). Moreover, the spectrum of a C*-algebra coincides with the spectrum of the f-algebra of
its self-adjoint elements~\cite{CoquandSpitters:cstar}.
We claim that the theory of
f-algebras is geometric. First, we observe that an f-algebra is precisely a
uniquely
divisible lattice ordered ring~\cite[p151]{coquand05},
since unique divisibility turns a ring into a $\Q$-algebra. The definition of
a lattice ordered ring is algebraic: it can be written using equations only.
The theory of torsion-free rings,
\ie\  $(nx=0\vdash_x x=0)$ for all $n>0$, is also algebraic.
The theory of divisible rings is obtained by adding infinitely many
geometric axioms $\top \vdash_x \exists_{y} n y=x$, one
for each $n>0$, to the algebraic theory of rings. A torsion-free divisible ring
is the same as a uniquely divisible ring: 
Suppose that $ny=x$ and $nz=x$, then $n(y-z)=0$, and so
$y-z=0$. We conclude that the theory of uniquely divisible lattice
ordered rings, \ie\ f-algebras, is geometric. In particular, $A\sa$ and hence $A^+$ are 
definable by a geometric theory. Secondly, the  relation $\approx$ in \er{Papp} is defined by an existential quantification, so that 
the generating lattice $A^+/\approx$ -- and hence by \er{Papp} also $L_A$ --
 is
preserved under inverse images of geometric morphisms.  This proves Lemma \ref{invlemma}.
\qed

Combining Lemma \ref{invlemma} with  Lemma \ref{prop:geometricformulae}, 
we obtain \er{Llocal} and hence Theorem \ref{thm:internalspectrum}.\qed

\medskip

For later we use, we put an important property of $L_A$ on record.
\begin{definition}\label{defnormal}
  A distributive lattice  is \emph{normal} if for all $b_1, b_2$ such that
  $b_1 \vee b_2 = \top$ there are $c_1, c_2$ such that $c_1 \wedge c_2 = \bot$
  and $c_1 \vee b_1 = \top$ and $c_2 \vee b_2 = \top$. A distributive lattice
is called \emph{strongly normal} if for all $a, b$ there exist $x, y$ such
that $a \leqslant b \vee x$ and $b \leqslant a \vee y$ and $x \wedge y = \bot
.$
\end{definition}
\begin{lemma}
  The lattice $L_A$ is strongly normal, and hence normal.
\end{lemma}
This lemma is due to Coquand \cite[Thm.\ 1.11]{coquand05}, but we give a proof. 
\begin{proof}
First,   every strongly normal lattice is normal.  To prove this,   let $b_1 \vee b_2 = \top$ and choose $x, y$ such that  $b_1 \leqslant b_2 \vee
  x$, $b_2 \leqslant b_1 \vee y$, and $x \wedge y = \bot .$ Then $\top
  \leqslant b_1 \vee b_2 \leqslant (b_2 \vee x) \vee b_2 = b_2 \vee x$.
  Similarly, $\top = b_1 \vee y$.

Second, to check that $L_A$  is strongly normal,  it is enough to verify the defining property on the generators $\prop{D}_a$.
So we pick $a,b$ in $A\sa$. Then one has $\prop{D}_a \leqslant \prop{D}_{a-b} \vee \prop{D}_b$,  $\prop{D}_b \leqslant \prop{D}_{b-a} \vee \prop{D}_a$, 
and  $\prop{D}_{a-b} \wedge \prop{D}_{b-a} = \bot$.
 \qed\end{proof}

We now turn to the relation \er{eq:continuity}, which is to be imposed on $L_A$. 
It turns out that \er{eq:continuity}
 is a special case of a relation that can be defined on any distributive lattice $L$ by $x\ll y$ iff there exists $z$ such that $x\wed z=\bot$ and $y\vee z=\top$.\footnote{Banaschewski and Mulvey write that $x$ is `rather below' $y$ \cite{banaschewskimulvey06}, whereas Johnstone \cite{johnstone82} says that  $x$ is `well inside' $y$. The notation $\ll$ is usually reserved for the so-called `way below' relation, but this relation coincides with the `well inside' relation on compact regular locales (see~\cite[p.303]{johnstone82} and Theorem~\ref{coverBas}), so we feel entitled to identify them notationally.}  
\begin{lemma}\label{l23}
For all $\prop{D}_a,\prop{D}_b\in L_A$, the following are equivalent:\footnote{In what follows, one may take $q>0$ either in $\mathbb{Q}$ or in $\R$.}
\begin{enumerate}
\item There exists  $\prop{D}_c\in L_A$ such that $\prop{D}_{c}\vee\prop{D}_{a}=\top$ and
$\prop{D}_{c}\wedge \prop{D}_{b}=\bot$;
\item There exists  $q>0$ such that $\prop{D}_{b}\leqslant \prop{D}_{a-q}$.
\end{enumerate}
\end{lemma}
\begin{proof}
$1\Rightarrow 2$: By \cite[Cor 1.7]{coquand05} there exists $q>0$ such
that $\prop{D}_{c-q}\vee \prop{D}_{a-q}=\top$. Hence $\prop{D}_{c}\vee \prop{D}_{a-q}=\top$, so
$\prop{D}_{b}=\prop{D}_{b}\wedge (\prop{D}_{c}\vee \prop{D}_{a-q}) = \prop{D}_{b}\wedge \prop{D}_{a-q}\leqslant \prop{D}_{a-q}$.

$2\Rightarrow 1$: Choose $\prop{D}_c:=\prop{D}_{q-a}$.
\qed\end{proof}

Hence in what follows we write
 \beq 
 \prop{D}_b\ll
\prop{D}_a \mbox{ iff } \exists_{q >0}\,  \prop{D}_b\leqslant \prop{D}_{a-q}, \label{ssnot}\eeq
and note with Coquand \cite{coquand05} that in view of the above lemma the relation  \er{eq:continuity}
 just states that the frame $\CO(\Sg)$ is regular.\footnote{See \cite[III.1.1]{johnstone82} for this notion. Recall that by the general theory of Banaschewski and Mulvey \cite{banaschewskimulvey06}, the spectrum has to be a compact regular frame.} This leads to the following description. 
  
  For any distributive lattice $L$,  an ideal $I\in \mathrm{Idl}(L)$ is called {\it  regular} if $I\supseteq\twoheaddownarrow x$ implies $x\in I$, 
 where 
 \beq \twoheaddownarrow  x=\{
  y\in L \mid y\ll
x \}. \label{sscover}\eeq
Expressed in logical language, $I$ is therefore a regular ideal if
 \beq \forall_{y\in L}\, (y\ll x\Raw y\in I)\Raw x\in I,\label{loglan}\eeq
 and hence one has the frame $\mathrm{RIdl}(L)$ of regular ideals of $L$, defined by 
 \beq \mathrm{RIdl}(L)= \{ U \in\mathrm{Idl}(L)\mid ( \forall_{y\in L}\, y \ll x \Raw y\in U)\Raw x\in U\}; \label{ofcourse2}
 \eeq
  for the sake of completeness, 
 $U \in\mathrm{Idl}(L)$ as a predicate on $\CP(L)$ stands for $\bot\equiv 0\in U$ and
\begin{eqnarray}
x\in U,\,  y\leqslant x &\Raw& y\in U; \label{id1} \\
x,y\in U &\Raw& x\vee y \in U. \label{id2}
\end{eqnarray}
Any ideal  $U \in\mathrm{Idl}(L)$ can be turned into a regular ideal $\mathcal{A}U$ by means of the closure operation
$\mathcal{A}:DL\raw DL$ defined by  \cite{coquand:entail}
 \beq
 \mathcal{A}U=\{x\in L\mid \forall_{y\in L}\, y\ll x\Raw y\in U\},\label{clCC}\eeq
and the canonical map $f:L\raw \mathrm{RIdl}(L)$ is given  in  terms of \er{clCC} by \er{canmap}. 

Combining Theorem 27 in \cite{coquand:entail} (which states that the regular ideals in a normal distributive lattice form a compact regular frame) with  Theorem 1.11 in \cite{coquand05} (which applies this to the case at hand), we finally obtain:
\begin{theorem}
The Gelfand spectrum $\CO(\Sg)$ of a commutative unital \ca\ $A$ is isomorphic (as a frame) 
to  the frame of all regular ideals of $L_A$, \ie\ 
 \beq \CO(\Sg)\cong \{ U \in\mathrm{Idl}(L_A)\mid ( \forall_{\prop{D}_b\in L_A}\, \prop{D}_b \ll \prop{D}_a \Raw \prop{D}_b\in U)\Raw \prop{D}_a\in U\}. \label{ofcourse}
\eeq
In this realization,  the  canonical map $f:L_A\raw \CO(\Sg)$ is  given by 
\beq f(\prop{D}_a)= \{\prop{D}_c\in L_A\mid \forall_{\prop{D}_b\in L_A}\,\prop{D}_b\ll\prop{D}_c\Raw \prop{D}_b\leqslant \prop{D}_a\}.\label{canK}
\eeq
\end{theorem}

By construction, we then have
\beq
f(\prop{D}_a)\leqslant \bigvee\{f(\prop{D}_{a-q})\mid q>0\}.\label{bycon}
\eeq
For later use, also note that \er{canK} implies
\beq  f(\prop{D}_a)=\top\;\LRaw  \prop{D}_a=\top. \label{lateruse}
\eeq

We may now  equip $L_A$ with the covering relation
defined by \er{unicover}, given
 \er{ofcourse} and the ensuing map \er{canK}.\footnote{Alternatively, writing $\prop{D}_a \drie_0 U$ iff  $U\supseteq \twoheaddownarrow   \prop{D}_a$,
the covering relation $\drie$ is inductively generated by
 $\drie_0$, as explained in \cite{Coq4,vickers06}.
The triple $(L_A,\leqslant,\drie_0)$ is a {\it flat site} as defined in \cite{vickers06}.}
Consequently, by Proposition \ref{preciseprop}
one has 
\beq
\CO(\Sg)\cong \CF(L_A,\drie). \label{vw0}\eeq
This description becomes computable by the following two results. 
\begin{theorem}\label{coverBas}
  In any topos, the covering relation $\drie$ on $L_A$ defined by \er{unicover} with \er{ofcourse} and \er{canK}   is given by
  $\prop{D}_a \drie U$ iff for all  $q>0$ there exists a (Kuratowski)  finite $U_0\subseteq U$ such that $\prop{D}_{a-q}\leqslant\bigvee U_0$. (If $U$ is directed, this means that there exists $\prop{D}_b\in U$ such that $\prop{D}_{a-q}\leqslant \prop{D}_b$.)
\end{theorem}
\begin{proof}
The easy part  is the ``$\Law$'' direction: from \er{bycon} and the assumption we have
 $f(\prop{D}_a)\leqslant\bigvee f(U)$ and hence $\prop{D}_a \drie U$ by definition of the covering relation.
 
 In the opposite direction, assume $\prop{D}_a \drie U$ and take some $q>0$. 
 From (the proof of) Lemma \ref{l23},
$\prop{D}_{a}\vee \prop{D}_{q-a}=\top$, hence $\bigvee f(U)\vee f(\prop{D}_{q-a})=\top$.
Since $\CO(\Sigma)$ is compact, there is a finite $U_0\subset U$ for which  $\bigvee f(U_0)\vee f(\prop{D}_{q-a})=\top$, so that by \er{lateruse} we have  $\prop{D}_b\vee \prop{D}_{q-a}=\top$, with $\prop{D}_b=\bigvee U_0$. By \er{eq:spectrum2} we have $\prop{D}_{a-q}\wed \prop{D}_{q-a}=\bot$, and hence $$
\prop{D}_{a-q}=\prop{D}_{a-q}\wed\top=\prop{D}_{a-q}\wed(\prop{D}_b\vee \prop{D}_{q-a})=\prop{D}_{a-q}\wed\prop{D}_b\leqslant \prop{D}_b=\bigvee U_0.$$ \qed
\end{proof}

Thus  we have two alternative expressions for the spectrum:
\begin{eqnarray}
\CO(\Sg)&\cong& \{ U \in\mathrm{Idl}(L_A)\mid  \forall_{q>0}\,  \prop{D}_{a-q}\in U\Raw \prop{D}_a\in U\}, \label{ofcourse4}\\
& \cong&  \{ U \in \CP(L_A)\mid \prop{D}_a\drie U\Raw \prop{D}_a\in U\}.  \label{ofcourse3}
\end{eqnarray}
The first follows from \er{ofcourse}, the second from \er{defCFL} and \er{vw0}. 

To apply this to our functor topos $\TA$, we apply  the Kripke--Joyal semantics for the internal language of the topos $\TA$ (see \cite[\S VI.7]{maclanemoerdijk92}, whose notation we will use, and \cite[\S 6.6]{borceux3}) to the  statement 
$\prop{D}_a \drie U$. This is a formula $\phi$ with two free variables, namely $\prop{D}_a$ 
 of type $L_A$, and $U$ of type $\CP(L_A)\equiv \Om^{L_A}$. Hence in the forcing statement 
 $C\Vdash \phi(\al)$ in $\TA$, we have to insert
 $$ \al\in (\underline{L}_{\uA}\x \underline{\Om}^{\underline{L}_{\uA}})(C)\cong L_C\x \mathrm{Sub}(\underline{L}_{\uA|\uparrow C}),$$
 where $\underline{L}_{\uA|\uparrow C}$ is the restriction of the functor $\underline{L}_{\uA}:\CA\raw\Sets$ to $\uparrow\!C\subset\CA$.
 Here we have used \er{Llocal}, as well as  the isomorphism  \cite[\S II.8]{maclanemoerdijk92}
 \beq \underline{\Om}^{\underline{L}_{\uA}}(C)\cong\mathrm{Sub}(\underline{L}_{\uA|\uparrow C}).
 \label{Yoneda}
 \eeq
 Consequently, we have $\al=(D_c,\underline{U})$, where $D_c\in L_C$ for some $c\in C\sa$ (note the change of typefont between the formal variable
 $\prop{D}_a$ and the actual element $D_c$) and $\underline{U}:\, \uparrow\!C\raw\Sets$ is a subfunctor of 
 $\underline{L}_{\uA|\uparrow C}$. In particular, $\underline{U}(D)\subseteq L_D$ is defined whenever $D\supseteq C$, and
 the subfunctor condition on $\underline{U}$ simply boils down to $\underline{U}(D)\subseteq \underline{U}(E)$
 whenever $C\subseteq D\subseteq E$.
\begin{corollary}\label{coverBas2}
In the topos $\TA$  the cover $\drie$ of Theorem \ref{coverBas} may be computed locally, in the sense that  for any 
$C\in \CA$, $D_c\in L_C$ and $\underline{U}\in \mathrm{Sub}(\underline{L}_{\uA|\uparrow C})$,
one has\begin{center}
$C\Vdash \prop{D}_a \drie {U}(D_c,\underline{U})$ iff  $D_c \drie_C \underline{U}(C)$,
\end{center}
in that for all  $q>0$ there exists a finite $U_0\subseteq \underline{U}(C)$ such that ${D}_{c-q}\leqslant\bigvee U_0$.
  \end{corollary}  
\begin{proof}
For simplicity,   assume  that $\bigvee U_0\in U$, so that we may replace $U_0$ by $\prop{D}_b=\bigvee U_0$; the general case is analogous. We then have to inductively analyze the formula $ \prop{D}_a \drie U$, 
 which, under the stated assumption,  in view of Theorem \ref{coverBas} may be taken to mean
\beq
 \forall_{q>0}\, \exists_{\prop{D}_b\in L_A}\, (\prop{D}_b\in U\wed  \prop{D}_{a-q}\leqslant \prop{D}_b). \label{meaning}
\eeq
We now  infer from the rules for  Kripke--Joyal semantics in a functor topos that:\footnote{The first one follows from 
\cite[Prop.\ 6.6.10]{borceux3} and a routine computation. The others are obvious from either  \cite[\S VI.7]{maclanemoerdijk92}
or \cite[\S 6.6]{borceux3}.}
 \begin{enumerate}
\item  $C\Vdash(\prop{D}_a\in U)(D_c,\underline{U})$ iff   for all $D\supseteq C$ one has $D_c\in \underline{U}(D)$; 
since $\underline{U}(C)\subseteq \underline{U}(D)$, this happens to be the case iff
$D_c\in \underline{U}(C)$.
\item $C\Vdash(\prop{D}_b\leqslant \prop{D}_a)(D_{c'},D_c)$ iff $D_{c'}\leqslant D_c$ in $L_C$.
\item $C\Vdash(\exists_{\prop{D}_b\in L_A}\, \prop{D}_b\in U\wed  \prop{D}_{a-q}\leqslant \prop{D}_b)(D_c,\underline{U})$
iff there is $D_{c'}\in \underline{U}(C)$ such that $D_{c-q}\leqslant D_{c'}$.
\item  $C\Vdash(\forall_{q>0}\, \exists_{\prop{D}_b\in L_A}\, \prop{D}_b\in U\wed  \prop{D}_{a-q}\leqslant \prop{D}_b)(D_c,\underline{U})$ iff for all $D\supseteq C$ and all $q>0$ there is $D_d\in \underline{U}(D)$ such that $D_{c-q}\leqslant D_d$,
where  $D_c\in L_C$ is seen as an element of $L_D$ through the injection $L_C\hookrightarrow L_D$ of 
Theorem \ref{thm:internalspectrum}, and  $\underline{U}\in \mathrm{Sub}(\underline{L}_{\uA|\uparrow C})$
is seen as an element of $\mathrm{Sub}(\underline{L}_{\uA|\uparrow D})$ by restriction.
 This, however,  is true at all $D\supseteq C$ iff it is true at $C$, because $\underline{U}(C)\subseteq \underline{U}(D)$ and hence one can take $D_d=D_{c'}$ for the $D_{c'}\in L_C$ that makes the condition true at $C$. \qed
 \end{enumerate}
\end{proof}
This brings us to our recipe for computing the spectrum in $\TA$  locally:
\begin{theorem}\label{finaltheorem}
The spectrum $\CO(\ulS)$ of $\uA$ in $\TA$ can be computed as follows:
\begin{enumerate}
\item At $C\in\CA$, the set $\CO(\ulS)(C)$ consists of those subfunctors
$\underline{U}\in\mathrm{Sub}(\underline{L}_{\uA|\uparrow C})$  such that  for all $D\supseteq C$ and all $D_d\in L_D$ one has $D_d\drie_D \underline{U}(D)\Raw D_d\in \underline{U}(D)$.  
\item In particular, at
 $\C\cdot 1$, the set $\CO(\ulS)(\C\cdot 1)$ consists of those subfunctors
$\underline{U}\in\mathrm{Sub}(\underline{L}_{\uA})$  such that  for all $C\in\CA$ and all $D_c\in L_C$ one has $D_c\drie_C \underline{U}(C)\Raw D_c\in \underline{U}(C)$.
\item The condition that $\underline{U}=\{\underline{U}(C)\subseteq L_C\}_{C\in\CA}$ be a subfunctor of $\underline{L}_{\uA}$ comes down to the requirement that $\underline{U}(C)\subseteq \underline{U}(D)$ whenever $C\subseteq D$.
\item The map $\CO(\ulS)(C)\raw \CO(\ulS)(D)$ given by the functoriality of $\CO(\ulS)$ whenever $C\subseteq D$ is given by truncating an element $\underline{U}:\uparrow\!C\raw\Sets$ of $\CO(\ulS)(C)$ to $\uparrow\!D$.
\item The external description of  $\CO(\ulS)$ is the frame map
\beq
 \pi_{\Sg}^*:\CO(\CA)\raw \CO(\ulS)(\C\cdot 1), \label{JTmap}
 \eeq
given  on the basic opens $\uparrow\! D\in \CO(\CA)$  by
\begin{eqnarray}
 \pi_{\Sg}^*(\uparrow\! D)=\ch_{\uparrow D}&:&  E\mapsto \top\:\: (E\supseteq D);\nn \\
& & E\mapsto \bot\:\: (E\nsupseteq D),\label{defpi}
\end{eqnarray}
where the top and bottom elements $\top,\bot$ at $E$ are given by $\{L_E\}$  and $\emptyset$, respectively. 
\end{enumerate}
\end{theorem}
\begin{proof}
By \er{ofcourse3}, $\CO(\ulS)$ is the subobject of $\underline{\Om}^{\underline{L}_{\uA}}$
defined by the formula $\phi$ given by
\beq  \forall_{\prop{D}_a\in L_A}\, \prop{D}_a\drie U\Raw \prop{D}_a\in U,\label{90}\eeq
whose interpretation in $\TA$ is an arrow from $\underline{\Om}^{\underline{L}_{\uA}}$ to $ \underline{\Om}$.
In view of  \er{Yoneda}, we may identify an element $\underline{U}\in \CO(\ulS)(C)$ with a subfunctor of $\underline{L}_{\uA|\uparrow C}$, and by \er{90} and Kripke--Joyal semantics
in functor topoi (see, in particular, \cite[\S IV.7]{maclanemoerdijk92}), we have 
$\underline{U}\in \CO(\ulS)(C)$  iff  $C\Vdash \phi(\underline{U})$, with 
$\phi$ given by \er{90}. UnfoldiIng this using the rules for Kripke--Joyal semantics and using
Corollary \ref{coverBas2} (including part 1 of its proof), we find that 
$\underline{U}\in \CO(\ulS)(C)$ iff
\beq  \forall_{D\supseteq C}\, \forall_{D_d\in L_D}\,  \forall_{E\supseteq D}\, D_d\drie_E \underline{U}(E)\Raw D_d\in \underline{U}(E),\label{strong}\eeq
where $D_d$ is regarded as an element of $L_E$.
This condition, however, is equivalent to  the apparently weaker condition
\beq
\forall_{D\supseteq C}\, \forall_{D_d\in L_D}\,D_d\drie_D \underline{U}(D)\Raw D_d\in \underline{U}(D);\label{weak}
\eeq
 condition \er{strong}  clearly implies \er{weak}, but the latter applied at $D=E$ actually  implies the first, since $D_d\in L_D$ also lies in $L_E$.

Items 2 to 4 are now obvious, and   the last 
follows by the explicit prescription for the external description of frames recalled in 
in Subsection \ref{Intext}. Note that each $\CO(\ulS)(C)$ is a frame in \Sets, inheriting  the frame structure of the ambient frame $\mathrm{Sub}(\underline{L}_{\uA|\uparrow C})$.
\qed
 \end{proof}
 
 An equivalent way to compute the spectrum, which derives from \er{ofcourse4} rather than \er{ofcourse3},  is as follows:  $\CO(\ulS)(\C\cdot 1)$ (and similarly all the other $\CO(\ulS)(C)$) consists of those subfunctors
$\underline{U}\in\mathrm{Sub}(\underline{L}_{\uA})$  such that  for all $C\in\CA$,
 $\underline{U}(C)$ is a regular ideal in $L_C$. 
 
 To prove this, according to \er{ofcourse4}   the formula expressing that $U\in \CP(L_A)$ be a regular ideal is 
 \beq U\in \mathrm{Idl}(L)\wed \forall_{\prop{D}_a\in L_A}\, 
\forall_{q>0}\,  \prop{D}_{a-q}\in U \Raw\prop{D}_{a}
\in U,\label{laatste}\eeq
where the  condition $U\in \mathrm{Idl}(L)$ is spelled out in \er{id1} and \er{id2}. The locality of this first condition and of the conjunction in \er{laatste} being almost trivial, we concentrate on the second term, calling it $\phi$ as usual. We then find that
$\C\cdot 1\Vdash \phi(\underline{U})$ 
  iff  for all $C\in\CA$,  all $D_c\in \underline{U}(C)$, and all $D\supseteq C$ one has:
   if $D_{c-q}\in \underline{U}(E)$ for  all $q>0$ at all $E\supseteq D$, then $D_c\in \underline{U}(D)$. Now the antecedent automatically holds at all $E\supseteq D$ iff it holds at $D$, and similarly the
if \ldots then statement holds  at all $D\supseteq C$ if it holds at $C$.
\subsection{\emph{Daseinisation} map}
\label{AppA:Das}
Our next aim is to construct the {\it Daseinisation} map \er{Da1}, which, read as a frame map,
for fixed $a\in A\sa$ is
\beq
\underline{\dl}(a)\inv:  \CO(\underline{\IR}) \raw \CO(\ulS).\label{Da2}\eeq
We will  use the realization \er{vw0} of the spectrum $\CO(\ulS)$ of $\uA$
as the frame $\CF(\underline{L}_{\uA},\underline{\drie})$ defined in the preceding subsection.
The second frame we deal with is that of the interval domain $\CO(\IR)$, {\it cf.}\ Subsection \ref{subsec:ID}. Following \cite{negri}, we construct the
 interval domain as a frame $\CF(\Q\x_{<}\Q,\coveredd)$ defined by a
 covering relation.
  Here the pertinent meet semilattice $\Q\x_{<}\Q$
 consists  of pairs $(p,q)\in\Q\x\Q$ with $p<q$, ordered by inclusion
 (\ie\ $(p,q)\leqslant (p',q')$ iff $p'\leqslant p$ and $q\leqslant q'$), with a bottom element $\bot$ added.
  The
 covering relation $\coveredd$ is defined by $\bot \coveredd U$ for all $U$ and
$(p,q) \coveredd U$ iff for all rational $p',q'$ with $p<p'<q'<q$ there exists $(p'',q'')\in U$ with
 $(p',q') \leqslant (p'',q'')$.
In \Sets\ one easily verifies the frame isomorphism
\beq \CF(\Q\x_{<}\Q,\coveredd)\cong\CO(\IR),\label{negriscott}\eeq
so that, in particular, we may regard $\CO(\IR)$ as a subset of the power set $\CP(\Q\x_{<}\Q)$.
\begin{proposition}\label{IRext}
The functor $\CO(\uIR)$ internalizing the interval domain in $\TA$ is given by
\beq
\CO(\uIR) \cong \CF(\underline{\Q\x_{<}\Q},\underline{\coveredd}).
  \label{negriscott2}
\eeq
Explicitly, we have
\beq \CO(\uIR)(\C\cdot 1)\cong \{\underline{S}\in\mathrm{Sub}(\underline{\Q\x_{<}\Q})\mid
  \underline{S}(C)\in \CO(\IR)\, \mbox{ for all } C\in\CA\},\label{IRframe}
  \eeq
where $\CO(\IR)\subset \CP(\Q\x_{<}\Q)$ through \er{negriscott}, as just explained. 
 Furthermore, $\CO(\uIR)(C)$ is the truncation of \er{IRframe} to $\uparrow\! C$ (cf.\ Theorem \ref{finaltheorem}), and the functorial map
 $\CO(\uIR)(C)\raw \CO(\uIR)(D)$ whenever $C\subseteq D$ is given by truncation.
Finally, the external description of $\CO(\uIR)$ is given by the frame map 
\beq
 \pi_{\IR}^*:\CO(\CA)\raw \CO(\uIR)(\C\cdot 1), \label{JTmaps}
 \eeq
where $\pi_{\IR}^*$ is given by a  formula similar to \er{defpi}.
\end{proposition}
\begin{proof}
This follows from a computation analogous to but simpler than the proof of 
Theorem \ref{finaltheorem}, combined with the remark following \er{ddrr}
and the observation that the condition that $\underline{S}:\CA\raw\CP(\Q\x_{<}\Q)$ 
in the right-hand side of \er{IRframe} be a subfunctor of  $\underline{\Q\x_{<}\Q}$ means that we may identify $\underline{S}$ with a monotone function from $\CA$ to $\CO(\IR)$. \qed
\end{proof}
We now give the external description
\beq
\dl(a)\inv:  \CO(\underline{\IR})(\C\cdot 1) \raw \CO(\ulS)(\C\cdot 1)\label{Da3}\eeq
of our \emph{Daseinisation} map \er{Da2}. 
 In view of~\er{vw0} and~\er{negriscott2}, we will define \er{Da3} as a frame map
\beq
  \dl(a)\inv:
  \CF(\underline{\Q\x_{<}\Q},\underline{\coveredd})(\C\cdot 1)\raw
  \CF(\underline{L}_{\uA},\underline{\drie})(\C\cdot 1).
  \label{daF}
\eeq
Internalizing Proposition \ref{approxmap} to $\TA$,
 we proceed by constructing a continuous map
\beq
  \underline{d}(a): (\underline{L}_{\uA}, \underline{\drie}) \to
  (\underline{\Q\x_{<}\Q},\underline{\coveredd}),
  \label{dwhole}
\eeq
for in that case we may put
\beq
  \dl(a)\inv=\CF(\underline{d}(a))(\C\cdot 1). \label{Fd}
\eeq
By definition, as
 a map in the functor topos $\TA$ the continuous map  $\underline{d}(a)$   is a natural transformation
  \beq \underline{d}(a)^*:\underline{\Q\x_{<}\Q}\raw\underline{\CP(L_A)}=\underline{\Om}^{\underline{L}_{\uA}}
  \label{dastar}
  \eeq
 with components $\underline{d}(a)^*_C: \underline{\Q\x_{<}\Q}(C) \raw \underline{\Om}^{\underline{L}_{\uA}}(C)$.
One has $\underline{\Q\x_{<}\Q}(C)\cong \Q\x_{<}\Q$, so by \er{Yoneda}
 the $\underline{d}(a)^*_C$ are maps
\beq 
\underline{d}(a)^*_C: \Q\x_{<}\Q\raw\mathrm{Sub}(\underline{L}_{\uA|\uparrow C}).
\label{fips}
\eeq
 By naturality,  $\underline{d}(a)^*_C$ is determined by
$\underline{d}(a)^*_{\C\cdot 1}: \Q\x_{<}\Q\raw\mathrm{Sub}(\underline{L}_{\uA})$ as
\beq  \underline{d}(a)^*_C(r,s)(D)=\underline{d}(a)^*_{\C\cdot 1}(r,s)(D), \label{allC}\eeq
for all  $D\supseteq C$, so $\underline{d}(a)^*$ is determined by 
$\underline{d}(a)^*_{\C\cdot 1}$. Using the description of the
 lattice $\underline{L}_{\uA}(C)$  by Theorem \ref{thm:internalspectrum}, we may now define
\beq
\underline{d}(a)^*_{\C\cdot 1} (r,s):C \mapsto \{{D}_{f-r} \wedge
 {D}_{s-g}\mid f,g\in C\sa, f \leqslant a \leqslant g\},
   \label{defdapp}
\eeq
which is indeed a subset of $\underline{L}_{\uA}(C)=L_C$, as required. 
\begin{lemma}\label{lemmal}
The map \er{dwhole} defined by \er{dastar}, \er{allC} and 
 \er{defdapp} is  continuous (in the sense of Definition \ref{def:contmap}, internalized to $\TA$). 
 \end{lemma}
 \begin{proof}
First, we claim that $\underline{d}$ is continuous
iff each $\underline{d}(a)_C$ is. Indeed, with regard to
 the first condition in Definition \ref{def:contmap} this is obvious;
for the second {\it cf.}\ \cite[Prop.\ I.8.5]{maclanemoerdijk92}, and for the third this is true because
both covering relations are described locally in $C$ ({\it cf.}\ Corollary  \ref{coverBas2}).
By Proposition \ref{approxmap},  continuity of $\underline{d}$, in turn, would mean that \er{daF} is well defined as a frame map. 

Thus what remains is to verify that each map  $\underline{d}(a)_C$ is continuous in the sense
of  Definition \ref{def:contmap}. This is indeed the case; we spare the readers the details.\footnote{These will appear in the PhD Thesis of the first author.}  \qed
\end{proof}

We now compute the associated frame map \er{daF}.
The map \er{dastar} induces a map $\mathrm{Sub}(\underline{\Q\x_{<}\Q})\raw\mathrm{Sub}(\underline{\Om}^{\underline{L}_{\uA}})$ as the left adjoint of the pullback in the opposite direction
 (see, e.g., \cite[Exercise I.10]{maclanemoerdijk92}), which by composition with $\cup$ yields a map  $\mathrm{Sub}(\underline{\Q\x_{<}\Q})\raw\mathrm{Sub}(\underline{L}_{\uA})$. The latter restricts to a map $ \CF(\underline{\Q\x_{<}\Q},\underline{\coveredd})(\C\cdot 1)\raw
  \CF(\underline{L}_{\uA},\underline{\drie})(\C\cdot 1)$, which by definition is the map \er{daF}
  and hence gives the external description \er{Da3} of our {\it Daseinisation} map.\footnote{According to \er{apeq}, this map
   is just the component of $\CF(\underline{d}(a))$ at $\C\cdot 1$. This component, however, determines $\CF(\underline{d}(a))$ as a whole, since $\CF(\underline{d}(a))(C)$ is just the restriction
of  $\CF(\underline{d}(a))(\C\cdot 1)$ to  the truncation of each subfunctor $\underline{S}$ in \er{IRframe}  to $\uparrow\! C$.} This is a frame map by construction; see
Lemma \er{lemmal} and Proposition \er{approxmap}.  The associated 
 locale map $\dl(a):\ulS\raw\uIR$  is our version of the D\"{o}ring--Isham {\it Daseinisation} map. It is unenlightening to write it down explicitly, but we give an appealing special case in Subsection \ref{sec:Das}.
\subsection{Localization of integrals}
Finally, to compute the interpretation of the locale of integrals we
may proceed analogously to  the case of the spectrum.
 The free distributive lattice satisfying the relations in
Section~\ref{ss:th-integrals} may alternatively be defined by an entailment
relation~\cite{coquand:entail}. Consequently, it suffices to describe when
$\wedge A \vdash \vee B$ in the lattice. As proved
in~\cite{coquand05, Coquand/Spitters:integrals-valuations}, this holds if
a positive combination of elements in $A$ is below a positive combination of
elements in $B$ - in symbols, if there are $r_i,s_j>0$ and $a_i$ in
$A$ and $b_j$ in $B$ such that $\sum r_i a_i \leqslant \sum s_j b_j$. This
is an existential quantification over finite subsets of an
f-algebra. The construction of taking the (Kuratowski) finite
powerset is geometric, see \eg\ \cite{Vic:LocTopSp}. So existential
quantification over it is preserved by geometric morphisms. Applying this to the
internal C*-algebra and applying Lemma~\ref{prop:geometricformulae} we
obtain ({\it cf.}\ Theorem \ref{thm:internalspectrum}):
\begin{proposition}
  The interpretation of the lattice generating the locale of integrals
  of the internal C*-algebra is given by the functor assigning to
  each commutative subalgebra $C$ the lattice generating the integrals
  on $C$. If $C\subseteq D$, then the inclusion maps generators of the
  lattice for $C$ to generators of the lattice for $D$ and preserves
  relations. The covering relation for the space of integrals is also
  interpreted locally.
\end{proposition}

A similar statement holds for
valuations;
see~\cite{vickersintegration,Coquand/Spitters:integrals-valuations}.
Vickers~\cite{PPExp} uses a presentation of locales which is similar to formal
topology, but which is tailored for geometric reasoning.

\section{Related work}
\label{sec:doringisham}
The present article was to a considerable extent motivated by the fundamental work of 
Butterfield and Isham \cite{butterfieldisham1,butterfieldisham2}
and D\"{o}ring and 
Isham~\cite{doringisham1,doringisham2,doringisham3,doringisham4,doeringisham:review}. We
refrain from a full comparison,  but restrict ourselves to what we see as the  key points.

As to Butterfield and Isham, our reformulation of the Kochen--Specker Theorem is in their spirit, but we feel our version is more powerful, especially from a logical perspective: our statement that a certain locale has no points has a logical interpretation in terms of (the lack of) models of a certain geometric theory, whereas the original reformulation \cite{butterfieldisham1} merely claims that some presheaf lacks global sections (i.e.\ points). 

Compared with D\"{o}ring and Isham,  our overall programme and philosophy, as explained in the Introduction, are quite different
from theirs:   our ambitions are limited to finding a spatial notion of quantum logic (although we do hope that locales in topoi might provide a generalized concept of space that will be useful in quantum gravity). The principal technical differences between the two approaches lie in
  our use of:
  \begin{enumerate}
  \item covariant functors (instead of contravariant ones);
\item  C*-algebras (instead of
von Neumann algebras);
\item
 locales (instead of Stone spaces);
\item  internal reasoning and the associated use of Kripke--Joyal semantics;
\item states as internal integrals and the correspondence between integrals and valuations (\ie\ measures defined on open sets).\footnote{An analogous external result has meanwhile been found by D\"{o}ring \cite{Doeringmeasure}.}
\end{enumerate}
 This has many technical advantages, which has made it possible to obtain our main results
(see Subsection \ref{mainresults}). 
Conceptually, the two programs in question overlap to the effect that the Gelfand spectrum
$\CO(\underline{\Sigma})$ of the Bohrification $\uA$ of $A$ provides a pointfree realization of 
D{\"o}ring and Isham's notion of a  \emph{state object} in a topos, whereas the  interval domain 
$\CO(\underline{\IR})$ realizes their \emph{quantity object}, again in the sense of  pointfree topology internalized to a suitable topos.\footnote{In fact, our use of pointfree techniques leads to topoi of covariant functors just as inevitably as the more conventional methods in
\cite{butterfieldisham1,butterfieldisham2, butterfieldisham3,doringisham1,doringisham2,doringisham3,doringisham4}
lead to topoi of presheaves.}
These objects are linked by {\it observables}, which define arrows from
the state object to the quantity object. Thus for each $a\in A\sa$, our  {\it Daseinisation} $\dl(a): \ulS\raw\underline{\IR}$ is an observable
in the sense of D\"{o}ring and Isham.  
Restricted to the special case $A=B(H)$, our construction resembles the single example
of such a topos that both  Butterfield, Hamilton and Isham \cite{butterfieldisham3} and D{\"o}ring and Isham \cite{doringisham2,doringisham3} give, namely that of presheaves over the preorder category of commutative von  Neumann subalgebras of $B(H)$ (ordered by inclusion).

{\small
\bibliographystyle{plain}
\bibliography{cstartopos}
}
\end{document}